\documentclass[format=manuscript,screen=true,timestamp]{acmart} 

\AtBeginDocument{%
  \providecommand\BibTeX{{%
    \normalfont B\kern-0.5em{\scshape i\kern-0.25em b}\kern-0.8em\TeX}}}

\setcopyright{acmcopyright}
\acmYear{2022}
\acmDOI{XXXXXXX.XXXXXXX}

\usepackage{url}
\usepackage{lscape}
\usepackage{xspace}
\usepackage{multirow}
\usepackage{enumitem}
\usepackage{xcolor}
\usepackage{subfig}
\usepackage{amsfonts}
\usepackage{comment}
\DeclareUnicodeCharacter{00A0}{ }

\usepackage{longtable}
\usepackage{rotating}
\usepackage{caption}
\usepackage{booktabs}
\usepackage{graphicx}
\usepackage{afterpage}

\usepackage[acronym]{glossaries}

\newacronym{vae}{VAE}{Variational Auto-Encoder}
\newacronym{ae}{AE}{Auto-Encoder}
\newacronym{am}{AM}{Attention Mechanism}
\newacronym{tf}{TF}{Transformer}
\newacronym{gan}{GAN}{Generative Adversarial Network}
\newacronym{cl}{CL}{Contrastive Learning}
\newacronym{avcl}{AVCL}{Audio-Visual Correlation Learning}
\newacronym{ml}{ML}{Machine Learning}
\newacronym{al}{AL}{Adversarial Learning}
\newacronym{ai}{AI}{Artificial Intelligence}
\newacronym{mlp}{MLP}{Multilayer Perceptrons}
\newacronym{cnn}{CNN}{Convolutional Neural Network}
\newacronym{dnn}{DNN}{Deep Neural Networks}
\newacronym{mmml}{MMML}{Multimodal Machine Learning}
\newacronym{av}{AV}{Audio-Visual}
\newacronym{rl}{RL}{Representation Learning}
\newacronym{dl}{DL}{Deep Learning}
\newacronym{soa}{SOTA}{State-of-the-Art}
\newacronym{ed}{E-D}{Encoder-Decoder}
\newacronym{mfcc}{MFCC}{Mel Frequency Cepstral Coeficients}
\newacronym{dct}{DCT}{Discrete Cosine Transform}
\newacronym{lms}{LMS}{Log-Mel Spectrograms}
\newacronym{rnn}{RNN}{Recurrent Neural Networks}
\newacronym{lstm}{LSTM}{Long Short Term Memory Networks}
\newacronym{llm}{LLM}{Large Language Models}
\newacronym{ar}{AR}{Action Recognition}
\newacronym{nn}{NN}{Neural Networks}
\newacronym{cca}{CCA}{Canonical Correlation Analysis}
\newacronym{ce}{CE}{Cross-Entropy}
\newacronym{midi}{MIDI}{Musical Instrument Digital Interface}
\newacronym{bert}{BERT}{Bidirectional Encoder Representations from Transformers}
\newacronym{pca}{PCA}{Principal Component Analysis}
\newacronym{vivit}{ViViT}{Video Vision Transformer}
\newacronym{vit}{ViT}{Vision Transformer}
\newacronym{videomae}{VideoMAE}{Video Masked Autoencoder}
\newacronym{clip}{CLIP}{Contrastive Language Image Pretraining}
\newacronym{nce}{NCE}{Noise Contrastive Estimation}
\newacronym{mil-nce}{MIL-NCE}{Multiple Instance Noise Contrastive Estimation}
\newacronym{lsl}{LSL}{Lifted Structured Loss}
\newacronym{npl}{NPL}{N-Pair Loss}
\newacronym{ql}{QL}{Quadruple Loss}
\newacronym{tl}{TL}{Triplet Loss}
\newacronym{mms}{MMS}{Masked Margin Softmax}
\newacronym{sdr}{SDR}{Source-to-Distortion Ratio}
\newacronym{i}{Inf}{Informativity}
\newacronym{mrr}{MRR}{Mean Reciprocal Rank}
\newacronym{wer}{WER}{Word Error Rate}
\newacronym{ndb}{NDB}{Number of Statistically Different Bins}
\newacronym{fid}{FID}{Fréchet Inception Distance}
\newacronym{map}{mAP}{Mean Average Precision}
\newacronym{iou}{IoU}{Intersection-over-Union}
\newacronym{auc}{AUC}{Area Under the Curve}
\newacronym{ciou}{cIoU}{Consensus Intersection-over-Union}
\newacronym{stft}{STFT}{Short Time Fourier Transform}

\acmJournal{CSUR}

\begin{document}

\title[Deep Audio-Visual Correlation Learning: A Survey]{A Survey of Recent Advances and Challenges in Deep Audio-Visual Correlation Learning}

\author{Luís Vilaça}
\authornote{Luis was involved in this work during his internship from September 14, 2021 – March 13, 2022 and January 15, 2024, -July 12, 2024 in National Institute of Informatics (NII), Tokyo.}
\orcid{0000-0002-3640-7019}
\email{luis.m.salgado@inesctec.pt}
\affiliation{%
    \institution{INESC TEC, Porto,}
    \city{Porto}
    \country{Portugal}
}
\affiliation{%
    \institution{National Institute of Informatics}
    \city{Tokyo}
    \country{Japan}
}
\affiliation{%
    \institution{ISEP, Polytechnic of Porto, School of Engineering}
    \city{Porto}
    \country{Portugal}
}

\author{Yi Yu}
\authornote{Corresponding author.}
\orcid{0000-0002-0294-6620}
\email{yiyu@hiroshima-u.ac.jp}
\affiliation{%
    \institution{Graduate School of Advanced Science and Engineering, Hiroshima University}
    \city{Hiroshima}
    \country{Japan}
}
\affiliation{%
    \institution{National Institute of Informatics}
    \city{Tokyo}
    \country{Japan}
}

\author{Paula Viana}
\orcid{0000-0001-8447-2360}
\email{pmv@isep.ipp.pt, paula.viana@inesctec.pt}
\affiliation{%
    \institution{ISEP, Polytechnic of Porto, School of Engineering}
    \city{Porto}
    \country{Portugal}
}
\affiliation{%
    \institution{INESC TEC}
    \city{Porto}
    \country{Portugal}
}

\renewcommand{\shortauthors}{Luís Vilaça, et al.}

\begin{abstract}
    Audio-visual correlation learning aims to capture and understand natural phenomena between audio and visual data. The rapid growth of \acrlong*{dl} propelled the development of proposals that process audio-visual data and can be observed in the number of proposals in the past years. Thus encouraging the development of a comprehensive survey. Besides analyzing the models used in this context, we also discuss some tasks of definition and paradigm applied in AI multimedia. In addition, we investigate objective functions frequently used and discuss how audio-visual data is exploited in the optimization process, i.e., the different methodologies for representing knowledge in the audio-visual domain. In fact, we focus on how human-understandable mechanisms, i.e., structured knowledge that reflects comprehensible knowledge, can guide the learning process. Most importantly, we provide a summarization of the recent progress of \acrfull*{avcl} and discuss the future research directions.
\end{abstract}



\maketitle

\section{Introduction}
\label{section:introduction}

A wide variety of multimedia information and data such as image, text, visual data, and audio is aggregated on the Internet over time, bringing opportunities to build knowledge from the structure hidden in such heterogeneous data through \acrfull*{dl} methods. Understanding the knowledge and structure hidden in natural \acrfull*{av} phenomena requires the ability to process multiple signals that compose the concept of what we are experiencing.\par

Capturing latent semantics, i.e., correspondence or correlation, from different signals is complex due to the natural characteristics and temporal correlations occurring at ``high levels'' of perception \cite{yu2012automatic}. Thus, the complexity of the problem increases because deep models typically use low-level data. For instance, speech recognition models attempt to relate phonemes and visemes (lip pose and motions) using raw pixels and audio spectrograms/waveforms of their corresponding sounds. It is naturally complex to bridge them due to the difference in meaning between different representation spaces (semantic gap). All application use cases that we studied face this issue. In addition, the complexity also increases due to inconsistent distributions and heterogeneous representations, which create a gap between modalities. These two issues compromise the correlation of audio and visual data in common spaces \cite{shah2014advisor} and require multimedia technology and \acrfull*{ml} methods to extract semantic relations between them.\par

\begin{figure*}
	\centering
	\includegraphics[width=\textwidth]{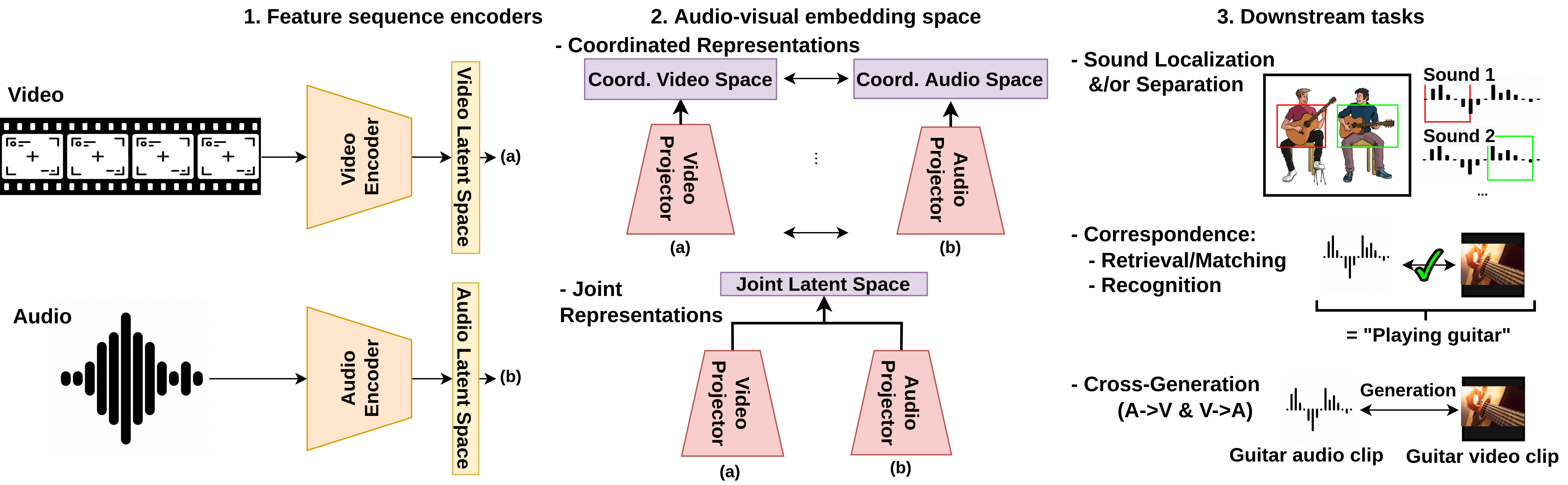}
	\caption{\acrlong*{av} correlation learning and subsequent downstream tasks: current proposals for \acrfull*{avcl} often use a similar pipeline but are not limited to it. Compressed latent representations are obtained using (pre-trained) feature extractors (encoders in orange) for audio (b) and visual data (a). Afterward, they embed representations into a joint space by coordination or fusion through separate projectors (mapping functions). The goal of these common latent spaces (in purple) is always to discriminate different semantics and maximize correlations between modalities.}
	\Description{\acrlong*{av} correlation learning and subsequent downstream tasks: current proposals for \acrfull*{avcl} often use a similar pipeline but are not limited to it. Compressed latent representations are obtained using (pre-trained) feature extractors (encoders in orange) for audio (b) and visual data (a). Afterward, they embed representations into a joint space by coordination or fusion through separate projectors (mapping functions). The goal of these common latent spaces (in purple) is always to discriminate different semantics and maximize correlations between modalities.}
	\label{fig:generic-approach}
\end{figure*}

\subsection{Problem Definition}
\label{section:problem-definition}

The emergence of \acrfull*{dl} and available audio-visual datasets has facilitated the development of \acrfull*{avcl}. \acrshort*{avcl} seeks to relate both audio and visual data through \acrfull*{dnn} into their feature sub-spaces\footnote{In our study we refer to representations as the embedded representations obtained from encoder models that can have one (feature vectors) or more dimensions (feature maps).} where the joint space composed by their aggregation preserves the shared information contained in each modality and emphasize their commonalities. In a nutshell, we are interested in data consisting of tuples ${x^{(n)} = (x_1^n, x_2^n)}_{n=1}^N$, where $x_1$ and $x_2$ are features extracted for audio and video respectively and $d_1$ and $d_2$ can be of arbitrary sizes and dimensions.
$$
	{x^{(n)} = (x_1^n, x_2^n)}_{n=1}^N, \quad {x_{1}} \in {\mathbb{R}^{d_{1}}}, {x_{2}} \in {\mathbb{R}^{d_{2}}}
$$
Representations are bridged into a common space $S$, which can be shared or not:
\begin{enumerate}
	\item Shared: $S_{12}$ is obtained through $h(S_1, S_2)$, which establishes the common and shared representation space between modalities.
	      $$
		      S_{12} = h(f_{1 \rightarrow 12}(x_1), g_{2 \rightarrow 12}(x_2))\\
	      $$
	\item Separate: $S_{1}$ and $S_{2}$ are obtained through $f(x_1; W_{1})$ and $g(x_2;W_{2})$.
	      $$
		      S_1 = f_{1 \rightarrow 12}(x_1) \quad S_2 = g_{2 \rightarrow 12}(x_2)\\
	      $$
\end{enumerate}

The weights of $f$, $g$, and $h$ are adjusted through back-propagation to maximize the correlation between modalities. The ``joint'' space can be obtained through feature fusion or coordination, where the final output consists of single or parallel representations. As illustrated in Figure \ref{fig:generic-approach}, all application scenarios propose to learn better representations for their specific goals. Usually, these proposals attempt to achieve their goals by constraining the outputs to have desired statistical properties or linking relevant features/subcomponents between them to synchronize semantic contexts \cite{MViewSurvey,MMMLSurvey}. Both feature fusion and coordination can leverage these methodologies, as described in this paper.\par

The learning problem can be established as a supervised, self-supervised, or unsupervised task. The selection of the most suitable learning framework depends on how the modalities, e.g. audio and video, are used and if auxiliary information is available or not. For instance, exploring relations within tuples $(x_{1}^{n_1}, x_{2}^{n_2})_{n_1 = n_2}$ and assuming each pair as having specific semantics, i.e. pseudo-labels, allows extracting correlations in a self-supervised way. In contrast, it can also be desirable to relate modalities as a whole $(x_{1}^{n_1}, x_{2}^{n_2})_{n_1 = n_2 \vee n_1 \neq n_2}$ and take into account supervised information (i.e., labels) regarding each pair of representations (audio and visual data). In the unsupervised setting, we observed many examples focusing on reconstruction between modalities (and modified versions of each modality), where the latent spaces can also be used for \acrshort*{avcl}. Additionally, we often use two terms relative to the learning process that help discriminate the type of relations extracted from both modalities: intermodal and intramodal. Intermodal refers to information shared between modalities, i.e., commonalities, while intramodal refers to specific information contained in each modality.\par

Within our research, we have observed that adversarial learning, reconstruction or weighted pooling processes through models such as \acrfull*{gan} \cite{zheng2021adversarial,seo2020hmtl}, \acrfull*{ae} \cite{recasens2021broaden,zhu2021learning} and models that explore multiple attention mechanisms \cite{zheng2021deep,min2021cross} have been investigated in recent years to improve the capability of \acrshort*{avcl} methods. However, the current panorama of the \acrshort*{soa} is saturated with attention-based methods (i.e., transformers) due to its excellent performances in large-scale scenarios. Therefore, we find it necessary to review not only recent progress on large-scale attention-based models but also provide a systematic view of the field of \acrshort*{avcl}.\par

\subsection{Contribution and Relevance to existing surveys}
\label{subsection:surveys}

Our research analyzed numerous published surveys focusing on \acrshort*{av} processing and multimodal learning. The following sections provide a brief analysis of them and summarize this article's contributions.\par

\subsubsection{\textbf{Analysis of existing surveys}}

Existing surveys that cover \acrshort*{av} processing introduce generic applications, core ideas, and theoretical concepts of \acrfull*{mmml} \cite{MViewSurvey,MMMLSurvey,liang2022foundations}. \cite{summaira2021recent} reviews the \acrshort*{mmml} field from the application level and lacks descriptive content for each method. More recently, a survey by Liang \textit{et. al.} \cite{liang2022foundations} proposed a taxonomy for the main research challenges within \acrshort*{mmml} with evidences from \acrshort*{avcl} but without focusing entirely on audio-visual data.\par

Dedicated to \acrshort*{av} learning we analyzed the following works: \cite{DAVLearningSurvey,wei2022learning,shi2021survey,ngiam2011multimodal}. \cite{ngiam2011multimodal} provides a benchmark study of various approaches for \acrshort*{av} speech recognition. Specifically, the authors emphasize the impact of isolated modalities in the shared representation scenario, i.e., what happens when one modality is unavailable. Nevertheless, they only cover one application use case, and the paper is rapidly getting outdated (the field made advances since then). More recently, \cite{DAVLearningSurvey} and \cite{wei2022learning} provide new taxonomies and extensively cover the field, but their analyses only include overviews at an application level without detailing how they are solved. In some cases, works can fit the descriptions of multiple categories in the proposed taxonomies \cite{wei2022learning}, which fails at grouping different methodologies. We believe this promotes the need for a study that compares and categorizes the different methodologies used in \acrshort*{av} learning over all its application use-cases. This is relevant because methods that target different applications have similar properties. Thus, their similarities and differences should be highlighted and discussed. While existing surveys are valuable, the emergence of a large volume of audio-visual content learning methods makes it necessary to conduct a comprehensive comparison of these works, highlighting their similarities and differences. Our survey fills this gap by offering a comprehensive study of deep correlation learning between audio and visual contents. This is the main contribution of our survey paper compared with others available.\par

We also reviewed survey papers that focus on different methodologies for representing knowledge in multimodal domains \cite{yang2021multiple,liang2022foundations}. The combination of multiple modalities regarding depth of concept relationships and raw signal perception is discussed. This is relevant for our study since we aim to cover how knowledge is represented and correlated in the \acrshort*{av} domain. This is related to the research challenges introduced by \cite{liang2022foundations}.\par

\cite{yang2021multiple} presents a taxonomy for the field, dividing it into deep and canonical knowledge representation. The first covers deep models for encoding raw signals into abstract representations. The latter assumes highly abstract concepts as inputs and concentrates on methods for extracting relationships from and among them. The authors also discuss the lack of explainability of deep models used for learning representations. This is relevant for current methodologies of the \acrshort*{soa} due to the number of proposals that rely on self-supervised learning. The authors believe that explainability can be relieved by introducing comprehensible knowledge in the learning process (i.e., similar to structured knowledge used in self-supervised learning settings). Moreover, it can also provide the means to evaluate the proposed models thoroughly.


\subsubsection{\textbf{Contributions}}

We address the natural complementary relationship between vision and audio that makes them suitable for exploiting underlying semantics. By analyzing the latest works in the field, we identified that the proliferation of attention-based models and objective functions based on self-supervised learning have provided a new bloom for the field, as can be observed by the high performances demonstrated in Table \ref{tab:methods-final}. Therefore, this survey paper focuses on \acrshort*{avcl} methodologies regarding feature extraction, objective functions, used datasets, and evaluation metrics. We propose a structure of the field that systematically groups proposals according to the methods used to correlate audio-visual representations. We aim to study how knowledge is represented and leveraged between audio and visual data on all application use cases illustrated in Figure \ref{fig:generic-approach}. Specifically, we review the models, objective functions/learning settings, and datasets. The main contributions of our article are listed below:

\begin{enumerate}
	\item Systematic categorization of recent proposals according to the models and learning frameworks used. We believe this could support new researchers in quickly studying how audio-visual data and their respective semantics can be exploited and extended for their research projects.
	\item The different techniques for \acrshort*{avcl} are categorized using brief details of their methodologies (and corresponding references). This can help new researchers to \acrshort*{avcl} to have a general view of the field.
	\item We provide discussions of benchmark datasets and a global list of all datasets used for \acrshort*{av} processing.
	\item We cover how structured knowledge (e.g., proxy tasks for self-supervised frameworks) is injected in \acrshort*{avcl} models and discuss future alternatives and challenges for the field.
\end{enumerate}

\section{Feature Extraction}

This section provides a brief analysis and discussion of the most used methods for extracting features from raw \acrshort*{av} data, i.e., ${x_{a}}$ and ${x_{v}}$. Since \acrshort*{avcl} uses these descriptions/features to extract knowledge from paired data, we believe it is essential to understand how \acrshort*{av} information is encoded.\par

\subsection{Audio Encoding}
\label{subsection:audio-feature-extrc}

From the taxonomy of \cite{audiosignals2019}, audio signals can be divided into three categories due to their very distinct characteristics (i.e., frequency range and envelope):
\begin{enumerate}
    \item Speech: characterized by a smooth envelope, i.e., transition between audio events, and is mostly present in an audible range from 100Hz to 7kHz;
    \item Music: exhibits dependencies between its simultaneous sound sources in time and frequency (their occurrence follows constraints in time and frequency) and mostly provides repetitive patterns in an audible range that can go from 40Hz up to 19.5kHz;
    \item Ambient Sounds: characterized by multiple independent sound sources, with or without periodicity. Some sounds have representations sparse in time and frequency due to the lack of periodicity. In contrast with the other two scenarios, ambient sounds are the most variable and have a broad audible range covering all others.
\end{enumerate}

As such, audio representations used for \acrshort*{avcl} must capture structures and dependencies in time and frequency. \acrshort*{soa} models for deep audio analysis use intermediate representations as input \cite{audiosignals2019,panns2020}. This pre-processing step provides better descriptions of characteristics in both dimensions and matches the input requirements, i.e., input shape, of these models. The most used representations are:\par

\begin{itemize}
    \item Spectrograms: compressed representations in the time-frequency domain (i.e., heterogeneous dimensions). They have the downside of losing phase, creating challenges for reconstruction;
          \begin{itemize}
              \item \acrfull*{mfcc}: Magnitude/power projected to a reduced set of frequency bands, converted to a log scale (perceptually relevant audio scale), and compressed with the \acrfull*{dct};
              \item \acrfull*{lms}: Same processing steps as \acrshort*{mfcc}, but without compression from the \acrshort*{dct}; 
              \item Constant-Q: High and low frequencies have lower and higher resolution (bandwidth), respectively; Reduces calculations in higher frequencies due to the reduction in frequency resolution;
          \end{itemize}
    \item \acrfull*{midi}: standard used to encode symbolic music.
\end{itemize}


\acrshort*{soa} models can use raw waveforms (AW), increasing computation and data needs. These models need to extract features from long data sequences without sampling, leading to spectrogram-based representations. These compress temporal information but lose some details. LMS is favored over noise-sensitive MFCCs within these representations \cite{tyagi2005desentitizing}. MFCCs are obtained via DCT and a filtering process, which selects cepstral coefficients relevant for speech audio but its dependent on the spectral form \cite{tyagi2005desentitizing}.\par


We selected AudioSet \cite{audioset2017} as our baseline dataset for comparing audio encoding models, due to its popularity. In Section \ref{appendix:audioset-proposals} we present its test performances for AudioSet. We categorized feature extraction methods by learning type and intermediate representations used, and detailed the spectrogram's computation parameters, audio embedding size, and kernel size and stride of the first layer for the audio waveform. This information aids in understanding the problem complexity through time-frequency representation dimensions.\par

1-D \acrshort*{cnn}s can process raw waveforms without performance advantages over other methods that process 2D inputs, like \acrshort*{lms} \cite{panns2020}. The rise of 2D \acrshort*{cnn}s at ImageNet has popularized \acrshort*{lms} due to its structure. It's a 2D matrix of frequency power evolution over time, that can be processed like images. In some cases, raw waveforms and spectrogram-based inputs can be used together to boost performance \cite{cl-audio2021,avcl2021}. Instead of solely depending on \acrshort*{cnn}s to learn spatial relations, frequency and time can be modelled separately. This method, using only \acrshort*{lms} as input, obtains \acrshort*{soa} performances \cite{eranns2021}. It further exploits frequency and time relations for downstream applications. However, \acrshort*{cnn}s' fixed receptive field limits the temporal context, encouraging small time-frames per sound clip (e.g., \textasciitilde 1ms for VGGish \cite{panns2020}). Many applications average all 1ms segments in longer audio clips, compressing information and adding noise. For larger audio sequences, \acrshort*{cnn}s can be paired with \acrfull*{rnn} or \acrfull*{lstm} to model temporal context in latent space \cite{flstm2015,tflstm2016}, at the cost of heavy computational models.\par

More recently, the Transformer architecture \cite{ast2021,mbt2021,li2023audioformer} and Attention Mechanisms increased further the benchmark results in AudioSet by allowing to analyze longer sequences and leverage their long-range global context. They have the issue of having a quadratic complexity. However, it can be reduced by: 1) using an auxiliary low-dimensional array \cite{perceiver2021}; 2) transferring knowledge from big models into smaller ones through knowledge distillation \cite{chen2022beats,schmid2022efficient,huang2022mavil}; 3) reduce the input sequence size \cite{hts-at2021,passt2021}\par

\begin{figure*}
    \centering
    \subfloat[Audio Spectrogram Transformer. Courtesy of \cite{ast2021}]{
        \includegraphics[width=0.47\textwidth]{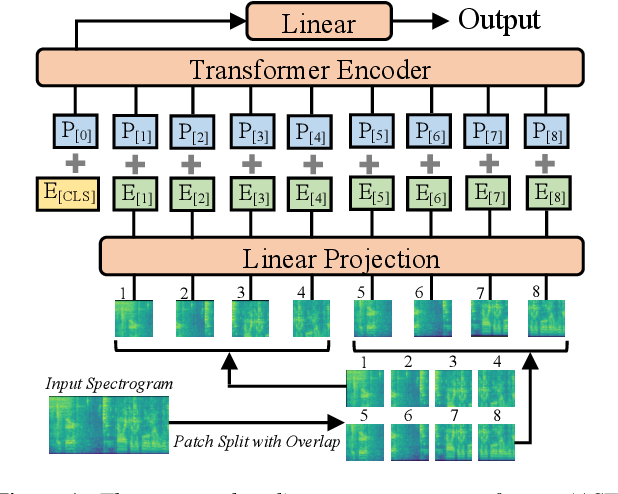}
        \label{fig:ast}
    }
    \quad
    \subfloat[PaSST. Courtesy of \cite{passt2021}]{
        \includegraphics[width=0.47\textwidth]{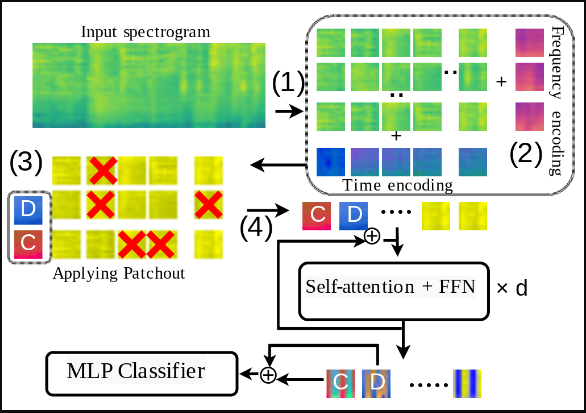}
        \label{fig:passt}
    }
    \quad
    \subfloat[HTS-AT. Courtesy of \cite{hts-at2021}]{
        \includegraphics[width=0.9\textwidth]{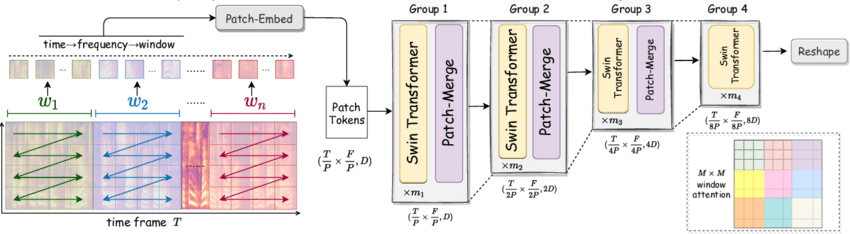}
        \label{fig:hts-at}
    }
    \caption{Transformations applied to audio spectrograms for Transformers: (a) overlap patching approach proposed by \cite{ast2021} that allows feeding 2-D inputs into the Transformer through a linear projection; (b) frequency positional encoding proposed by \cite{passt2021}; (c) patching order proposed by \cite{hts-at2021} to keep different frequency bins at the same time frame together in the input sequence.}
    \Description{Transformations applied to audio spectrograms for Transformers: (a) overlap patching approach proposed by \cite{ast2021} that allows feeding 2-D inputs into the Transformer through a linear projection; (b) frequency positional encoding proposed by \cite{passt2021}; (c) patching order proposed by \cite{hts-at2021} to keep different frequency bins at the same time frame together in the input sequence.}
    \label{fig:transformers-audio}
\end{figure*}



Transformers work with all mentioned intermediate forms, accepting raw waveforms and \acrshort*{midi} data without alteration. However, 2-D representations like spectrograms require modifications. Figure \ref{fig:transformers-audio} illustrates the different methodologies for dealing with 2-D inputs. Various methods handle 2-D inputs, such as splitting the 2D spectrogram into overlapping patch sequences and adding learnable position embeddings (Figure \ref{fig:ast}), as done by several researchers \cite{ast2021,li2023audioformer,passt2021,glass2022contrastive,hts-at2021}. Patches are ordered time-wise, and frequency bins can be arranged adjacently to capture frequency relations. Some have extended positional encoding to the frequency dimension to enhance efficiency on shorter audio clips and better capture relations between frequencies (Figures \ref{fig:hts-at} and \ref{fig:passt}) \cite{hts-at2021,passt2021}.\par


Transformers have seen many proposals use the learning framework from \acrfull*{bert} for natural language processing, involving masked language reconstruction \cite{bert2018,roberta2019,hubert2021}. This method uses incomplete sentences and generates complete versions. It can be applied to audio by patching the spectrogram and feeding it to the Transformer model \cite{li2023audioformer,arnab2022audiovisual,glass2022contrastive}, or optimizing models for translation between modalities \cite{hubert2021}, like speech-to-text. Masked learning can also be used with discrete label prediction \cite{chen2022beats}, introducing entropy into the learning process, similar to dropout. Additionally, some proposals use knowledge distillation to reduce resource needs \cite{schmid2022efficient,chen2022beats}. This method conditions a smaller model's weights using a larger model's predictions, aiming to match their performance, but it also propagates errors from the larger model.\par



In Section \ref{appendix:audioset-proposals}, on Table \ref{tab:audiosetTest}, we can observe that top methods on AudioSet use the Transformer architecture and can support multiple modalities for improved performance. Adding visual data is common to enhance the \acrfull*{map} score. Key aspects when developing an audio embedding solution include positional encoding methodology and feeding intermediate representations to the model. Top approaches, as seen in Table \ref{tab:audiosetTest}, use foundational models and create inductive biases, like patching near timesteps or adding frequency domain positional encodings, through self-supervised learning from large-scale datasets.\par

\subsection{Visual Data Encoding}
\label{section:video-feature-extrc}

\begin{figure*}[!ht]
    \centering
    \subfloat[Late Fusion Methods. Courtesy of \cite{gool2016temporal}]{
        \includegraphics[width=0.47\textwidth]{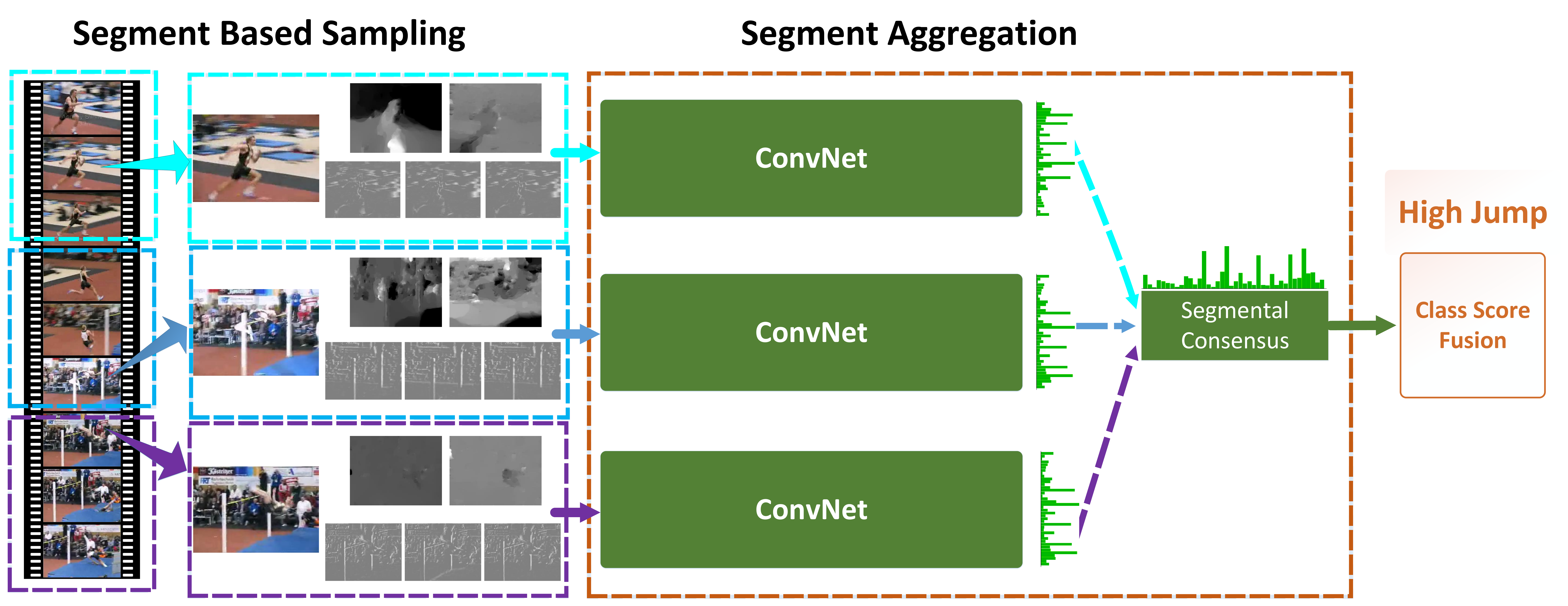}
        \label{fig:videof-latef}
    }
    \quad
    \subfloat[Long Temporal Dynamics. Courtesy of \cite{kalfaoglu2020late}]{
        \includegraphics[width=0.47\textwidth]{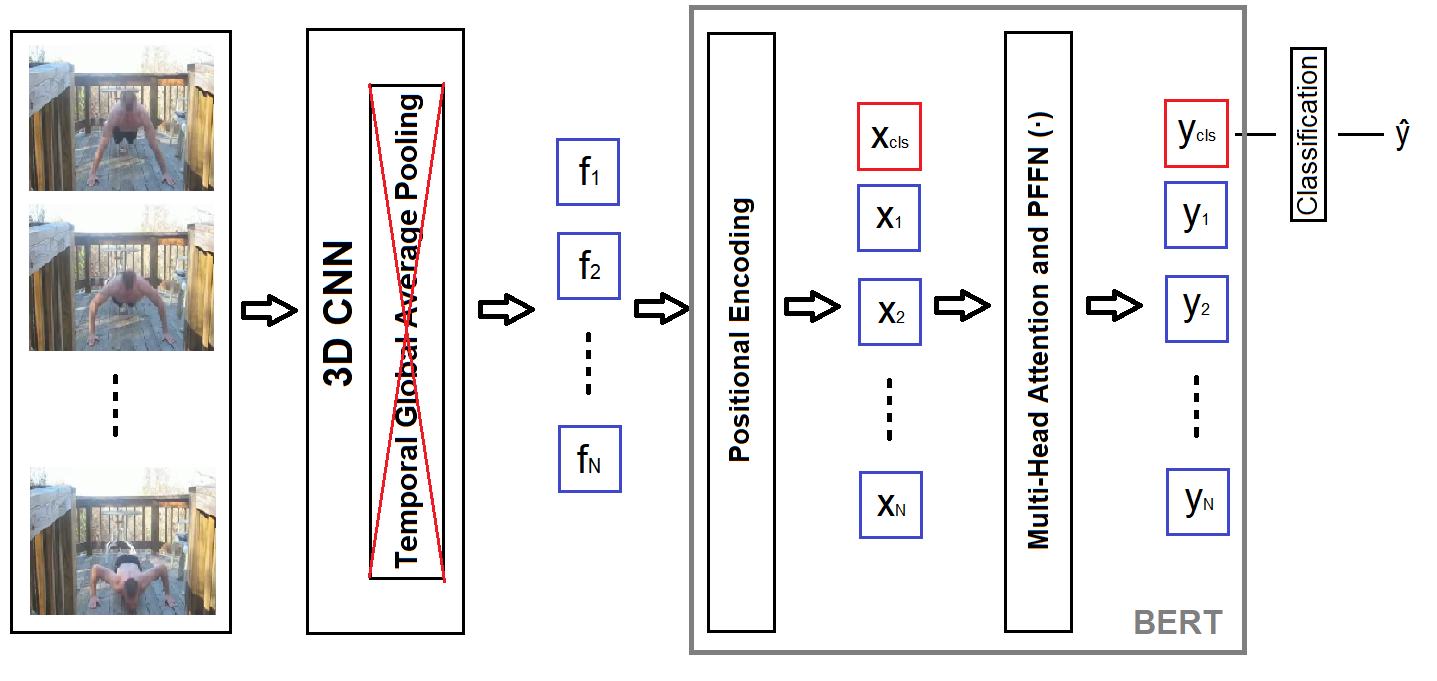}
        \label{fig:videof-long}
    }
    \quad
    \subfloat[Two-Stream Methods. Courtesy of \cite{carreira2017quo}]{
        \includegraphics[width=0.47\textwidth]{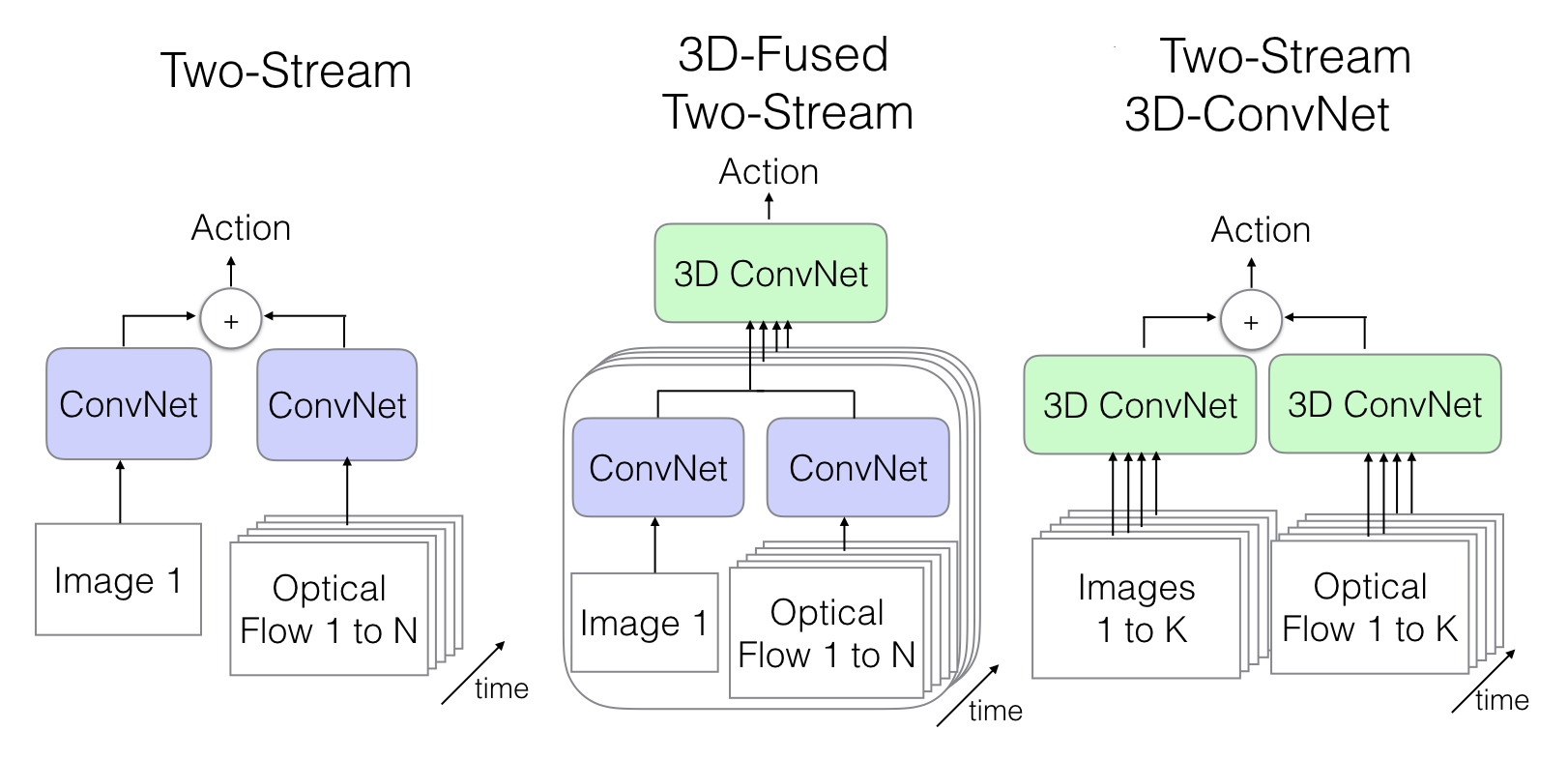}
        \label{fig:videof-twostream}
    }
    \quad
    \subfloat[Image/Frame-based Methods. Courtesy of \cite{zha2015exploring}]{
        \includegraphics[width=0.47\textwidth]{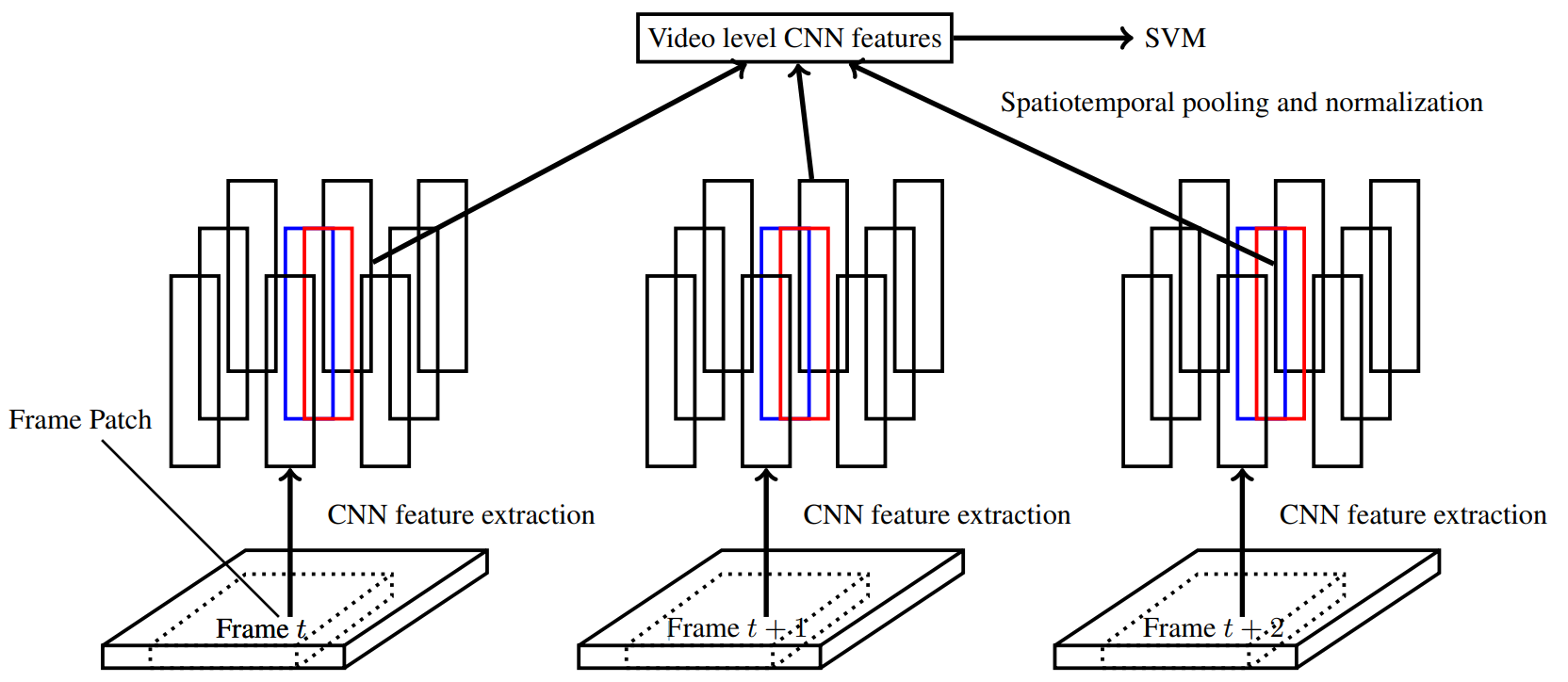}
        \label{fig:videof-frame}
    }
    \quad
    \subfloat[3-D Convolutional Methods. Courtesy of \cite{carreira2017quo}]{
        \includegraphics[width=0.25\textwidth]{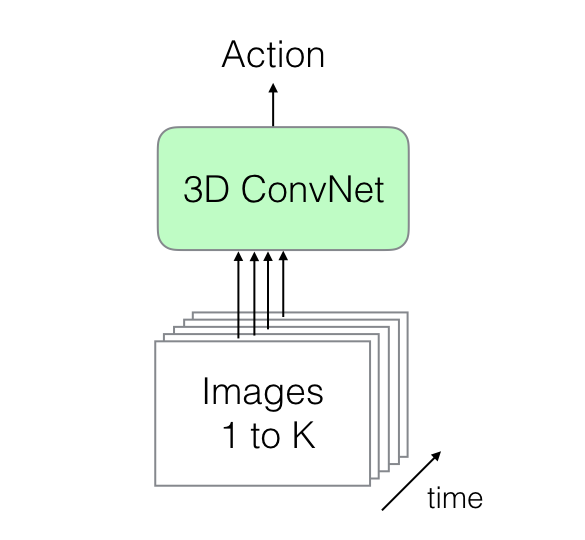}
        \label{fig:videof-c3d}
    }
    \caption{Main approaches for Feature Extraction in Video Frame Sequences}
    \Description{Main approaches for Feature Extraction in Video Frame Sequences}

    \label{fig:videofextraction}
\end{figure*}

Apart from Youtube-8M \cite{abu2016youtube8m}, there are no commonly used benchmarks for general video classification. Therefore, in this analysis, we will consider the task of \acrfull*{ar} due to its popularity and because it is one of the most common classification tasks where video frame sequences are analyzed. Nevertheless, \acrshort*{ar} datasets tend to be human-centric, and \acrshort*{soa} methods tend to infer actions by typically leveraging motion-related features \cite{wu2017deep}. Therefore, our analysis of visual feature extraction models is based on the following video classification architectures from which \acrshort*{ar} models commonly derive. We gathered performances for the top-2 benchmarking datasets for action recognition, HMDB-51, and UCF-101, as illustrated in Table \ref{tab:actionRecMethods} (Section \ref{appendix:actionrecognition-proposals}). We present results with the two most common evaluation settings, fine-tuning and linear evaluation, and order them using the mean of both results. In Figure \ref{fig:videofextraction}, we summarize the identified approaches for feature extraction on visual data (video).\par



Image/frame-based methodologies (Figure \ref{fig:videof-frame}) treat video clips as a collection of frames, using models for image classification to gain feature representation. This is then averaged or dimensionally reduced over the temporal axis.\par


Two-stream models decompose videos into spatial and temporal components (Figure \ref{fig:videof-twostream}). They use two branches: RGB (Visual component) and Optical Flow (Temporal component) \cite{hong2019contextual,qiu2019learning}. These models require optical flow extraction, which can be costly, but some approaches estimate this from the RGB frames \cite{piergiovanni2019representation}. Another limitation is that the temporal stream is solely responsible for capturing features from motion \cite{wu2017deep}. Temporal Segment Networks (TSN) \cite{gool2016temporal} extend two-stream models by dividing longer videos into segments, each classified separately, with the final classification merging the segment scores \cite{gowda2021smart,zhang2019cooperative}.\par


3D convolutional networks exploit spatial and temporal information with 3-dim kernels (Figure \ref{fig:videof-c3d}). These are complex to train due to the increased number of parameters with an expanded temporal dimension. To address this, some proposals replace 3D convolutions with separate spatial and temporal convolutions, reducing parameters and computational cost \cite{tran2018closer,carreira2017quo}.\par


Models focused on long temporal dynamics use \acrfull*{rnn}, \acrfull*{lstm} or Transformers for the temporal axis \cite{ng2015beyond,wu2017deep}. While sequence models are commonly placed on top of features from image classifiers, attention-based models reduce temporal complexity and favor parallel computation, leading to a rise in Transformers replacing \acrshort*{rnn}s/\acrshort*{lstm}s (Figure \ref{fig:videof-long}).\par


Top models capture temporal correlations between frames \cite{tao2020self,sarkar2021self,luo2020video,dave2022tclr,qian2022multimodal} and their relation with other modalities \cite{radford2021learning,wu2022revisiting,kahatapitiya2023victr}, with Transformers excelling in this aspect. Initially applied to images, this architecture divided images into patches, transformed them into embeddings, and used them in a standard transformer model \cite{dosovitskiy2020image}. This approach drives the best-performing unimodal and multimodal proposals \cite{feichtenhofer2021large,radford2021learning,wu2022revisiting}.\par

\begin{figure*}
    \centering
    \subfloat[ViViT Model Overview (``ViT applied to video frame sequences''). Courtesy of \cite{arnab2021vivit}]{
        \includegraphics[width=0.5\textwidth]{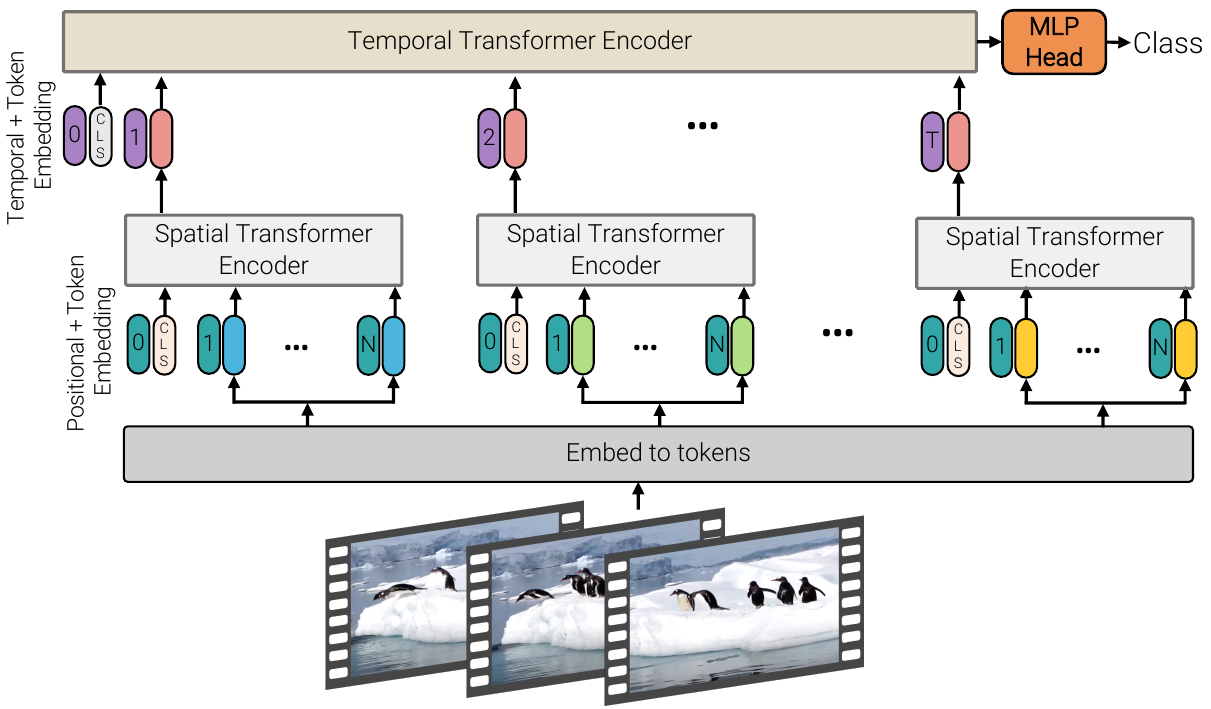}
        \label{fig:videof-vivit}
    }
    \quad
    \subfloat[Extract frames and linearly embed them in non-overlapping ``tubes'' that span the temporal dimension. Courtesy of \cite{arnab2021vivit}]{
        \includegraphics[width=0.3\textwidth]{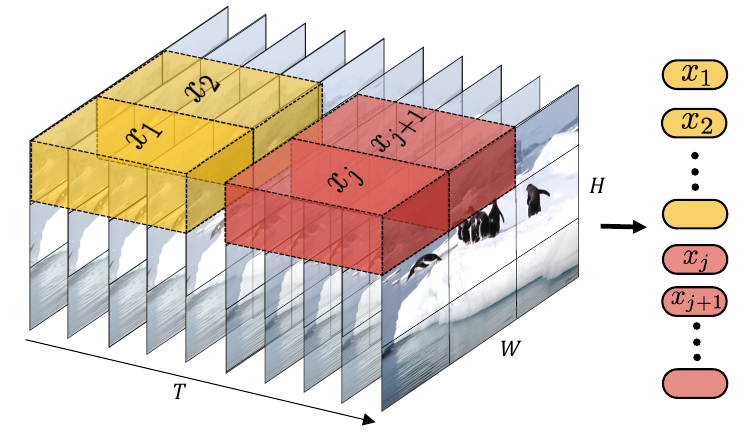}
        \label{fig:videof-vivit-posenc}
    }
    \quad
    \subfloat[ViT Model Overview. Courtesy of \cite{dosovitskiy2020image}]{
        \includegraphics[width=0.65\textwidth]{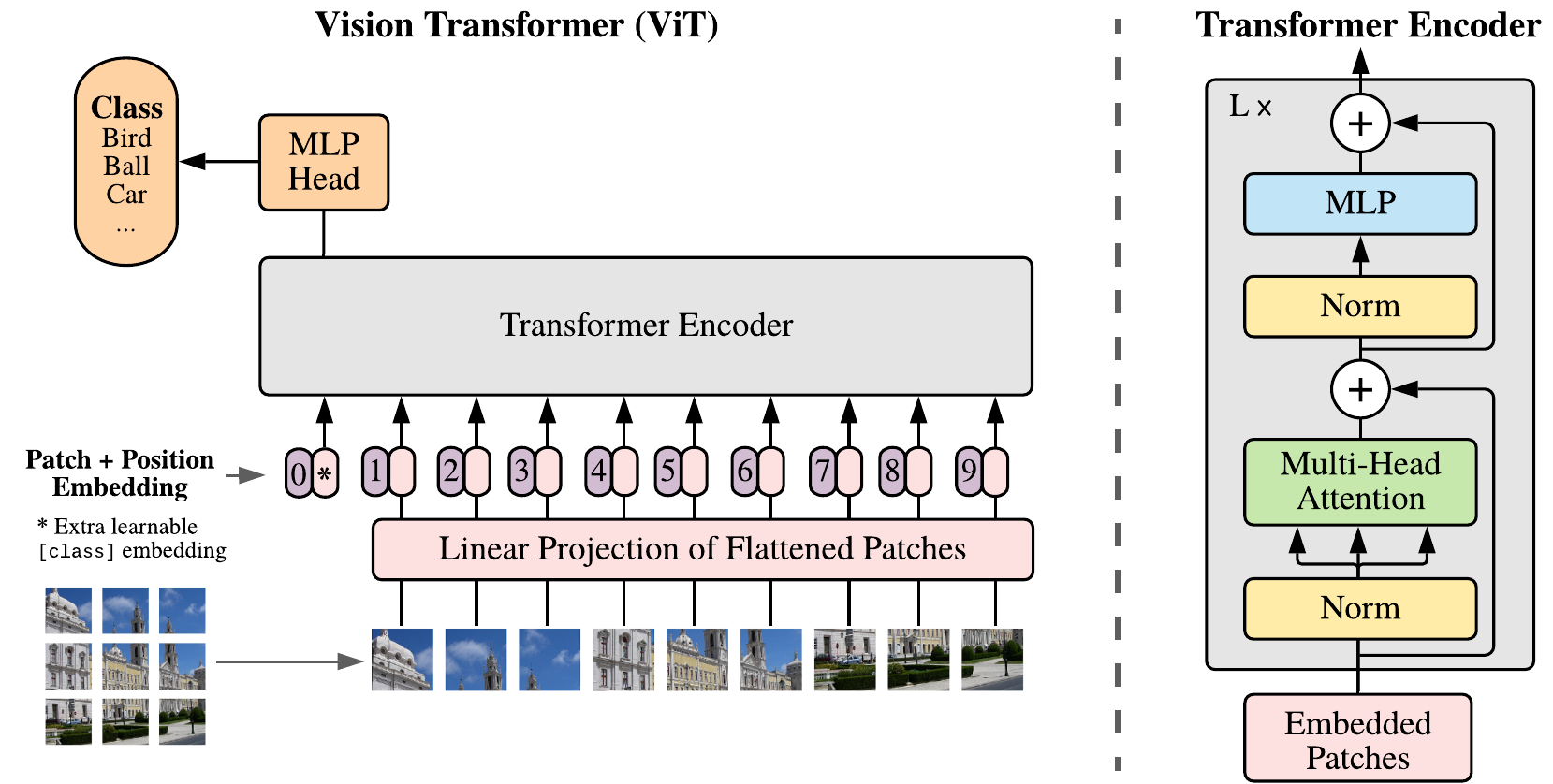}
        \label{fig:videof-vit}
    }
    \caption{Differences between the ViT and ViViT models. Example of an adaptation of the transformer model to video frame sequences. In addition to spatial patches, it also receives ``temporal patches''.}
    \Description{Differences between the ViT and ViViT models. Example of an adaptation of the transformer model to video frame sequences. In addition to spatial patches, it also receives ``temporal patches''.}
    \label{fig:videof-transformer}
\end{figure*}

\acrfull*{videomae} and \acrfull*{vivit} extend the ViT architecture for video data, using visual embeddings from different video segments and applying masking strategies temporally (Figure \ref{fig:videof-transformer}) \cite{tong2022videomae,arnab2021vivit}. Extensions propose new masking schemes, motion trajectory reconstruction, and knowledge distillation from larger models or modalities \cite{wang2023videomaev2,sun2023masked,wang2023masked,sarkar2022xkd}. VideoMAE2 is currently the best performer on average \cite{wang2023videomaev2}.\par

Text can also be exploited as supervision \cite{radford2021learning,wu2022revisiting,kahatapitiya2023victr,lin2023match}. For instance, \acrfull*{clip} proposed to learn text/image representations from large-scale datasets \cite{radford2021learning}. The approach can be applied to videos by selecting the most relevant frames \cite{wu2022revisiting,kahatapitiya2023victr}. Both approaches predict which caption matches the audiovisual input. \acrshort*{clip} and \acrshort*{videomae} show promising results for HMDB-51 and UCF-101. Despite not topping the UCF-101, \acrshort*{clip} is a top performer using only one frame (+ \acrshort*{vit}) \cite{wu2022revisiting,radford2021learning}. These results are due to self-supervised learning frameworks leveraging large data amounts without supervision.\par


Limited supervision learning needs inductive biases for relevant representations. Temporal cohesion and correlations, learned via pairs of relevant and non-relevant embeddings, are crucial in videos \cite{tao2020self,qian2022multimodal,sarkar2021self,qian2021spatiotemporal,hu2022self}. This is the starting point to encourage features to be distinct across the temporal dimension \cite{dave2022tclr}. The model should also learn semantic relations between video frame sequences, exploitable through additional semantic information \cite{denize2022similarity,qian2021spatiotemporal,feichtenhofer2021large,hu2022self}.\par

Late fusion models (5) combine multiple decisions for a final prediction (Figure \ref{fig:videof-latef}). Inputs can include predictions from video segments \cite{gool2016temporal} or high-level information \cite{shen2021self}.\par

Video processing is computationally heavy due to high frame rates and extensive data per frame. Methods to reduce complexity include video sampling strategies \cite{gowda2021smart,zhi2021mgsampler,wang2023differentiable}. Dynamic sampling rates can be achieved by leveraging motion \cite{lin2022action,zhi2021mgsampler}, high-level information like objects and people \cite{wang2020symbiotic}, using statistical methods \cite{zhi2021mgsampler}, and relevance predictions \cite{korbar2019scsampler,wang2023differentiable,gowda2021smart} using features extracted within the temporal domain. In the UCF-101 and HMDB-51 context, SMART is the only dynamic video sampling example, selecting the top $n$ frames with high discriminative scores \cite{gowda2021smart}. It achieves top-5 scores on both datasets using ten frames, improving performance by 15\% and 4\% on HMDB-51 and UCF-101, respectively. SMART's potential with attention-based models remains unexplored. These results shed some light on the potential impact of frame sampling on \acrshort*{ar} and video processing.\par


The presented models are all viable options for obtaining visual embeddings for videos, and we believe that the choice of feature extractor for \acrshort*{avcl} applications must be made according to the ratio between performance and input size (i.e., temporal length, number of frames).\par


\section{Learning Audio-Visual Correlation}
\label{section:learningAVCL}

In \acrlong*{avcl}, audio and video representations can be obtained directly from existing pre-trained models \cite{zhang2019deep,zeng2018audio}. However, this results in representation spaces with different dimensions, thus hindering the extraction of relations between modalities. Due to inconsistent distributions and representations between them, we have to learn a common space to bridge this gap and further measure and maximize their correlation. In other words, since our input data ($x_1$ and $x_2$) belong to sub-spaces with different sizes, cross-correlating audio and visual data is most commonly done by using separate projections ($f(x_1; W_1)$ and $g(x_2; W_2)$) \cite{zhen2019deep,zeng2021learning}. Nevertheless, this is not the only scenario that we should consider. Often, feature maps from different modalities can be correlated using attention-based approaches without requiring any projection. However, these sub-spaces should have similar dimensions.

Generally, Deep \acrshort*{avcl} contains mainly two steps:\par
\begin{enumerate}
    \item Extract temporal information from frame-level audio and visual features, where $x_a$ and $x_v$ are fed into two separate encoders. To better learn sequential dependencies between audio and visual data, recent works, such as the attention mechanism, focus on synchronizing semantic information between modalities and minimizing the heterogeneity gap simultaneously. We called them multimodal encoding models;
    \item Objective functions optimize these two ``sub-networks'' to maximize the correlation between modalities in the ``joint'' space through back-propagation. According to different kinds of audio-visual data exploited in the deep learning stage, various learning frameworks (i.e., objective functions) are used in existing proposals.
\end{enumerate}

In this section, we address the recent advances for step (1) and (2) and discuss some tasks and paradigms in Deep \acrshort*{avcl}. In Table \ref{tab:methods-final}, we provide a brief summary of these \acrshort*{soa} \acrshort*{avcl} methods.\par

\subsection{Multimodal Encoding Models}
\label{subsection:models}

Various encoding models are proposed to capture and synchronize semantic temporal information between audio and visual data, which facilitates the correlation of audio-visual sub-spaces.\par

\subsubsection{\textbf{Attention}}
\label{subsection:attention-model}

\begin{figure}
    \centering
    \subfloat[][Co/Cross-Attention. Courtesy of \cite{duan2021audio}]{
        \includegraphics[width=0.47\textwidth]{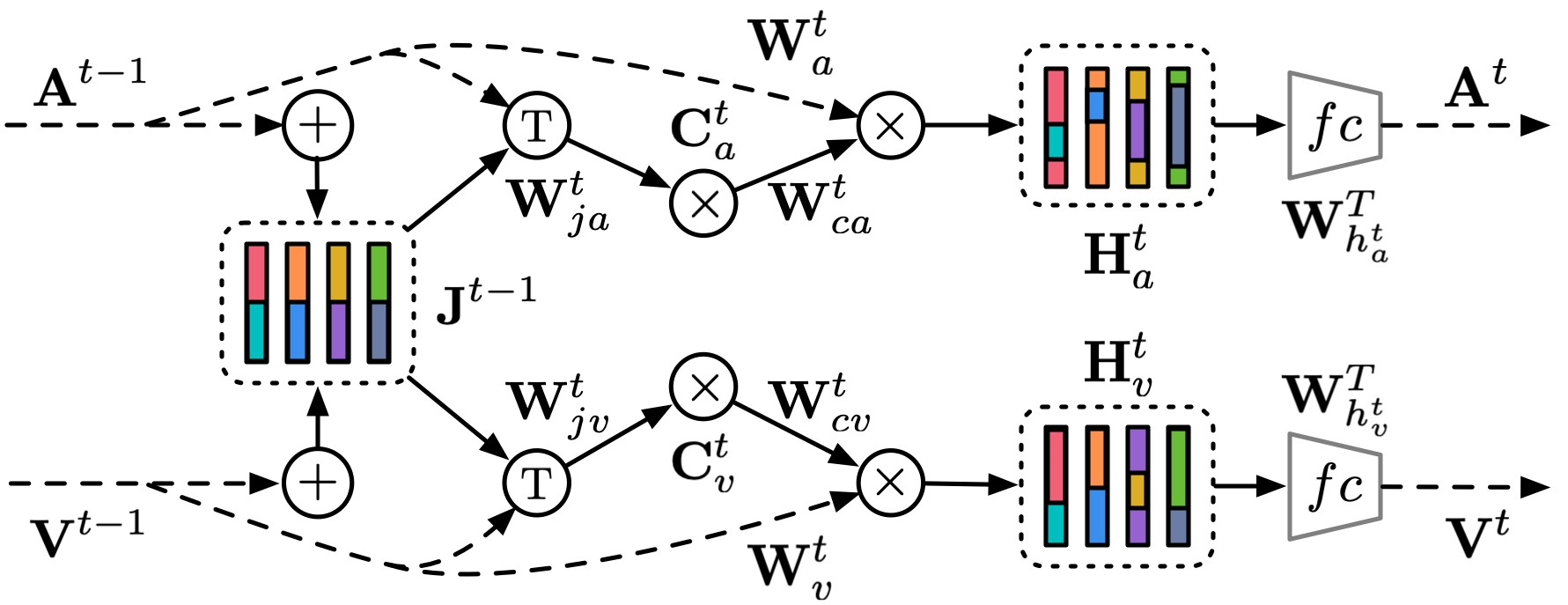}
        \label{fig:coattention}
    }
    \quad
    \subfloat[][Co/Cross-Attention using the Transformer Encoder. Courtesy of \cite{cheng2020look}]{
        \includegraphics[width=0.47\textwidth]{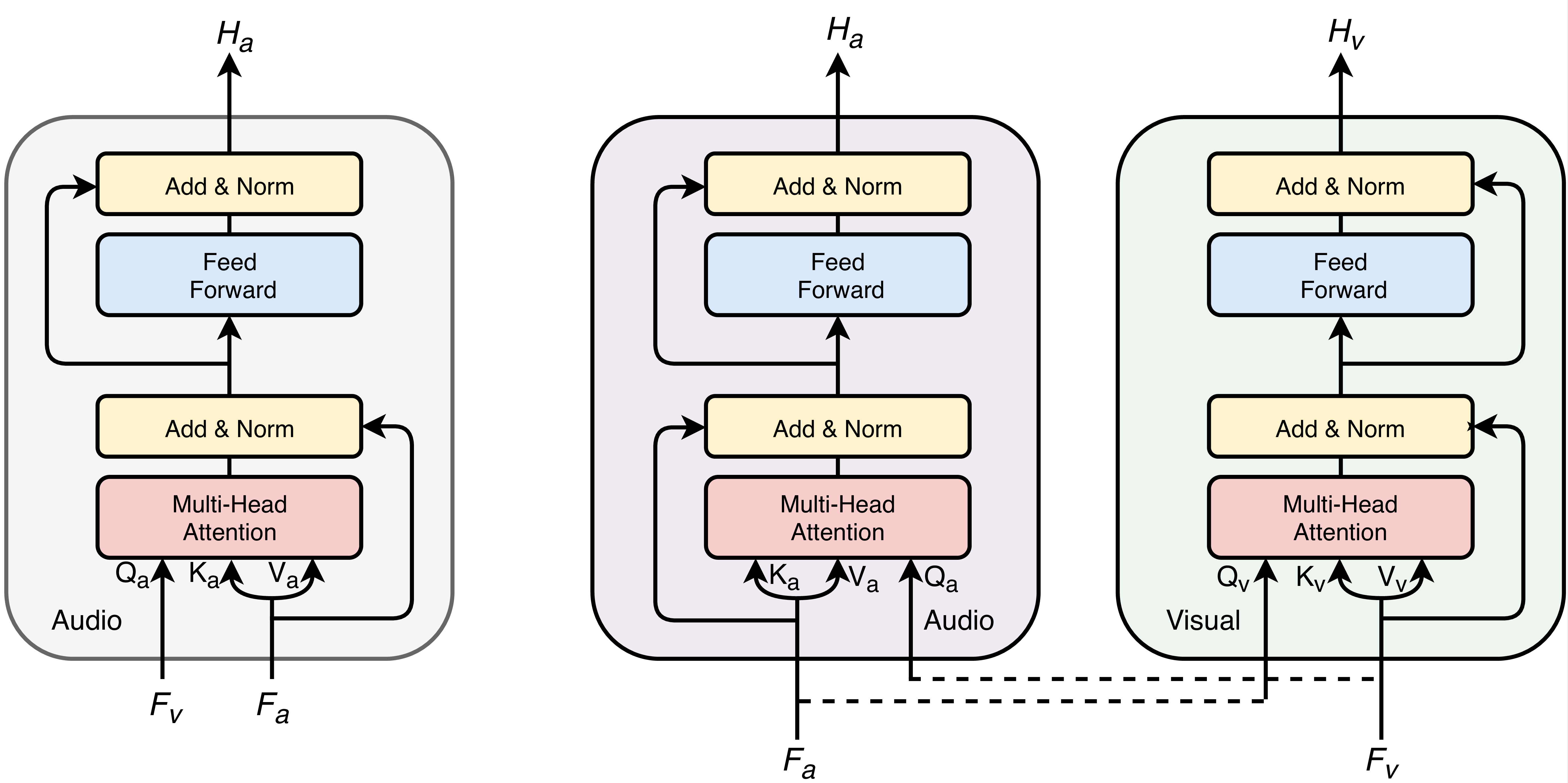}
        \label{fig:transformer-coattention}
    }
    \quad
    \subfloat[][Self-Attention. Courtesy of \cite{attention-is-all-need}]{
        \includegraphics[height=0.3\textwidth]{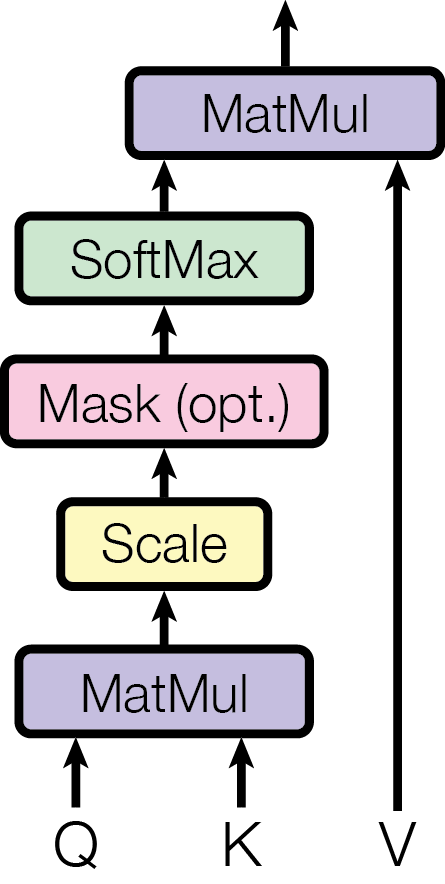}
        \label{fig:selfattention}
    }
    \quad
    \subfloat[][Multi-Head Self-Attention. Courtesy of \cite{attention-is-all-need}]{
        \includegraphics[height=0.3\textwidth]{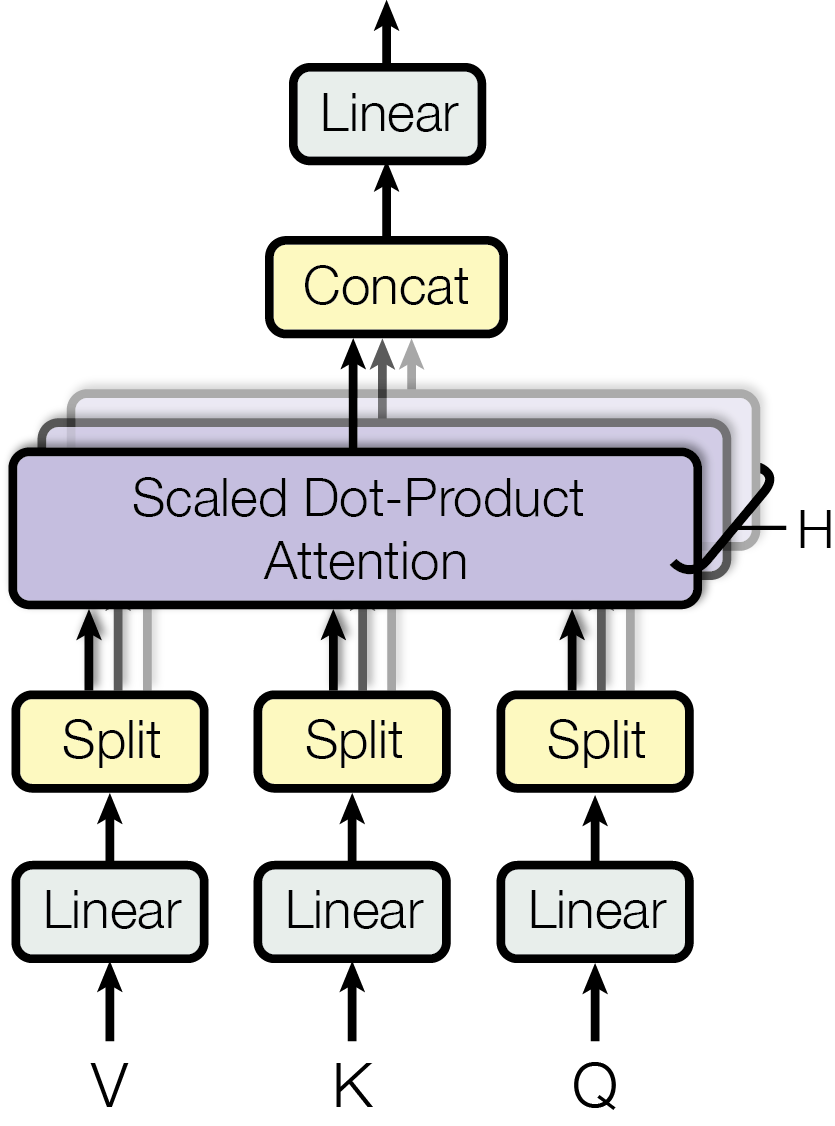}
        \label{fig:multiheadattention}
    }
    \quad
    \subfloat[][Transformer Model. Courtesy of \cite{attention-is-all-need}]{
        \includegraphics[height=0.5\textwidth]{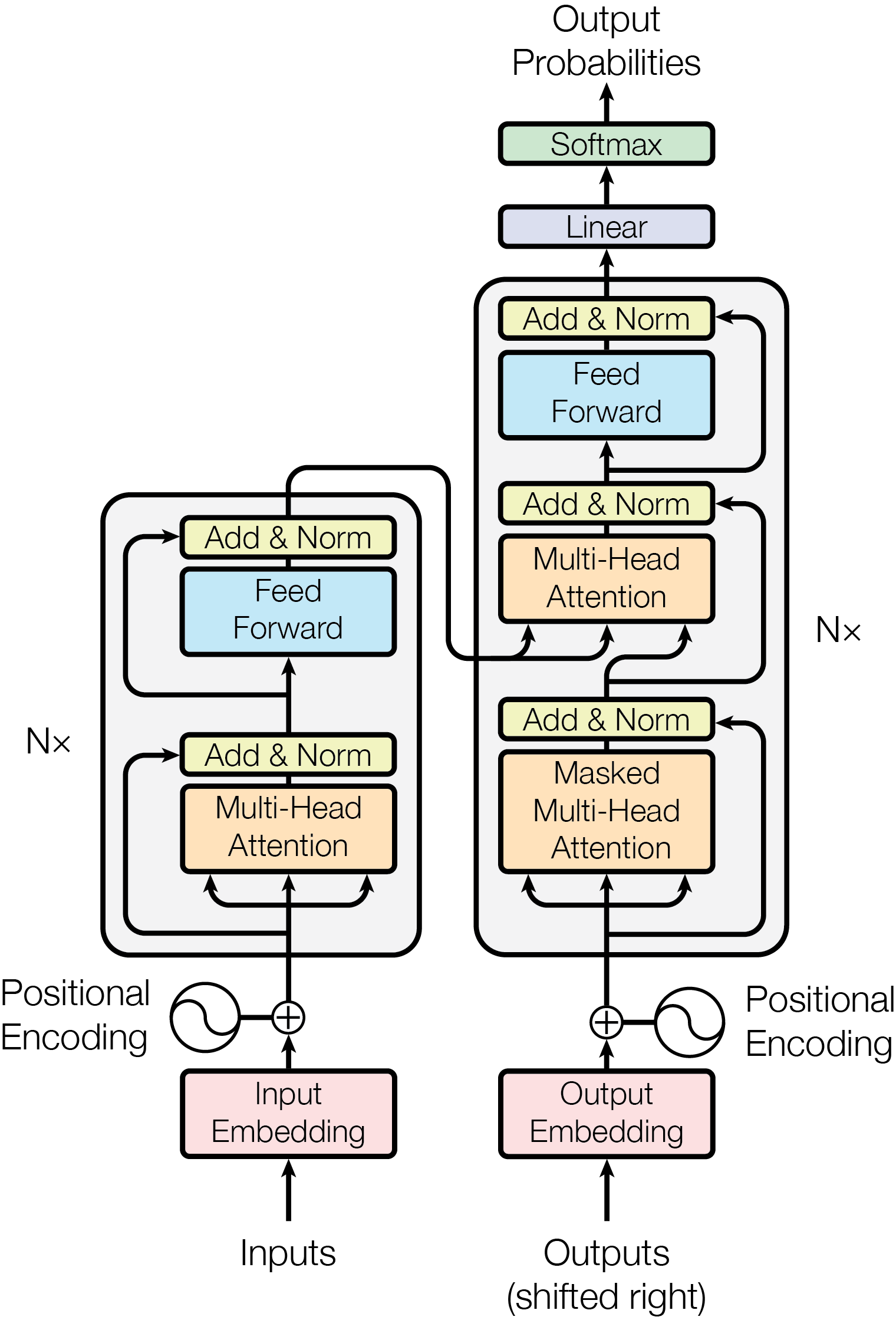}
        \label{fig:transformer}
    }
    \caption{Attention mechanisms used for \acrshort*{avcl}: (a) Co/Cross-attention consists in attending one modality conditioned on the other, e.g., use audio to attend visual representations. For the symbols, \textcircled{+} denotes concatenation, \textcircled{x} indicates dot-product between audio and visual features, \textcircled{t} represents the transpose operation; (b) Co/Cross-attention using the encoder blocks from the Transformer model, where parallel modalities are used to attend each other using multi-head self-attention; (c) Self-attention is used to extract relationships within its input sequence. The input representation ``attends to itself''. This method is used in the Transformer decoder to allow each position to attend to all up to and including that element in the sequence, thus preserving its auto-regressive properties. It is implemented inside self-attention (e.g., dot-product attention) by masking out all values in the input; (d) Multi-head self-attention consists of several attention layers running in parallel; (e) Transformer model, which involves an encoder (on the left) and a decoder (on the right).
    }
    \Description{Attention mechanisms used for \acrshort*{avcl}: (a) Co/Cross-attention aims to use parallel modalities to attend one another, e.g., use audio to attend visual representations. For the symbols, \textcircled{+} denotes concatenation, \textcircled{x} indicates dot-product between audio and visual features, \textcircled{t} represents the transpose operation; (b) Co/Cross-attention using the encoder blocks from the Transformer model, where parallel modalities are used to attend each other using multi-head self-attention; (c) Self-attention is used to extract relationships within its input sequence. The input representation ``attends to itself''. This method is used in the Transformer decoder to allow each position to attend to all up to and including that element in the sequence, thus preserving its auto-regressive properties. It is implemented inside self-attention (e.g., dot-product attention) by masking out all values in the input; (d) Multi-head self-attention consists of several attention layers running in parallel; (e) Transformer model, which involves an encoder (on the left) and a decoder (on the right).
    }
    \label{fig:attentionmechanisms}
\end{figure}

In neural networks, the effect of attention aims to enhance some parts of the input data while attenuating other parts. It can be utilized for strengthening semantic information between modalities to align audio and visual sequences through a dynamic pooling process that allows learning relations between sub-elements (i.e., segments of audio or visual locations). It consists of a dynamic weighted sum of vectors with scalar values (i.e., probabilities) obtained through an interaction/score function (e.g., addition, dot-product, scaled dot-product). Therefore, it allows obtaining context from different sources (within or between modalities) to encode a given input (e.g., using $x_a$ to encode $x_v$). Attention is often computed using three kinds of vectors: Queries (Q), Keys (K), and Values (V). We calculate attention for the Qs, using Ks, which subsequently try to filter the results in Vs. This is analogous to a retrieval system, where we use the Q to filter the K and want to obtain the best results in V. We identified the following methodologies, which are illustrated in Figure \ref{fig:attentionmechanisms}:\par

\paragraph{\textbf{\textit{i. Co/Cross-Attention}}} (Figure \ref{fig:coattention} and \ref{fig:transformer-coattention}) enables the learning of pairwise attention, which can be used to explore inter-modal correlations by encoding a feature vector with context from another. Typically, a residual connection is added in the attended feature vector, which can be seen as adding missing relevant information from one modality to another. Co-attention allows learning correlations between multiple pairs of data \cite{zheng2021deep,duan2021audio,xuan2021discriminative,lee2021crossattentional,lee2021crossattentional,afouras2020self,tian2018audio}. Nevertheless, the context used to encode one representation can also be drawn from features extracted at multiple scales (i.e., different framerate) \cite{wu2019unified}, different timesteps \cite{ma2020contrastive,yang2019dual} or between different sub-elements (e.g., image regions or patches) \cite{lin2021exploiting,gan2020music}. In addition, memory banks can leverage context from previous training steps, which can further improve the quality of the context used \cite{nam2017dual}. Therefore, we can easily infer that relations between different sources of information can be obtained through attention. For \acrshort*{avcl}, it is used to identify the most important patterns that characterize an audio-visual event (alignment). Furthermore, it can also be used to coordinate/align the weights of models used to extract features for each modality \cite{min2021cross}. In other words, co/cross-attention can be used to coordinate/align representation spaces for different modalities.\par

\paragraph{\textbf{\textit{ii. Self-Attention}}} (Figure \ref{fig:selfattention}) relates different positions of a sequence to calculate a representation of this sequence, which can be used to explore the sequential structure of each modality (inter-modal), and it can be combined with co-attention to weight intra and inter-modal relations \cite{cheng2020look}. Moreover, this kind of attention can be used to extract relations from joint (multimodal) representations, which allows to weigh inter-modal relations. On the other hand, when applied to single modalities, it allows to weigh intra-modal features. It is the base of the \acrfull*{tf} architecture, which is applied in two different ways: multi-head self-attention (encoder) and masked self-attention (decoder). They both consist of the same process of adding several parallel self-attention projections. However, masked self-attention limits the context of each one to avoid attention bias (i.e., any query in the decoder only attends to all positions up to the current one).\par

\begin{figure*}[!ht]
    \centering
    \subfloat[][Early summation. Courtesy of \cite{xu2022multimodal}]{
        \includegraphics[width=0.1\textwidth]{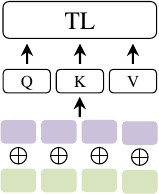}
        \label{fig:transformer-fusion-earlysum}
    }
    \quad
    \subfloat[][Early concatenation. Courtesy of \cite{xu2022multimodal}]{
        \includegraphics[width=0.2\textwidth]{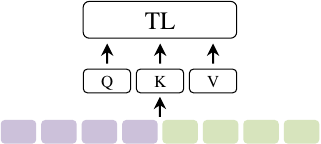}
        \label{fig:transformer-fusion-earlyconcat}
    }
    \quad
    \subfloat[][Hierarquical fusion (multi-stream on the left and one-stream on the right). Courtesy of \cite{xu2022multimodal}]{
        \includegraphics[width=0.2\textwidth]{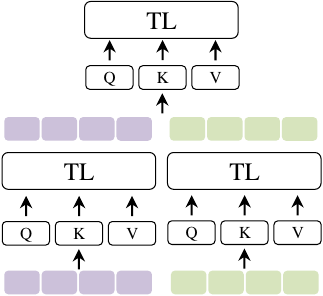}
        \includegraphics[width=0.2\textwidth]{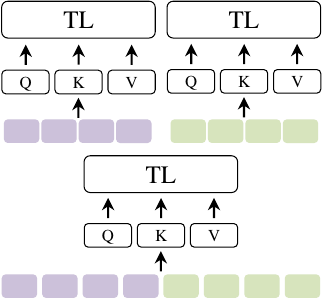}
        \label{fig:transformer-fusion-hiearattent}
    }
    \quad
    \subfloat[][Co/Cross-attention. Courtesy of \cite{xu2022multimodal}]{
        \includegraphics[width=0.2\textwidth]{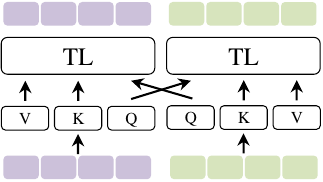}
        \includegraphics[width=0.2\textwidth]{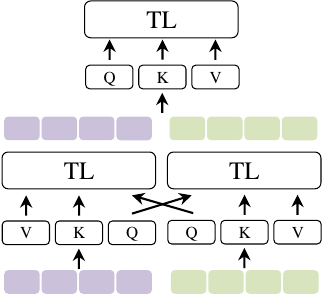}
        \label{fig:transformer-fusion-crossatt}
    }
    \caption{Interactions between multiple modalities using Transformers: (a) Early summation consists in combining multiple modalities through a weighted sum at each token position; (b) Early concatenation consists in concatenating the input token sequences, such that the multimodal token positions are attended as a whole; (c) Hierarchical fusion, where Transformer layers are combined hierarchically; (d) Co/Cross-attention consists in exchanging the query (Q) embeddings in a cross-stream manner. For the symbols, Q is the query embedding, K is the key embedding, V is the value embedding, and T is the Transformer Layer. Different colors correspond to different modalities.
    }
    \Description{Interactions between multiple modalities using Transformers: (a) Early summation consists in combining multiple modalities through a weighted sum at each token position; (b) Early concatenation consists in concatenating the input token sequences, such that the multimodal token positions are attended as a whole; (c) Hierarchical fusion, where Transformer layers are combined hierarchically; (d) Co/Cross-attention consists in exchanging the query (Q) embeddings in a cross-stream manner. For the symbols, Q is the query embedding, K is the key embedding, V is the value embedding, and T is the Transformer Layer. Different colors correspond to different modalities.
    }
    \label{fig:transformer-multimodal}
\end{figure*}

\paragraph{\textbf{\textit{iii. Transformer}}} is an encoder-decoder model with a self-attention mechanism as well as positional encoding, which can be exploited to capture semantic information between audio and visual data within the sequential structure of its input. \acrshort*{tf}s are formed by an encoder (i.e., multi-head) and a decoder (i.e., masked), as illustrated in Figure \ref{fig:transformer}, and an encoder-decoder module. The \acrshort*{tf} encodes the output from the decoder using the encoder context. Similar to co-attention, the \acrshort*{tf} encoder can be replaced by different feature extractors or features from sub-elements of each modality (e.g., past and future frames or different patches within the same frame) and leverage the implicit co-attention mechanism between modalities \cite{gan2020foley,lin2020audiovisual,morgado2020learning}. Furthermore, \acrshort*{tf}s can also be used to relate cross-translations between modalities of the same media asset to align/correlate them in sequence-to-sequence generation \cite{morgado2020learning}. In Figure \ref{fig:transformer-multimodal}, we illustrate the proposed combinations of interactions between modalities that using Transformers.\par

\textbf{Discussion:} Attention mechanisms allow fusing and coordinating data from different sub-spaces by selectively weighting the importance of each element in each one, where redundant and relevant information is respectively removed or added \cite{zhu2021leveraging,chen2021localizing,zhu2020visually}. The context for weighting the representations can be obtained from different combinations or features of $x_a$ and $x_v$ or their sub-elements. Thus, establishing where (and when) to draw context depends on the available data and the target application but opens a pathway for many possible extensions based on attention to audio-visual alignment. Nevertheless, attention requires an interaction/score function that fits the constraint length of each subspace and expresses the intended correlation between its inputs. For instance, addition requires inputs with equal lengths and accumulates the growth of both vectors, while the dot product can accept mixed sizes but capture the directional growth of one into the other. Therefore, defining the interaction function is crucial for many applications when attention is exploited. Specifically, for \acrshort*{av} applications, attention can be used to provide global \cite{morgado2020learning}, local \cite{min2021cross} and context between modalities \cite{seo2020hmtl,cheng2020look} to the extracted embeddings.\par

As far as we know, we could not find an in-depth study on how different combinations of attention mechanisms impact learning multimodal embeddings for different paradigms (i.e., supervised, self-supervised). We could only find an in-depth analysis of the different multimodal settings for Transformers, but without numerical results on downstream tasks \cite{xu2022multimodal}. We provide Figure \ref{fig:transformer-multimodal} as a reference for the different types of multimodal interactions that can be used with Transformer models. Early concatenation is often preferred over summation because each modality can be encoded using the context of the other \cite{mohamed2022learning}. Nevertheless, summation does not increase the problem's computational complexity, whereas concatenation increases. Hierarchical attention allows preserving the independence of each unimodal subspace and their interactions \cite{glass2022uavm,kanazawa2021choreographer}. Finally, cross-attention does not cause higher complexity and allows extracting relations between modalities \cite{chen2021vset}. However, it fails to learn from each independent subspace, which often requires applying self-attention separately to each modality beforehand.\par

From an application point-of-view, attention can be used to link audio events and their corresponding spatial locations in video feeds (sound localization - localization in the spatial domain) \cite{ramaswamy2020see,morgado2020learning} to separate multiple sound sources by analyzing visual data (sound separation) \cite{gao2019co,zhao2019sound,zhao2018sound} and to condition the generation between modalities (e.g., visual data to audio) \cite{fanzeres2021sound,mira2021end,gan2020foley}. This is achieved by weighting paired occurrences such that events in one modality can be translated/mapped to others. For instance, what visual cues allow understanding when different sound sources start.\par

On a final note, similarly to the attention mechanism (co and self-attention), \acrshort*{tf}s can also be used to learn inter and intra-modal relationships. Higher performances are generally achieved using parallel connections associated with masked and multi-head self-attention modules \cite{akbari2021vatt}. However, it should also be noted that the number of parameters of these models is substantially large. Some studies have already shown that the number of parameters should escalate with $N$ \cite{hahn2020theoretical} to achieve a perfect accuracy/cross-entropy on the elements in a sequence of length $N$.\par

\subsubsection{\textbf{Auto-Encoders}}
\label{subsubsection:aes}

\begin{figure*}[!ht]
    \centering
    \subfloat[][\acrshort*{ae}s with individual latent spaces. Courtesy of \cite{cao2016correlation,recasens2021broaden}]{
        \includegraphics[width=0.67\textwidth]{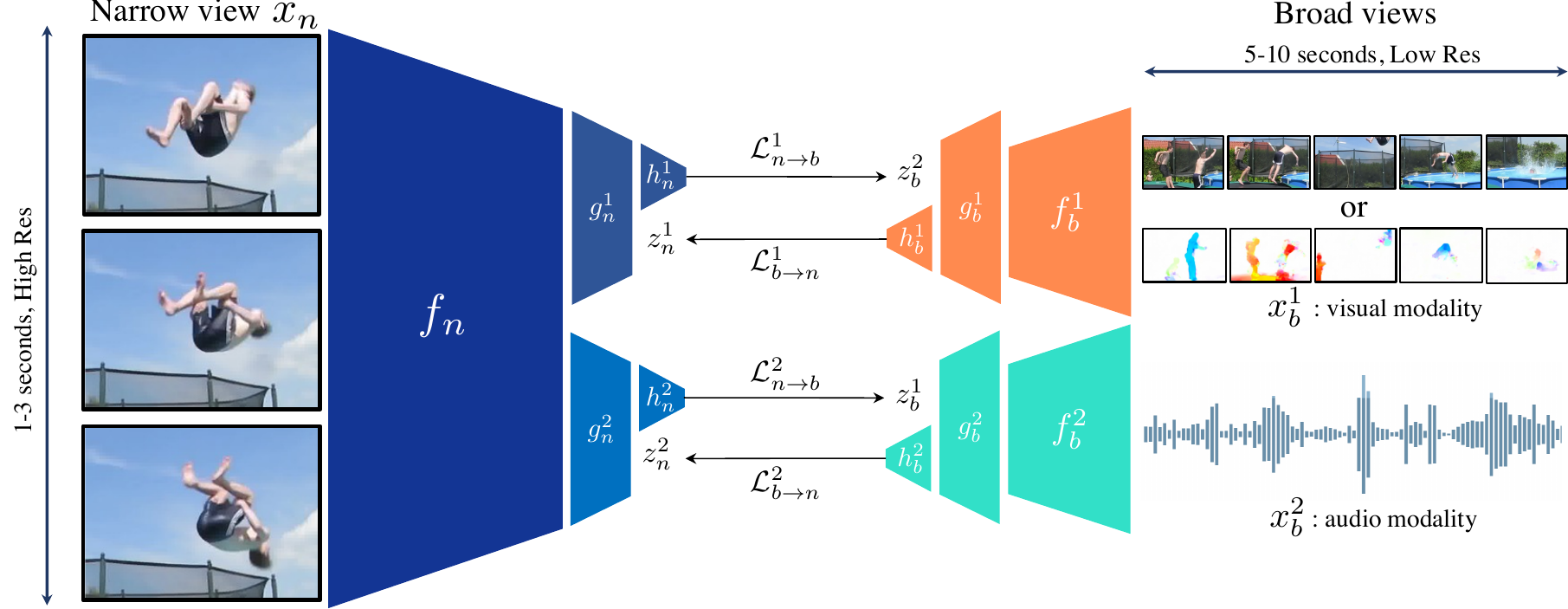}
        \includegraphics[width=0.3\textwidth]{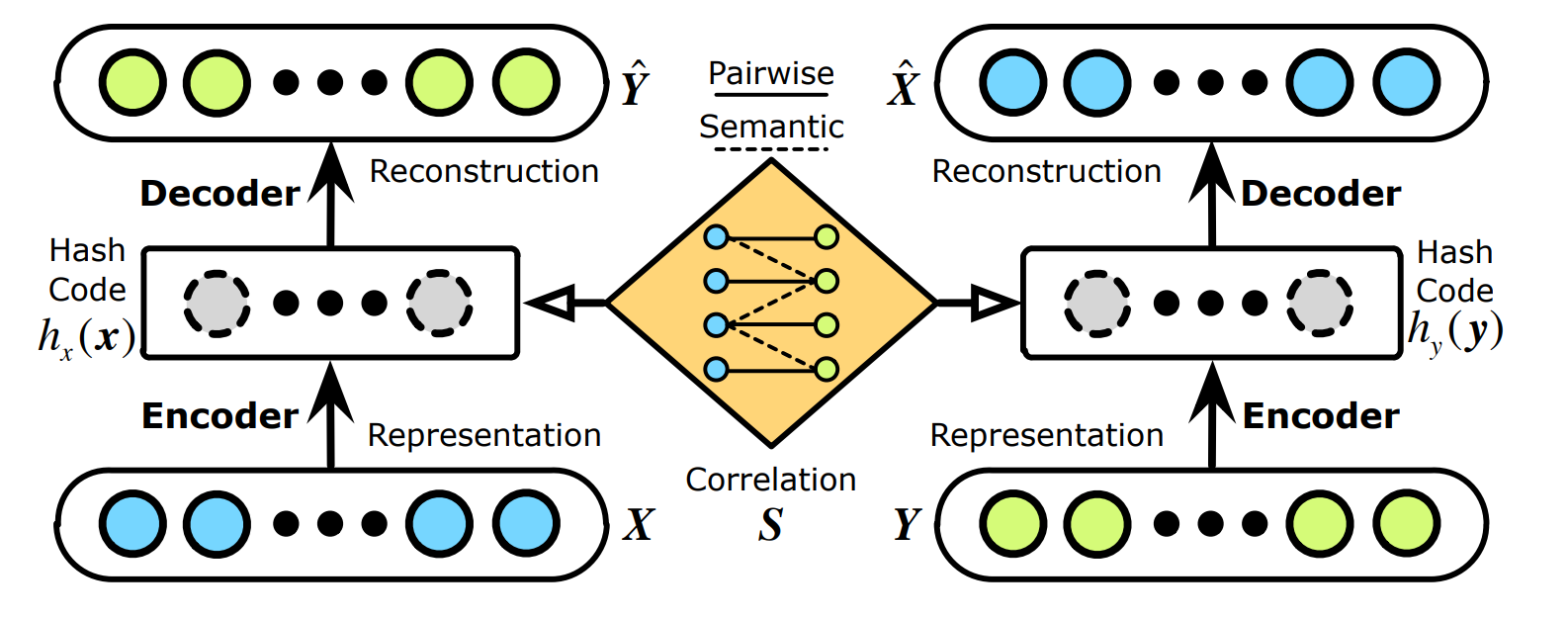}
        \label{fig:aesindivspaces}
    }
    \quad
    \subfloat[][\acrshort*{ae}s with joint latent spaces. Courtesy of \cite{zhao2019sound,wang2018associative}]{
        \includegraphics[width=0.67\textwidth]{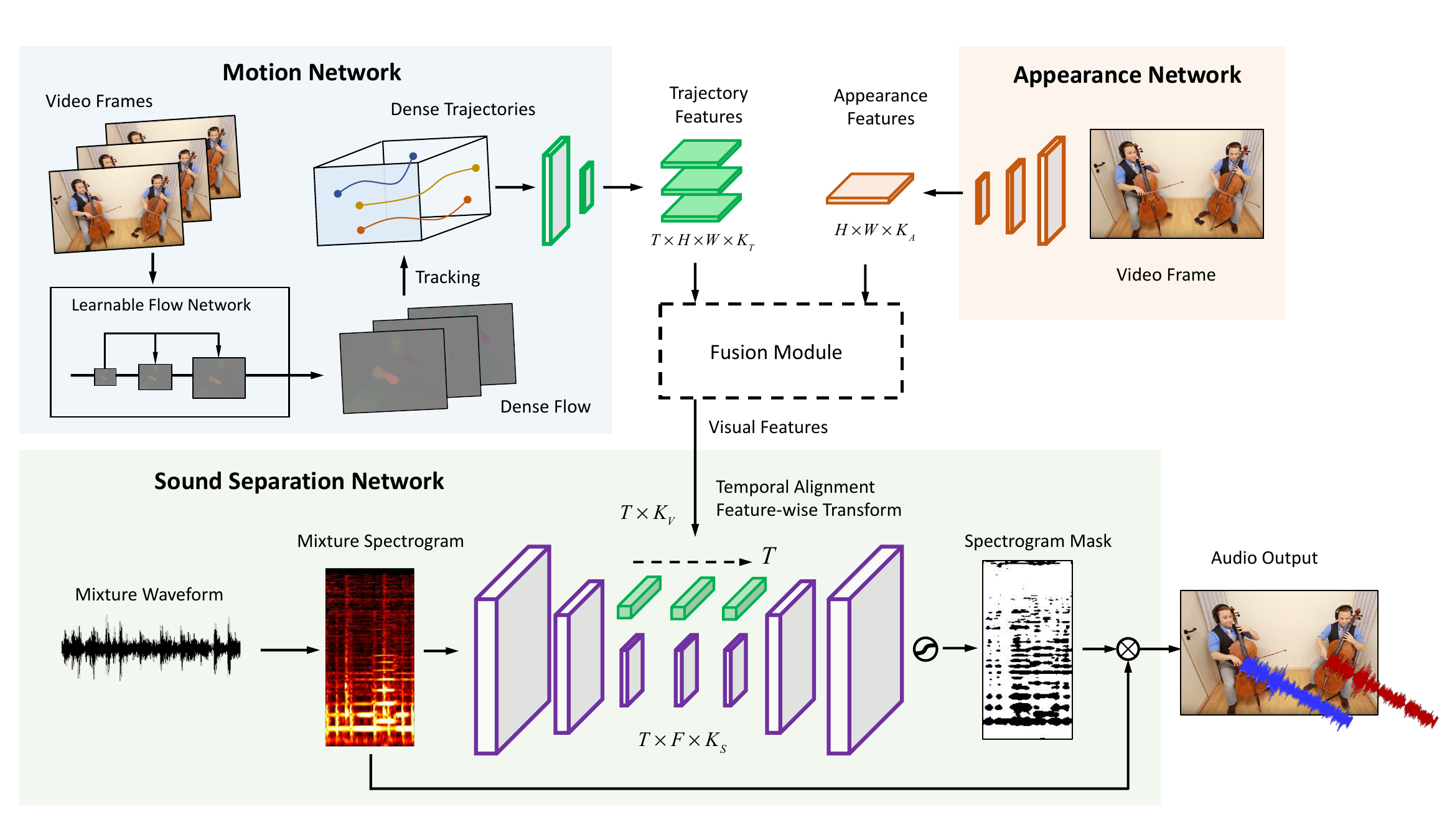}
        \includegraphics[width=0.3\textwidth]{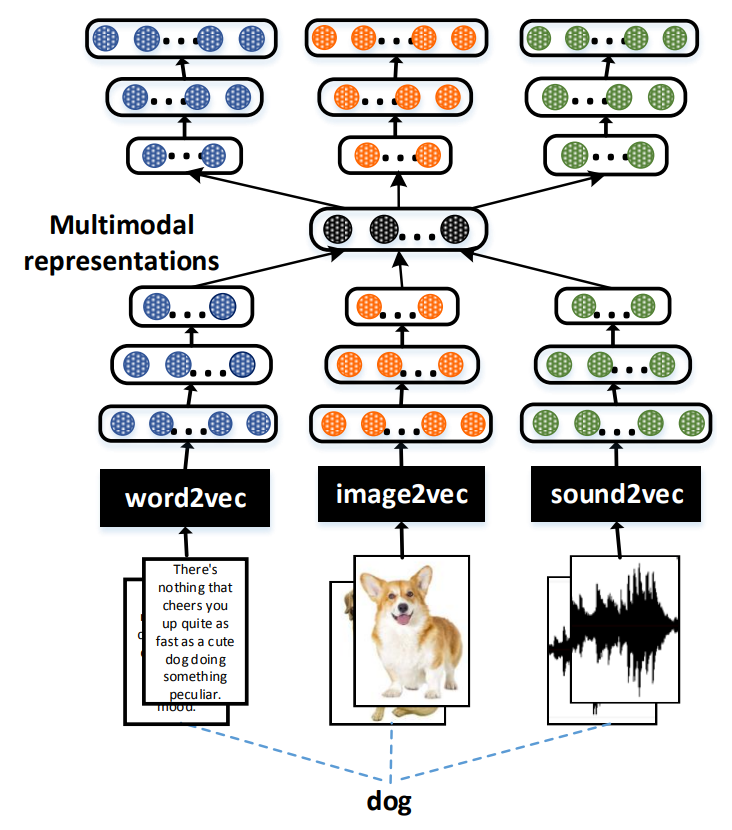}
        \label{fig:aesjointspaces}
    }
    \quad
    \subfloat[][Variational \acrshort*{ae}s. Courtesy of \cite{zhu2021learning}]{
        \includegraphics[width=0.8\textwidth]{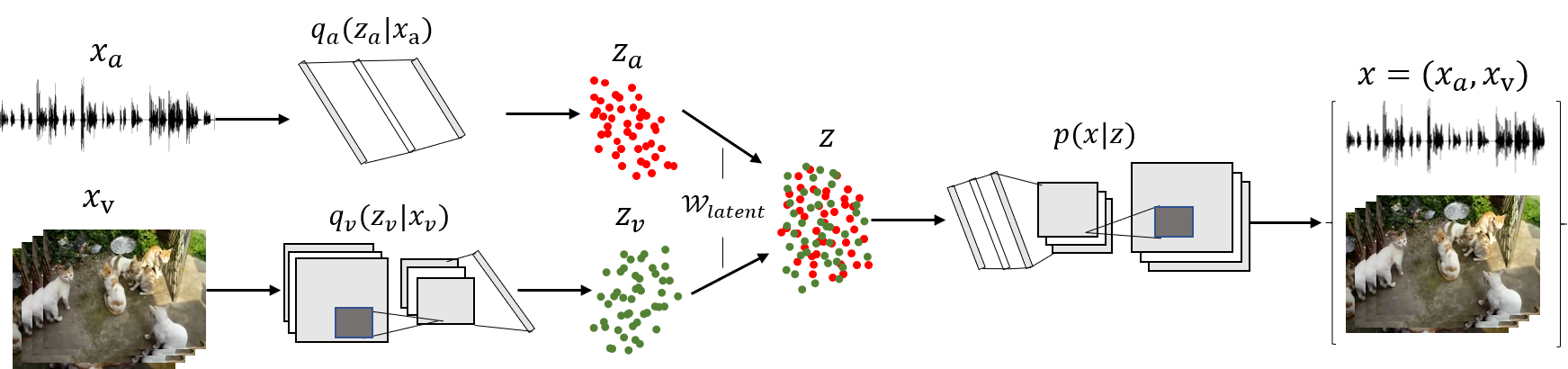}
        \label{fig:variationalAEs}
    }
    \caption{Approaches based on AEs for \acrshort*{avcl}. In (a), we present two examples where the authors proposed to leverage correlations between unimodal sub-spaces and their reconstructions. (b) also illustrates two examples of \acrshort*{ae}s that rely on a shared/single latent space for reconstructing all modalities. (c) illustrates the use of Variational \acrshort*{ae}s to reconstruct multiple modalities conditioned on multiple inputs to learn their correlations.}
    \Description{Approaches based on AEs for \acrshort*{avcl}. In (a), we present two examples where the authors proposed to leverage correlations between unimodal sub-spaces and their reconstructions. (b) also illustrates two examples of \acrshort*{ae}s that rely on a shared/single latent space for reconstructing all modalities. (c) illustrates the use of Variational AEs to reconstruct multiple modalities conditioned on multiple inputs to learn their correlations.}
    \label{fig:aesforavcl}
\end{figure*}

\acrshort*{ae}s are unsupervised encoder-decoder models that create latent representations through reconstruction. The encoder $f_{\theta_1}(x)$ computes the latent space $h=f_{\theta_1}(x)$ from the input $x$ and the decoder computes its reverse mapping, $\hat{x}=g_{\theta_2}(h)$. Therefore, the latent representation reflects the structural distribution of the original data (similar to summarizing/reducing dimensionality). We identified the following methodologies (Figure \ref{fig:aesforavcl}) using \acrshort*{ae}s for \acrshort*{av} applications:\par

\begin{enumerate}
    \item[i.] Using a shared latent representations obtained:
        \begin{enumerate}
            \item[(a)] through concatenation of uni-modal latent spaces \cite{wang2018associative};
            \item[(b)] through attention between uni-modal latent spaces \cite{zhao2019sound,zhao2018sound}.\par
        \end{enumerate}
    \item[ii.] Injecting supervised semantic context using the latent space for classification \cite{rajan2021robust};
    \item[iii.] Imposing similarity constraints between separate latent spaces \cite{cao2016correlation};
    \item[iv.] Injecting additional information through attention in the latent spaces to condition the reconstruction \cite{gao2019co,lin2021exploiting,hu2021dmman};
    \item[v.] Leveraging similarities between reconstructions and cross-reconstructions \cite{recasens2021broaden,wang2018associative,rajan2021robust};
    \item[vi.] \acrfull*{vae} reconstruction conditioned by multiple inputs \cite{zhu2021learning,zhang2021variational,sadeghi2019audio};
    \item[vi.] Leveraging subsequent reconstructions \cite{pham2019found}.
\end{enumerate}

\textbf{Discussion:} Through reconstruction, \acrshort*{ae}s are an efficient way to reduce the dimensionality of input features while preserving the same distribution of the original data, which can be applied for creating individual latent spaces of the same size (can be used as projectors) and balancing the information between coordinated sub-spaces (Figure \ref{fig:aesindivspaces}). For instance, the discrepancy between modalities can be reduced by \acrshort*{ae}s when the input features are obtained with different feature extractors (different information compression levels). This helps to balance the performance scores in applications where only one modality is given as input (e.g., cross-modal retrieval, retrieval in one modality using the other).\par

Instead of strictly defined relations between modalities, using a shared latent space in reconstruction has the advantage of implicitly allowing each modality to leverage insight from itself and others (Figure \ref{fig:aesjointspaces}). This results in a lower heterogeneity gap than an approach with separate latent spaces. We have seen consistently better results than traditional methods based on \acrfull*{cca} and Deep \acrshort*{cca} \cite{wang2018associative}. Moreover, shared latent spaces in \acrshort*{ae} models make the input share the same context and apply constraints that affect all modalities simultaneously.\par

\begin{figure*}[!ht]
    \centering
    \subfloat[][MAE Architecture. Courtesy of \cite{he2022masked}]{
        \includegraphics[width=0.5\textwidth]{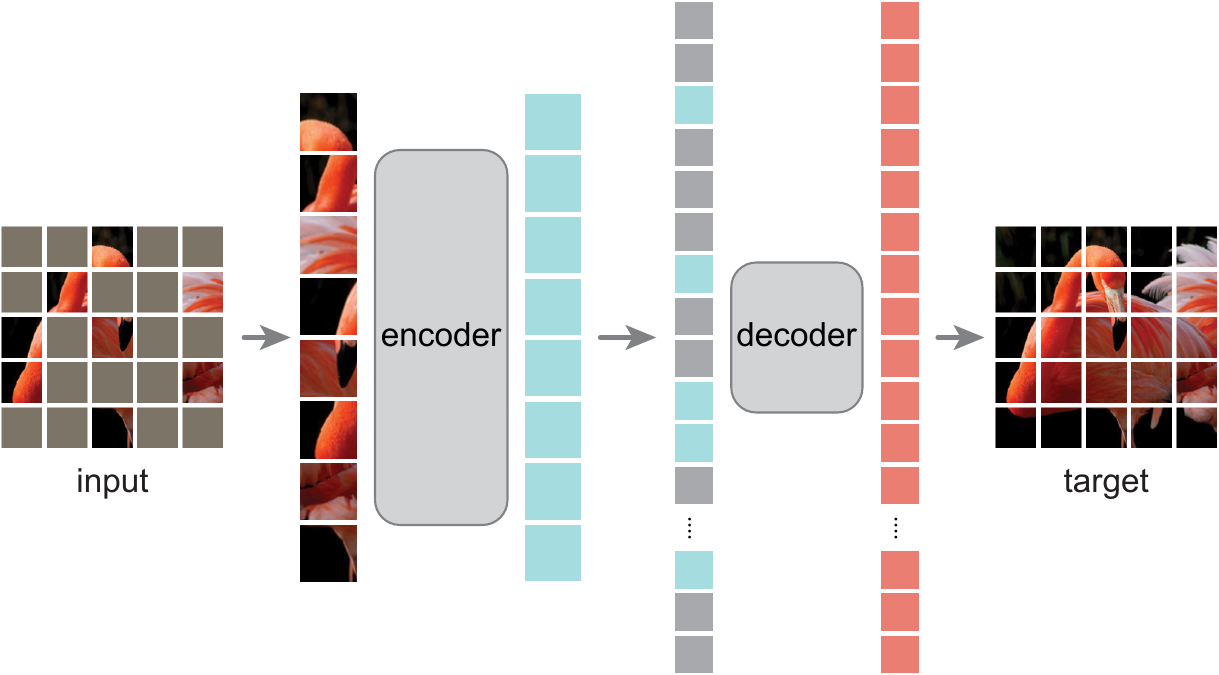}
        \label{fig:mae}
    }
    \quad
    \subfloat[][VideoMAE Architecture. Courtesy of \cite{tong2022videomae}]{
        \includegraphics[width=0.9\textwidth]{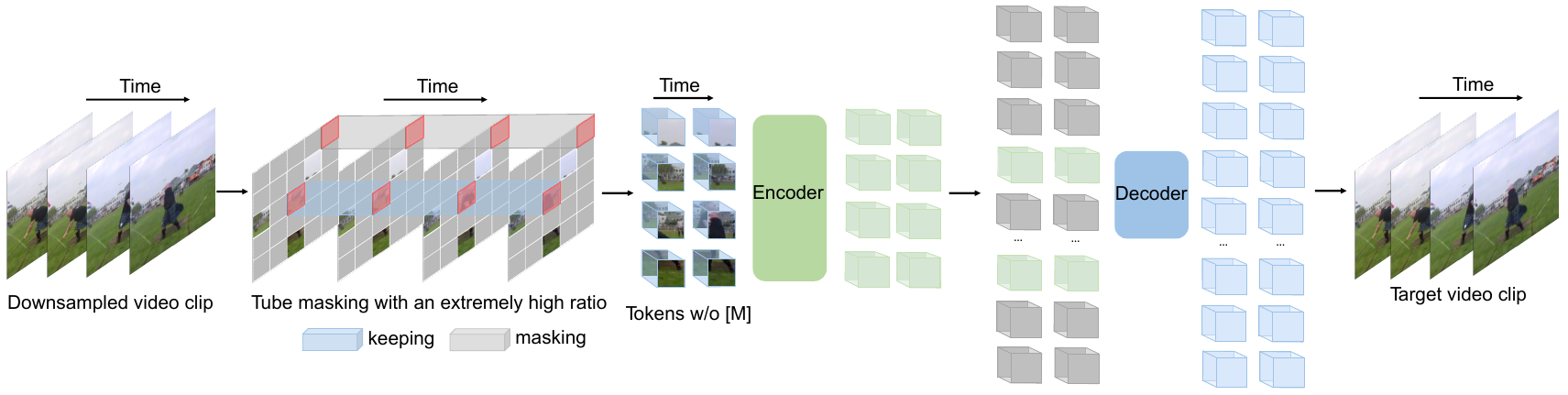}
        \label{fig:videomae}
    }

    \caption{Masked Auto-Encoder Architectures for Image and Video Frame sequences}
    \Description{Masked Auto-Encoder Architectures for Image and Video frame sequences}
    \label{fig:mae-model}
\end{figure*}

Recently, masked sequence modeling proposed in NLP \cite{bert2018,hubert2021} has been extensively used in audio-visual models \cite{bao2021beit,zhou2021ibot,wang2022bevt,wei2021masked,sun2023masked,tong2022videomae,wang2023videomaev2,huang2022mavil,arnab2022audiovisual}. It consists in selectively masking some elements within a sequence of data, e.g. frames of a video clip, and train deep models that can reconstruct the original input. This methodology can be used together with the TF architecture and obtain representations that allow to achieve \acrshort*{soa} results in downstream tasks. In Figure \ref{fig:mae-model}, we illustrate the schematic of the Masked Auto-Encoder architecture \cite{he2022masked} applied to images and video frame sequences. As referred in Section \ref*{section:video-feature-extrc}, these models can be used to reconstruct original videos (only visual data) by masking 3D regions (temporal axis) \cite{tong2022videomae,wang2023videomaev2}. Masked modeling improves the understanding of global context, resulting in better reconstructions and more valuable latent representations.\par

Some works proposed reducing the vector embeddings' size using Vector Quantization (VQ). VQ involves mapping the input vectors to a discrete set of codes from a predefined codebook \cite{van2017neural}. This allows for improving the quality of the generated data samples and the computational efficiency of the model. This model builds upon the VAE architecture but applies a vector quantization layer to map the latent representations into the discrete codes using a ``codebook''. The ``codebook'' is a fixed number of entries (vector) representing the original high-dimensional raw data. The encoder transforms the input, which is subsequently mapped to the closest corresponding vector in the codebook. Afterward, the decoder uses the set of quantized vectors to reconstruct the original information. This allows us also to create interpretable/structured latent spaces because the ``codebook'' is known a priori. The ``codebook'' can be learned or defined by hand. However, training the VQ-VAE model is more expensive (computationally) than AEs because it also introduces a loss term that encourages embeddings for each entry close to the codebook entries. Depending on the size of the dataset, VQ-VAEs are attractive solutions due to their compression capabilities. Still, they suffer from limited generalization for big datasets due to the fixed code size.\par

As far as we know, this model's most relevant application is to preprocess audio clips. Audio tokenization consists of learning discrete representations for audio/speech \cite{dhariwal2020jukebox,hubert2021,chiu2022self,zhu2022quantized}. When pairing the VQ approach with autoregressive models, such as the Transformer, it is possible to obtain high-quality outputs \cite{dhariwal2020jukebox,hubert2021,chiu2022self,chen2022beats,li2023audioformer}. Nevertheless, we still have not seen relevant applications of VQ in Deep \acrshort*{avcl}.\par

\subsubsection{\textbf{Generative Adversarial Networks}}
\label{subsubsection:gan-model}

\acrfull*{gan} are composed by two components: a generator $G$ and a discriminator $D$. The generator $G$ learns to produce target samples as similar to the real ones as possible to confuse the discriminator $D$, which attempts to distinguish generated samples from the real ones, keeping itself from being confused. Instead of directly mapping a latent representation from the data as \acrshort*{ae}s, \acrshort*{gan}s learn to implicitly map latent random representations to ``samples belonging to the training set'', which try to narrow the difference between distributions of real (i.e., training set) and generated data.\par

\begin{figure}
    \centering
    \subfloat[][Example: GAN for heterogeneity gap minimization (generation between modalities) (i. \& ii.). Courtesy of \cite{zheng2021adversarial}]{
        \includegraphics[width=0.57\textwidth]{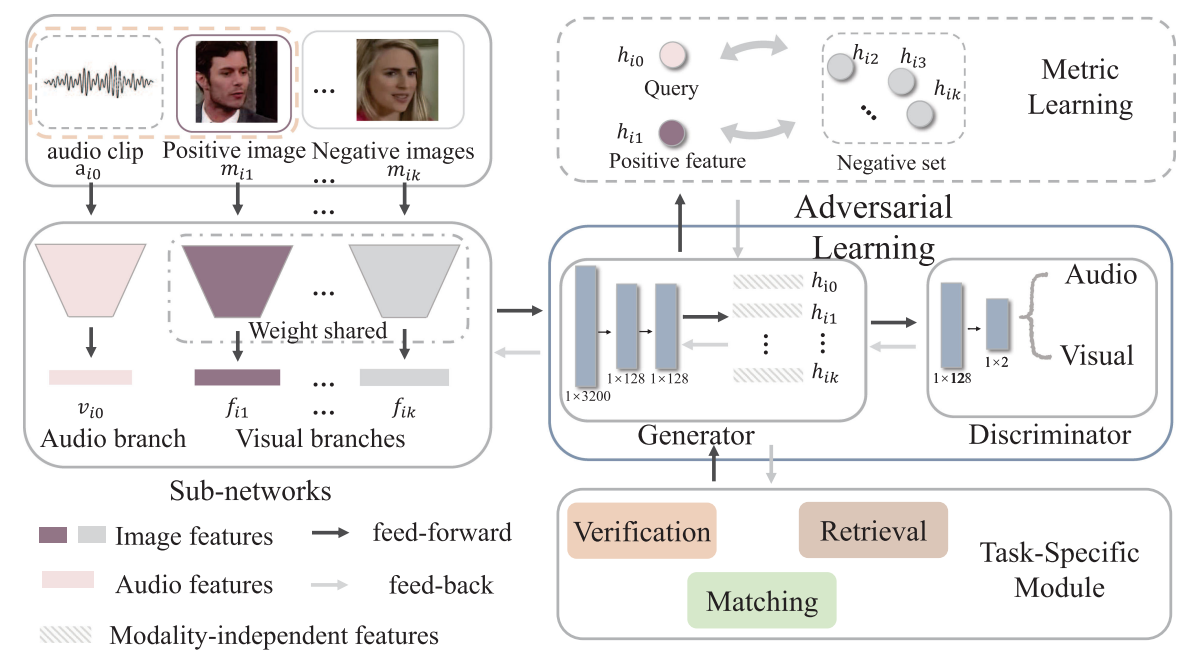}
        \label{fig:gan12}
    }
    \quad
    \subfloat[][Example: joint optimization between adversarial and other functions (iv.). Courtesy of \cite{mira2021end}]{
        \includegraphics[width=0.37\textwidth]{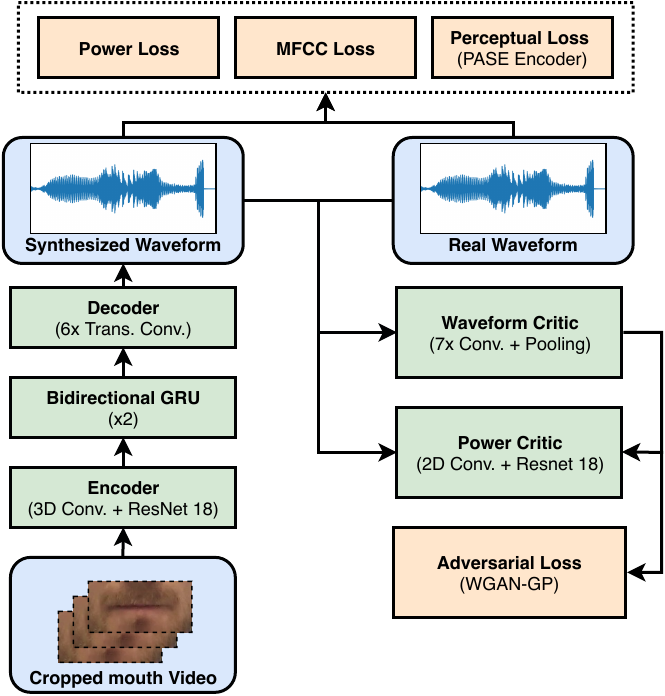}
        \label{fig:gan4}
    }
    \quad
    \subfloat[][Example: condition the discriminator's decision and the generator's input (iii.). Courtesy of \cite{athanasiadis2020audio}]{
        \includegraphics[width=0.7\textwidth]{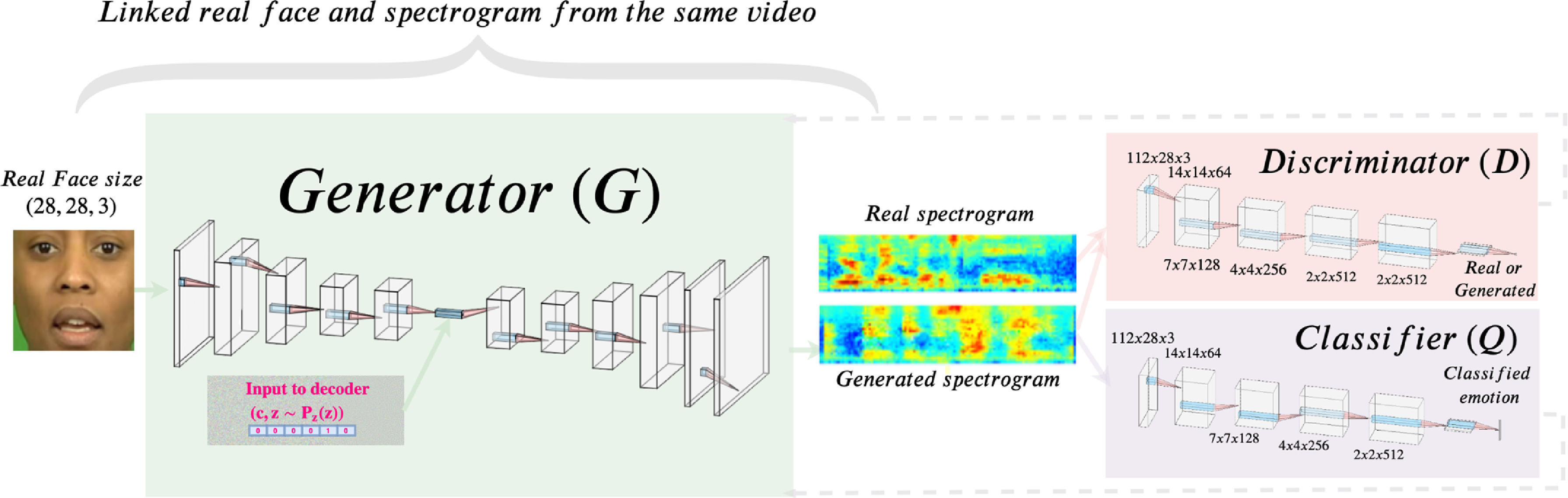}
        \label{fig:gan3}
    }
    \caption{Examples of GANs applied for \acrshort*{avcl}: (a) illustrates an example of performing cross-reconstruction using GANs; (b) consists of optimizing a GAN model together with other loss functions. In this example, the authors applied functions that evaluate the quality of the reconstructed output; (c) consists of conditioning the latent space that is fed to the discriminator, thus injecting semantics in the reconstruction process.}
    \Description{Examples of GANs applied for \acrshort*{avcl}: (a) illustrates an example of performing cross-reconstruction using GANs; (b) consists of optimizing a GAN model together with other loss functions. In this example, the authors applied functions that evaluate the quality of the reconstructed output; (c) consists of conditioning the latent space that is fed to the discriminator, thus injecting semantics in the reconstruction process.}
    \label{fig:ganforavcl}
\end{figure}

For multimodal data, \acrshort*{gan}s can be used to translate different modalities. Moreover, their data generation process can also be conditioned using additional information, allowing them to learn a multimodal model. Therefore, as generative models, \acrshort*{gan}s can be used to learn representations, allowing to encode (multimodal) data. We identified the following methodologies based of \acrshort*{gan}s, which are also illustrated in Figure \ref{fig:ganforavcl}:\par

\begin{enumerate}
    \item[i.] Heterogeneity gap minimization through adversarial learning \cite{zheng2021adversarial,seo2020hmtl};
    \item[ii.] Using GANs and AEs for generation between modalities \cite{mira2021end,fanzeres2021sound,athanasiadis2020audio};
    \item[iii.] Condition the discriminator by injecting context semantics in the form of:
        \begin{enumerate}
            \item[i.] Class-labels \cite{athanasiadis2020audio};
            \item[ii.] Latent representations of other modalities \cite{fanzeres2021sound,zhu2022quantized};
            \item[iii.] Classifiers' prediction on generated samples \cite{athanasiadis2020audio}.
        \end{enumerate}
    \item[iv.] Joint optimization between adversarial and other loss functions \cite{seo2020hmtl,mira2021end}.
\end{enumerate}

\textbf{Discussion:} For retrieval between modalities with audio-visual data, the standard \acrshort*{gan}s adversarial scheme can be used to minimize the heterogeneity gap and obtain modality-independent representations \cite{zheng2021adversarial} or to transfer knowledge between modalities that can be used to learn coordinated representation spaces \cite{seo2020hmtl}. These methods are often complemented with deep metric learning frameworks \cite{zheng2021adversarial} or classification \cite{seo2020hmtl} to include semantic context. In contrast, from a generational point of view, we can also leverage reconstruction. The random vector is replaced by latent representations obtained from a model, typically an encoder-decoder, used for generating one modality with the other \cite{mira2021end,fanzeres2021sound,athanasiadis2020audio}. In this case, the main differences lie in exploring complementary information. The discriminator receives an input and tries to classify it as real or not, but its decision can be conditioned by adding information in the last layer before the classification. In addition, these models also allow to consider several sources of conditional information: latent representation (from \acrshort*{ae}s) \cite{fanzeres2021sound} or class-label information.\par

\acrshort*{gan}s perform generation of new samples in an unsupervised way, which reduces the dependency on training data and allows coping with scenarios of missing data. However, the randomness of the latent vector makes it hard to obtain the context semantics associated with it. This is an active research topic, and several works have been proposed to mitigate this through regularization or by adding auxiliary information. In addition, \acrshort*{gan}s suffer from training instability, requiring a careful tune of its hyper-parameters due to the adversarial nature of the process that makes the models' parameters oscillate. We believe that leveraging the generative nature of \acrshort*{gan}s (i.e., artificial data) will allow the development of more complex self-supervised frameworks, where the proxy tasks depend on samples generated with the learned data distribution. Thus, we expect to see their presence increase in future proposals for Deep \acrshort*{avcl}.\par

More recently, diffusion models have been applied to the audio-visual domain. Diffusion Probabilistic Models are generative models that learn to convert a simple Gaussian distribution into the original data distribution \cite{sohl2015deep}. It consists of forward diffusion and reverse denoising, where the input is corrupted with Gaussian noise (forward diffusion), and the model learns to remove the noise. Thus learning the data distribution of the input data.\par

Diffusion can be implemented to perform cross-generation \cite{ramesh2021zero,gu2022vector,zhu2022discrete}. However, proposals typically focus on unimodal generation while incorporating additional information in the denoising process. The process can be formulated as follows:
$$
    p_{\theta}(x_{t-1} | x_t, \textbf{c}) = \mathcal{N}(x_{t-1}; \mu_\theta (x_t, t, \textbf{c}), \Sigma_\theta (x_t, t)), \qquad \text{where} \quad \Sigma_\theta (x_t, t) = \beta_{t}I
$$
$p_{\theta}(x_{t-1} | x_t, x_0, \textbf{c})$ is the posterior distribution between ``denoising'' steps also conditioned using additional information. Examples include class labels \cite{dhariwal2021diffusion}, text prompts \cite{gu2022vector}, and information from additional modalities (e.g., audio if video generation is being performed) \cite{ao2023gesturediff,zhu2023taming,lee2023imaginary,choi2023diffv2s}. Several works apply the two diffusion steps over pre-trained latent spaces (latent diffusion models), where the latent space can be encoded from multiple modalities \cite{ao2023gesturediff,rombach2022high}. In these models, unimodal denoising is performed over the conditioned latent representations. In contrast, the joint distribution between AV can also be learned by denoising (reverse diffusion) samples from multiple modalities \cite{ruan2023mm}. However, it increases the computational complexity of the training loop.\par

Diffusion is at the forefront of current generative models. Despite their performances exceeding GAN-based models in text-to-image synthesis \cite{dhariwal2021diffusion}, there is little work on learning multimodal AV representations using diffusion models. Specifically, there is little exploration on the utilization of latent representations from forward diffusion for downstream tasks. We assume this is mainly due to the computational complexity associated with their training process as it requires multiple forward and backward passes through the model. The effectiveness of diffusion models and normalizing flows heavily relies on their iterative processes, which gradually construct a sample from random noise. This incremental refinement, involving sequential steps of evaluating large neural networks, leads to slower sampling speeds compared to single-step methods such as \acrshort*{gan}s and \acrshort*{vae}s \cite{ruan2023mm}. This inefficiency presents a challenge for real-time applications.

\subsection{Objective Functions}
\label{subsection:learning-frameworks}

Objective functions are used to guide the solution of optimization problems. In Deep \acrshort*{avcl}, it helps to adjust audio-visual embedding models through back-propagation to maximize the correlation between modalities in the joint sub-space. This can help to maintain and capture specific audio-visual semantics. Typically, several regularization terms constrain the embeddings to mitigate the heterogeneity gap and further increase correlation.\par

\textit{i. \textbf{\acrfull*{ce} and other distance-based extensions}} can be used \cite{he2019new,zhen2019deep} to achieve inter-class discrimination via a projection to a latent semantic sub-space. \acrshort*{ce} uses entropy as a measure of distance between the true and the predicted distribution as follows:
\begin{align}
    L(\hat{y}, y) = -\sum_{n=1}^{N} y_{n} \log(\hat{y_{n}}),
\end{align}
where $\hat{y}$ and $y$ are the softmax probability of the n class and ground truth, respectively. In contrast, distance-based proposals use distance measures between joint spaces with the same size as the one-hot encoded vectors as follows:
\begin{align}
    L(\hat{y}, y) = \frac{1}{n} \sum_{n=1}^{N} || \hat{y} - y ||_2^2
\end{align}
This methodology is used to minimize the distance between the \acrshort*{nn}'s output and the ground truth \cite{zhen2019deep}.\par

The distance losses (L1 or L2) can be used to approximate two different models. This approach is known as Knowledge Distillation (KD) and works by conditioning the model's weights of the student (smaller) model using the teacher's (bigger model) predictions/embeddings. The goal is to minimize the discrepancy between the predictions of both models. In other words, this learning paradigm allows the student model to exploit the ``rich knowledge'' from the teacher model. Based on this idea, the KD loss function often takes the form of:
\begin{align}
    L_{KD} = \frac{1}{n}\sum_{i=1}^{n} (\hat{a_i} - \hat{a_i}^{teacher})^2 + \frac{1}{n}\sum_{i=1}^{n} (\hat{v_i} - \hat{v_i}^{teacher})^2
\end{align}
where the distance between embeddings or predictions for both modalities, i.e., $a_i$ and $v_i$, is minimized. By using knowledge distillation, smaller and less complex models can be trained to achieve a performance close to the teacher. Thus allowing for a reduction in the amount of resources needed in terms of memory and computation.

\textbf{Discussion:} Despite being able to establish relations between semantics, they do not enforce any intra-class relation. To ensure intra-class consistency, some proposals use the center loss as regularization (L1 distance between the center of each class and each uni-modal embedding) \cite{zhen2019deep,he2019new}. Similarly, to align both modalities and reduce the heterogeneity gap, some works constrain them to have similar representations using the distance between embeddings (e.g., L2, cosine similarity) \cite{min2021cross}. All these terms are typically leveraged in joint optimization schemes (i.e., a weighted sum of linear components) for \acrshort*{avcl}.\par

\textit{v. \textbf{Correlation Loss (Corr. L.) and extensions}}: The correlation loss is an extension of the Deep CCA approach \cite{zhang2021variational,zeng2021learning}. It involves projecting both modalities into a latent space of the same size and maximizing their correlation. This is done by reducing the discrepancy between modalities and forcing the cross-correlation matrix to be as close to the identity matrix as possible. The loss function takes the form of:\par
\begin{align}
    L_{corr} = \frac{1}{n} \sum_{i=1}^{n} \frac{W_a^T\Sigma_{av}W_v}{\sqrt{W_a^T\Sigma_{aa}W_a . W_v^T\Sigma_{vv}W_v}},
\end{align}
where $W_a$ and $W_v$ are the weights associated with each projection, and $\Sigma_{aa}$, $\Sigma_{vv}$ and $\Sigma_{av}$ are the covariance and cross-correlation matrices.\par

Additionally, the Corr. L. can also take the form of the cross-entropy function \cite{zhen2019deep}. In this scenario, we must first calculate the cosine similarity between modalities. The similarities between samples are fed to a ``logistic regression model'' that assumes pseudo-labels: positive class (1) if the two input vectors are from the same class; otherwise, a negative class (0). This requires the model to have supervised information about the class.
\begin{align}
    L([u_a, u_v], y)_{corr CE} = \frac{1}{n^2} \sum_{i,j = 1}^{n} (\log(1+e^{A_{ij}}) - y A_{ij}) = -A_{ij} * y + log(1 + e^{A_{ij}})
\end{align}
where, $A_{ij} = \frac{1}{2}\cos(u_a, u_v)$. This is similar to the NCE loss function. Instead of maximizing correlation, we maximize the similarity between embeddings.\par

\textbf{Discussion:} Despite having the same goal of our task (\acrshort*{avcl}), the Corr. L. is mainly used as an additional loss term. It provides good results in small tasks \cite{zhen2019deep}, but it still requires the computation of the cross-correlation matrix between modalities. It does not escalate properly to a large dataset. Thus, the CE variation (inspired by NCE) was proposed \cite{zhang2021variational,zeng2021learning}. Similar to \cite{zbontar2021barlow}, Cross. L. (CE) can maximize the correlation between paired modalities. However, to add semantic context, we must rely on pseudo-labels as defined above (i.e., maximizing the correlation for each class separately). Corr. L. does not encourage the separation of classes. It only encourages them to be close. Therefore, it is important to refer to the extensions of NCE to find answers to this challenge.\par

\textit{v. \textbf{Adversarial Loss (Adv. L.) and extensions}:} The Adversarial Loss (Adv. L.) function is used to optimize \acrshort*{gan}s. The Generator ($G$) generates samples obtained from the training set, and the Discriminator ($D$) discriminates between real and artificially generated elements. The goal of $G$ is to minimize the difference between generated samples and real ones, while $D$ aims the opposite. Optimization is performed iteratively by updating the weights of one component while keeping the weights fixed on the other. This is called Adversarial Learning because both sub-models learn in opposite directions, i.e., they evolve alternatively.\par

The $G$ maps a latent representation $z$ consisting of a random distribution into a sample $x'$ belonging to the same space of the training data $x$, i.e., $G(z, \theta_g) = x'$. $G$ can also be conditioned, i.e., accept an embedding in input instead of $z$. The $D$ receives the real or generated data as inputs, $x$ or $x'$ respectively, and predicts the probabilities that each element belongs to the real distribution, i.e., $D(x, x', \theta_d) = p$. As the $G$ approaches the real data-generating distribution, the accuracy of $D$ should decrease in theory. However, this iterative process depends on $G$, meaning that $G$ and $D$ evolve together. Therefore, the standard \acrshort*{gan} is optimized in an adversarial way as follows:
\begin{align}
    \min_G \max_D V(G,D) = \min_G \max_D E_{x\sim p_{data}(x)} [log D(x)] + E_{z \sim p_{z}(z)} [(1-log D(G(z)))]
\end{align}

\textbf{Discussion:} These models can be extended by imposing correlation constraints and separating intra from inter-modal features \cite{peng2019cm}, adding a classification term to the standard adversarial loss (i.e., min-max loss) \cite{xu2019deep} or using metric learning approaches to learn common embedding spaces (e.g., triplet loss) \cite{xu2020joint,wang2017adversarial}.\par

\textit{v. \textbf{Evidence Lower Bound Loss (ELBO)}}: ELBO is a loss function used for probabilistic inference tasks with latent variables, such as \acrshort*{vae}s. Take as a visual example the representation of the \acrshort*{vae} model in Figure \ref*{fig:variationalAEs}. Let's consider only the scenario with one modality ($x_a$).\par

Assume we have our observed data ($x_a$) modeled using a random variable $Z$ ($Z_a$ and $Z$ are the same). Both random variables are distributed as a joint distribution $p(Z_a, Z;\theta)$, where $\theta$ parameterizes the distribution (often Gaussian). The latent variable model is defined as:
\begin{align}
    p(Z|x) = \frac{p(x|Z)}{p(x)}p(Z)
\end{align}
Estimating this probability distribution is intractable due to $p(x)$. We use a neural network with parameters $\theta$ to approximate the distribution $p(x|Z)$, giving us $p_\theta(x|Z)$. Given this, we want to maximize the likelihood of $p(x, Z;\theta)$. Thus we calculate:
\begin{align}
    log\int_{z}p_{\theta}(x,Z) dz
\end{align}
This idea is to perform statistical inference, which aims to infer the value of one random variable given the observed value of another distribution. In this example, we want to approximate the probability distribution $p_{\theta}(Z|x)$ with the distribution $q_\theta(Z|x)$. If we extend the likelihood function, we obtain the ELBO loss function:
\begin{align}
    L = log\int_{z}p_{\theta}(x,Z) dz = log \int_z p_\theta(x,Z)\frac{q_\theta(Z|x)}{q_\theta(Z|x)} dz = log \mathop{\mathbb{E}}_{Z \sim q_{\theta}(Z|x)}[\frac{p_{\theta}(x,Z)}{q_\theta(Z|x)}] >= \mathop{\mathbb{E}}_{Z}[log \frac{p_{\theta}(x,Z)}{q_\theta(Z|x)}]
\end{align}
Evidence (L) is the term we give to a likelihood function with fixed parameters. If we use the Kullback-Leibler (KL) divergence between $p(Z|x)$
and $q(Z|x)$ to minimize the gap between distributions we obtain:
\begin{align}
    KL(q(Z|x) || p(Z|x)) = log(p(x)) - \mathbb{E}_{Z \sim q_{\theta}(Z|x)} [\frac{p_{\theta}(x,Z)}{q_\theta(Z|x)}]
\end{align}
If we further arrange the equation, we obtain the following:
\begin{align}
    L = \mathbb{E}_{z \sim q_\theta(z|x)} [p_\theta(x|z)] - KL(q_\theta(z|x) || p(z|x))
\end{align}

\textbf{Discussion:} ELBO is a trade-off between the reconstruction accuracy (computed using cross entropy between real data and reconstructed data) against the complexity of the variational posterior. The KL term can be interpreted as a measure of similarity between distributions. This loss function is only employed in probabilistic latent models \cite{zhu2021learning,zhang2021variational,sadeghi2019audio}. It is used to maximize the correlation between modalities through reconstructions between modalities.\par

\textit{ii. \textbf{Hinge Loss (and extensions)}} are used in several works to increase compactness and separability at the same time because traditional \textit{softmax} losses cannot directly enforce them. Given a similarity function between embeddings (e.g., cosine similarity), it imposes a margin between correctly classified and misclassified samples. Comparing positive and negative relationships is allowed between embeddings, where the negative samples are drawn from each mini-batch, and the margin controls its amount. This approach can improve the capability of correlation learning by exploiting the comparison between different modalities of classified and misclassified samples. Margins can be fixed (hinge loss) to establish a fixed distance between matched and mismatched pairs' similarity. It is computed between two pairs of embeddings as follows:
\begin{align}
    L(c, i) = \sum_{(c,i) \in \beta} ( \max(0, \cos(c,i) + \alpha) + \max(0, \cos(c,i) + \alpha) ),
\end{align}
where $i,c$ are pairs with different modalities (i.e., audio and visual data from the same clip) \cite{merkx2019language}. $cos()$ represents the cosine similarity between embeddings, but other similarity functions can replace it.\par

Because the relationships between embeddings change during training, margins can be progressively adapted to fit those changes. Therefore, \cite{ilharco2019large,rouditchenko2020avlnet} proposed to use monotonically increasing margins (i.e., \acrfull*{mms}). Works that leverage \acrshort*{mms} typically use it bidirectionally between audio and visual data. Therefore, using a mini-batch of size $B$, it is computed as follows:\par

\begin{gather}
    L_{mms} = L_{av} + L_{va}, \nonumber \\
    L_{av} = -\frac{1}{B} \sum_{i=1}^B \log [\frac{e^{\cos(ii) - \sigma}}{e^{\cos(ii) - \sigma} + \sum_{i=1}^B M_{ij}e^{\cos(ij)}}], \qquad
    L_{va} = -\frac{1}{B} \sum_{i=1}^B \log [\frac{e^{\cos(jj) - \sigma}}{e^{\cos(jj) - \sigma} + \sum_{i=1}^B M_{ij}e^{\cos(ij)}}],
\end{gather}

where $M_{ij}$ controls whether both embeddings are positively associated (masking). The similarity between embeddings ($\cos$) is subject to a margin $\sigma$, scheduled to change during training according to a predefined scheduler. Nevertheless, margins can adapt to the current mini-batch by using the mean distance between positive and negative pairs \cite{monfort2021spoken}.\par

\textbf{Discussion:} In sum, margin-based approaches for \acrshort*{avcl} allow sampling negative pairs efficiently (randomly drawn), but the initial margin values must be carefully tweaked to avoid divergence. Moreover, dynamic margin heuristics can alleviate the issue of under or oversampling negative pairs but at the cost of creating different convergence rates for similar models.\par

\textit{iii. \textbf{\acrfull*{tl} (and extensions)}} is calculated with three samples at each training step (anchor - input, positive and negative). Using a projector with shared weights (for positive and negative samples), it aims to maximize the distance between the anchor (input) and the negative while minimizing its distance to the positive sample. It is computed as follows:
\begin{align}
    L_{triplet}(x,x^+,x^-) = \sum_{x \in \mathcal{X}} \max(0,d_{pos}-d_{neg}+\zeta), \quad d_{neg} = ||f(x)-f(x^-)||^2_2, \quad d_{pos} = ||f(x)-f(x^+)||^2_2,
\end{align}
where $\zeta$ is the minimum distance between the relations of matched and mismatched pairs. This approach requires selecting challenging negative samples for fair convergence, which slows training and increases computational demands.\par

The combination of positive and negative samples can be done in several ways. \cite{ramaswamy2020see} gathered only positives from the same media asset. Similarly, \cite{pretet2021cross} extended this idea and used it bidirectionally by leveraging each modality as an anchor separately and jointly optimizing the model. After learning correlations between different modalities, the authors could relate different media assets by measuring distances between embeddings. In contrast, the triplet loss can also leverage supervised information (using class labels as reference - anchors, and positive samples) to maximize semantic discrimination \cite{zeng2020deep}. In sum, adding semantic information depends on how anchors and positive examples are gathered.\par

\textbf{Discussion:} The triplet loss is limited to one-to-one relationships between samples. For this reason, several extensions were proposed to generalize the comparison with multiple negative or positive samples \cite{horiguchi2018face,he2019new} (\acrfull*{lsl}, \acrfull*{npl}, \acrfull*{ql}). However, they still have the burden of requiring a methodology for selecting the best negative candidates. Therefore, self-supervised learning paradigms such as contrastive learning and noise contrastive estimation have a computational advantage due to the simplicity of their sampling approaches.\par

\textit{iv. \textbf{\acrfull*{cl} (and extensions)}} takes a pair of inputs and minimizes the distance of samples from the same class (or from pairs of samples), maximizing it otherwise. In contrast to triplet loss and margin-based approaches, the negative samples are usually obtained randomly, reducing their dependency on negative sampling.

\begin{align}
    L_{contr}(x_a, x_v) = 1[y_i = y_j] \cos(x_a, x_v)^2 + 1[y_i \neq y_j]\max(0, \zeta - \cos(x_a, x_v))^2,
\end{align}

where $\zeta$ establishes a margin between samples of different classes. It can be viewed as an extension of margin-based loss functions, but the margin is only applied between positive and negative samples. However, in contrast to triplet loss, it only considers one pairwise relation at each training step.\par

Similarly to previous approaches, for audio-visual data, we identified that negative samples could be obtained in a self-supervised way by using paired modalities to learn correlations/alignment (i.e., only pairing audio and visual data when they describe the same media asset) \cite{hu2020curriculum}. In addition, supervised information can also be leveraged bidirectionally from audio and visual embeddings with the same semantics for ensuring semantic discrimination (i.e., from the same class) \cite{ma2020active}.\par

\textbf{Discussion:} According to the application scenario, negative and positive candidates can be obtained in various ways. For instance, for learning correlations between spatial locations, \cite{chen2021localizing} proposed to contrast image regions with audio segments to learn the association between a given sound and its visual representation. Similarly, for learning temporal alignment between events, \cite{patrick2020multi} used the augmented versions of both modalities (e.g., temporal shifts, augmentations in the embeddings). In sum, depending on the application and how pairs of samples are generated, the contrastive objective function can be used for learning the correlation between different modalities. Alternatively, semantic discrimination is added to the representations when this information is available.\par

\textit{v. \textbf{\acrfull*{nce} and extensions}} are a set of contrastive learning objectives used to distinguish positive and negative pairs of embeddings. Instead of using the original NCE function, current \acrshort*{soa} methods use a surrogate version that maximizes mutual information (InfoNCE). Assuming the input embedding $x_a$ (audio in this example) and a set of $N$ negative samples (usually taken from the opposing modality for alignment purposes - visual data), infoNCE is computed as:
\begin{align}
    L_{nce}(x_a, x_v) = - \log \frac{e^{\frac{\cos(x_a,x_v)}{\tau}} }{e^{\frac{\cos(x_a,x_v)}{\tau}} + \sum_{x'\in N} e^{\frac{\cos(x_a,x_v)}{\tau}}},
\end{align}
where N is the set of negative modality pairs for audio and $\tau$ is the temperature parameter to control the concentration of features in the representation space (it controls the contribution between small and large distances). For example, this framework allows the comparison with multiple negative elements and extensions of \acrshort*{nce} typically include the comparison with multiple positive instances (\acrfull*{mil-nce}) \cite{akbari2021vatt,mm-versatile2020}. Additionally, MIL-NCE can be extended to include multiple positive labels (Multi-Instance Multi-Label Learning (MIML)) \cite{gao2018learning}.\par

\textbf{Discussion:} Temporal co-occurrence is explored between modalities to increase their correlation. \cite{akbari2021vatt} and \cite{afouras2021self} leveraged positive samples from paired modality data and contrasted them with others. This has proven superior performance compared to the learning alignments/correlations approaches. Representations extracted using \acrshort*{nce} (and its extensions) reported top-1 accuracy in several benchmark datasets for audio-visual classification \cite{rouditchenko2020avlnet,akbari2021vatt}. However, \acrshort*{nce} is mainly used for alignment between modalities, and semantic discrimination is left to a posterior phase (as a downstream task) \cite{afouras2021self}.\par

\afterpage{
    \begin{landscape}
    \scriptsize
    \begin{longtable}[c]{llllllllllll}
        \caption{Overview of representative works for audio-visual correlation learning. For the acronyms, P denote projectors, Att denote attention mechanisms, AE denote auto-encoders, and GAN denote Generative Adversarial Networks.}
        \label{tab:methods-final}                                                                                                                                                                                                                           \\
        \hline
        \multicolumn{1}{c}{Paper}        & P.         & Att.       & AE         & GAN        & Loss            & Reg.                               & Learning       & Main Task         & Metrics     & Performance       & Benchmark Datasets             \\ \hline
        \endfirsthead
        \endhead
        \hline
        \endfoot
        \endlastfoot
        \cite{mohamed2022learning}       & X          & \checkmark & X          & X          & CE              & -                                  & Self-Sup.      & Generation        & WER         & 26.9              & LRS3                           \\
        \cite{chen2021vset}              & X          & \checkmark & \checkmark & X          & L1              & -                                  & Self-Sup.      & Generation        & SNR         & 3.14              & AVSpeech                       \\
        \cite{sadeghi2019audio}          & X          & \checkmark & X          & \checkmark & Adv. Loss       & -                                  & UnSup.         & Generation        & SDR         & 13                & NTCD-TIMIT                     \\
        \cite{zhu2022discrete}           & X          & \checkmark & \checkmark & X          & CE              & -                                  & UnSup.         & Generation        & FID         & 28.76             & MSCOCO                         \\
        \cite{fanzeres2021sound}         & X          & X          & \checkmark & \checkmark & Adv. Loss       & MSE                                & Self-Sup.      & Generation        & I           & 0.14              & VEGAS                          \\
        \cite{zhu2021learning}           & X          & X          & \checkmark & X          & ELBO            & Wasserstein                        & UnSup.         & Generation        & MRR         & 0.175             & AVE                            \\
        \cite{mira2021end}               & X          & X          & X          & \checkmark & Adv. Loss       & L1 + L2                            & UnSup.         & Generation        & WER         & 23.13 | 42.51     & GRID | LRW                     \\
        \cite{gan2020foley}              & X          & \checkmark & X          & X          & CE              & -                                  & Sup.           & Generation        & NDB         & 20                & MUSIC                          \\
        \cite{athanasiadis2020audio}     & X          & X          & X          & \checkmark & Adv. Loss       & Gauss. Noise + CE                  & UnSup.         & Generation        & FID         & 49.8 | 59.4       & CREMA-D | RAVDESS              \\
        \cite{zhu2022quantized}          & X          & \checkmark & X          & \checkmark & Adv. Loss       & L1                                 & UnSup.         & Generation        & ACC         & 26.7              & AIST++                         \\
        \hline
        \cite{rajan2021robust}           & X          & X          & \checkmark & X          & MSE             & CCA + CE                           & Sup.           & Recognition       & ACC         & 56.5(A) | 55.2(V) & SEW                            \\
        \cite{wu2021exploring}           & \checkmark & \checkmark & X          & X          & CL              & BCE                                & Sup.           & Recognition       & ACC         & 50.6              & LLP                            \\
        \cite{seo2020hmtl}               & X          & \checkmark & X          & \checkmark & Adv. Loss       & CE                                 & Self-Sup.      & Recognition       & ACC         & 65.2 | 40.1       & CMU-MOSI | IEMOCAP             \\
        \cite{zhang2019deep}             & X          & \checkmark & X          & X          & CE              & -                                  & Sup.           & Recognition       & ACC         & 62.5              & AFEW 8.0                       \\
        \cite{sun2023masked}             & X          & \checkmark & \checkmark & X          & L1              & -                                  & Self-Sup.      & Recognition       & ACC         & 96.5 | 78.0       & UCF-101 | HMDB-51              \\
        \cite{alwassel2019self}          & \checkmark & \checkmark & X          & X          & CE + Clust.     & -                                  & Self-Sup.      & Recognition       & ACC         & 95.5 | 68.9       & UCF-101 | HMDB-51              \\
        \cite{patrick2020multi}          & \checkmark & X          & X          & X          & NCE             & -                                  & Self-Sup.      & Recognition       & ACC         & 95.2 | 72.8       & UCF-101 | HMDB-51              \\
        \cite{sarkar2022xkd}             & \checkmark & \checkmark & X          & X          & L1 + KD         & -                                  & Self-Sup.      & Recognition       & ACC         & 94.1 | 69.0       & UCF-101 | HMDB-51              \\
        \cite{ma2020active}              & \checkmark & X          & X          & X          & NCE             & Act. neg. sampl.                   & Self-Sup.      & Recognition       & ACC         & 94.1 | 67.2       & UCF-101 | HMDB-51              \\
        \cite{recasens2021broaden}       & X          & X          & \checkmark & X          & L2              & -                                  & Self-Sup.      & Recognition       & ACC         & 93.2 | 69.9       & UCF-101 | HMDB-51              \\
        \cite{mm-versatile2020}          & \checkmark & X          & X          & X          & MIL-NCE + NCE   & -                                  & Self-Sup.      & Recognition       & ACC         & 91.8 | 67.1       & UCF-101 | HMDB-51              \\
        \cite{min2021cross}              & X          & \checkmark & X          & X          & NCE             & L2                                 & Self-Sup.      & Recognition       & ACC         & 90.3 | 61.1       & UCF-101 | HMDB-51              \\
        \cite{akbari2021vatt}            & \checkmark & \checkmark & X          & X          & MIL-NCE         & -                                  & Self-Sup.      & Recognition       & ACC         & 89.6 | 65.2       & UCF-101 | HMDB-51              \\
        \cite{cheng2020look}             & \checkmark & \checkmark & X          & X          & CE              & -                                  & Self-Sup.      & Recognition       & ACC         & 87.8 | 58.2       & UCF-101 | HMDB-51              \\
        \cite{qian2022multimodal}        & \checkmark & \checkmark & X          & X          & CL              & -                                  & Self-Sup.      & Recognition       & ACC         & 87.1 | 64.7       & UCF-101 | HMDB-51              \\
        \cite{sarkar2021self}            & \checkmark & \checkmark & X          & X          & CL              & -                                  & Self-Sup.      & Recognition       & ACC         & 74.8              & UCF-101                        \\
        \cite{xuan2021discriminative}    & \checkmark & \checkmark & X          & X          & L1              & -                                  & Self-Sup.      & Recognition       & ACC         & 79                & AVE                            \\
        \cite{gan2020cross}              & \checkmark & \checkmark & X          & X          & MIL-NCE         & -                                  & Self-Sup.      & Recognition       & ACC         & 78.3              & AVE                            \\
        \cite{lee2021crossattentional}   & \checkmark & \checkmark & X          & X          & MIL-NCE         & CE (sim.) + L1 (motion) + Hinge L. & Self-Sup.      & Recognition       & ACC         & 77.1              & AVE                            \\
        \cite{yang2019dual}              & \checkmark & \checkmark & X          & X          & CE              & BCE                                & Sup.           & Recognition       & ACC         & 47.8              & AVE                            \\
        \cite{huang2022mavil}            & X          & \checkmark & \checkmark & X          & KD              & CL                                 & Self-Sup.      & Recognition       & ACC         & 53.3              & AudioSet                       \\
        \cite{arnab2022audiovisual}      & X          & \checkmark & \checkmark & X          & CE              & -                                  & UnSup.         & Recognition       & ACC         & 51.8              & AudioSet                       \\
        \cite{glass2022uavm}             & \checkmark & \checkmark & X          & X          & CE              & -                                  & Sup.           & Recognition       & ACC         & 50.4              & AudioSet                       \\
        \hline
        \cite{rouditchenko2020avlnet}    & X          & \checkmark & X          & X          & MMS             & -                                  & Self-Sup.      & Retrieval         & Recall @10  & 49.9 | 67.9       & MSR-VTT | YouCook2             \\
        \cite{afouras2021self}           & \checkmark & X          & X          & X          & NCE + CE        & -                                  & Sup.           & Retrieval         & mAP @50     & 39.4 | 28.0       & VGG Sound | AudioSet           \\
        \cite{pretet2021cross}           & \checkmark & X          & X          & X          & TL              & -                                  & Self-Sup.      & Retrieval         & Recall @10  & 7.4               & MVD                            \\
        \cite{zheng2021adversarial}      & \checkmark & X          & X          & \checkmark & LSL + Adv. Loss & -                                  & Sup.           & Retrieval         & ACC         & 93.0              & VGGFace + VoxCeleb             \\
        \cite{zeng2020deep}              & \checkmark & X          & X          & X          & TL              & Corr. L. (align.)                  & Sup.           & Retrieval         & mAP         & 74.2              & VEGAS                          \\
        \cite{merkx2019language}         & \checkmark & \checkmark & X          & X          & Hinge           & -                                  & Sup.           & Retrieval         & Recall @10  & 52.3              & Flickr-8K                      \\
        \cite{ilharco2019large}          & \checkmark & \checkmark & X          & X          & MMS/TL          & -                                  & Sup.           & Retrieval         & Recall @10  & 52.6              & Flickr Audio Caption Corpus    \\
        \cite{he2019new}                 & X          & X          & X          & X          & CE              & Center + QL                        & Sup.           & Retrieval         & mAP         & 41.2              & CUB-200-2011 + Youtube Birds   \\
        \cite{horiguchi2018face}         & \checkmark & X          & X          & X          & NPL             & -                                  & UnSup.         & Retrieval         & mAP         & 2.0               & FVCeleb                        \\
        \cite{wu2019unified}             & \checkmark & \checkmark & X          & X          & L2              & L1                                 & Sup.           & Retrieval         & Recall @10  & 71.9              & MS-COCO                        \\
        \cite{cao2016correlation}        & X          & X          & \checkmark & X          & L2              & L2                                 & Sup.           & Retrieval         & mAP         & 35.5 | 71.7       & Wiki Dataset | Flickr          \\
        \cite{zhen2019deep}              & \checkmark & X          & X          & X          & CE              & Corr. L. (CE) + L2                 & Sup.           & Retrieval         & mAP         & 71.6 | 61.3       & Pascal Sentence | NUS-WIDE-10k \\
        \cite{zhang2021variational}      & X          & X          & \checkmark & X          & ELBO            & Corr. L. (CE) + Center L. + L2     & Sup.           & Retrieval         & mAP         & 81.2 | 35.        & VEGAS | AVE                    \\
        \cite{zeng2021learning}          & \checkmark & X          & X          & X          & Corr. L.        & L2 + Frob. Norm                    & Sup.           & Retrieval         & mAP         & 77.8 | 30.8       & VEGAS | AVE                    \\
        \hline
        \cite{chen2021localizing}        & \checkmark & \checkmark & X          & X          & NCE             & -                                  & Self-Sup.      & S. Local.         & AUC         & 0.573 | 0.590     & Flickr-SoundNet | VGG-SS       \\
        \cite{lin2020audiovisual}        & X          & \checkmark & X          & X          & CE              & -                                  & Self-Sup.      & S. Local.         & ACC         & 76.8              & AVE                            \\
        \cite{morgado2020learning}       & \checkmark & \checkmark & X          & X          & NCE             & -                                  & Self-Sup.      & S. Local.         & ACC         & 73.8 | 37.66      & UCF-101 | HMDB-51              \\
        \cite{ramaswamy2020see}          & \checkmark & \checkmark & X          & X          & CE/TL           & -                                  & Self-/(Un)Sup. & S. Local.         & ACC         & 74.8              & AVE                            \\
        \cite{duan2021audio}             & \checkmark & \checkmark & X          & X          & CE              & -                                  & Sup.           & S. Local.         & ACC         & 76.2              & AVE                            \\
        \hline
        \cite{lin2021exploiting}         & X          & \checkmark & \checkmark & X          & CE + L2         & -                                  & Self-Sup.      & S. Sep.           & STFT        & 0.865 | 1.448     & FAIR PLAY | YT-MUSIC           \\
        \cite{gao2018learning}           & \checkmark & \checkmark & X          & X          & MIML            & -                                  & Self-Sup.      & S. Sep.           & SDR         & 2.53              & AudioSet                       \\
        \cite{hu2021dmman}               & X          & \checkmark & \checkmark & X          & BCE             & -                                  & Sup.           & S. Sep.           & SDR         & 7.64              & Solo+Duet                      \\
        \cite{zhu2020visually}           & \checkmark & \checkmark & X          & X          & CE              & -                                  & Sup.           & S. Sep.           & SDR         & 8.25              & MUSIC                          \\
        \cite{gao2019co}                 & X          & \checkmark & \checkmark & X          & L1 + CE         & -                                  & Sup.           & S. Sep.           & SDR         & 7.64              & MUSIC                          \\
        \cite{zhao2019sound}             & X          & \checkmark & \checkmark & X          & CE              & -                                  & Self-Sup.      & S. Sep.           & SDR         & 8.31              & MUSIC                          \\
        \hline
        \cite{hu2020curriculum}          & X          & \checkmark & \checkmark & X          & CL + L1         & -                                  & Self-Sup.      & S. Sep. \& Local. & AUC/SDR     & 49.2 | 6.59       & Flickr-SoundNet/MUSIC          \\
        \cite{zhao2018sound}             & X          & \checkmark & \checkmark & X          & CE              & -                                  & Self-Sup.      & S. Sep. \& Local. & SDR         & 8.31              & MUSIC                          \\
        \cite{zhu2021leveraging}         & \checkmark & \checkmark & X          & X          & CE              & -                                  & Self-Sup.      & S. Sep. \& Local. & SDR         & 10.74             & MUSIC                          \\
        \cite{owens2022mix}              & \checkmark & \checkmark & X          & X          & infoNCE         & CL                                 & Self-Sup.      & S. Sep. \& Local. & AUC         & 40.6              & MUSIC                          \\
        \cite{afouras2020self}           & \checkmark & \checkmark & X          & X          & CE              & BCE                                & Self-Sup.      & S. Sep. \& Local. & ACC         & 7.2 (3 speakers)  & LRS2                           \\
        \cite{kanazawa2021choreographer} & X          & \checkmark & X          & X          & L2              & -                                  & Self-Sup.      & S. Sep. \& Local. & FDIk | FDIg & 35.35 | 12.39     & AIST++                         \\
        \cite{tian2018audio}             & \checkmark & \checkmark & X          & X          & CL + L1         & -                                  & Self-Sup.      & S. Sep. \& Local. & ACC         & 72.7              & AVE                            \\
        \hline
    \end{longtable}
\end{landscape}
}

\section{Discussion}
\label{section:discussion}


As presented in the previous sections, our research aims to compare, categorise, and identify shared properties between different methodologies used in \acrshort*{av} learning over all its application use cases. We determined that \acrshort*{soa} methodologies targeting different applications have common properties. Thus, we elaborated Section \ref{section:learningAVCL} to group and summarise them. As reflected in Table \ref{tab:methods-final}, recent research has intensively demonstrated that i) self-supervised learning methods can achieve more discriminative features as they allow leveraging large volumes of (unlabeled) data, and ii) attention-based methods are designed to improve the synchronisation between audio and visual sequences. After reviewing the audio-visual learning survey by H. Zhu \emph{et. al.} \cite{DAVLearningSurvey}, we categorized the \acrshort*{avcl} approaches into the following applications: Sound Separation \& Localization, Retrieval, Recognition, and Generation. We also classified the approaches based on their loss functions, regularisation components, model/architecture used, learning paradigm, and corresponding benchmarks. Model types were analysed as simple projectors (P.), attention mechanisms (Att.), auto-encoders (AE), and GANs.\par

The typical scenarios for self-supervised learning frameworks lose the definition of labels and create negative and positive samples in supervised or self-supervised ways. Self-supervised learning requires a considerable batch size, especially for methods that use in-batch samples (e.g., margin-based losses). This allows them to obtain diverse data embeddings at each step while covering a wide range of negative samples and enabling the model to learn meaningful representations. However, the time complexity of model training will increase with the batch size. Consequently, increasing their computational requirements hinders the widespread use of these \acrshort*{avcl} methods. Furthermore, the performance of self-supervised learning methods is limited in evaluating (zero-shot) classification or retrieval-based downstream tasks. Applying self-supervised learning approaches to a broader range of multimodal tasks is necessary. Simultaneously, various attention mechanisms can be developed to learn and understand the intra-modality and inter-modality correlation between audio and visual data, which helps to enhance fine-grained alignment and increase the performance of models. Moreover, attention and memory banks can tackle long-term temporal signals for better sequence prediction.\par

We argue that most observed proposals focus primarily on researching methodologies for ``representing knowledge in the \acrshort*{av} domain''. As such, they can also be classified within the field of multiple knowledge representation \cite{yang2021multiple}. Following the author's taxonomy, the revised methods lie within the deep knowledge representation, which focuses on encoding raw signals into abstract representations. The outputs from these methods are not explainable. Representations extracted using \acrshort*{soa} models cannot present information in understandable terms to humans. Therefore, we firmly believe that it is an issue that could be relieved by introducing human-understandable mechanisms, that reflect comprehensible knowledge, to guide the learning process, i.e., similar to structured knowledge used in self-supervised learning settings. In this context, we refer to structured knowledge as the different self-supervised methodologies/proposals for selecting pseudo-labels and their associated proxy tasks.\par

Currently, multimodal visual-language models have been proposed for video-language pretraining tasks \cite{huang2022clover,alayrac2022flamingo,zhu2023languagebind,xu2021videoclip,ye2022hitea,chen2023valor,wang2022image,radford2021learning}, which is far from the scope of our survey paper. However, they are important enough to mention. This kind of pretraining focuses on creating versatile multimodal models for tasks such as text-video retrieval or video question answering. For \acrshort*{avcl}, these approaches allow to leverage text as a reference signal to align audio and video \cite{alayrac2022flamingo}. Multimodal vision-language models trained with contrastive learning objectives outperformed other approaches in zero-shot benchmarks \cite{xu2021videoclip,radford2021learning}. In other words, a single \acrfull*{llm} conditioned on video allows us to obtain strong performances in many text-video downstream tasks while only using its text interface \cite{alayrac2022flamingo,wang2022image}. On the other hand, these proposals can also perform audio-visual alignment by ``blindly'' adding them as additional inputs on text generation models \cite{wang2022image,alayrac2022flamingo}. Such approaches are usually optimised via masked language learning, which focuses on randomly masking certain words in a sentence and requiring the model to predict those masked words based on the surrounding context. The context can contain multimedia content, such as audio or video. Moreover, masked language modelling focuses heavily on local context rather than the global understanding of each item. Thus, it can lead to suboptimal performances for some downstream tasks. Therefore, we expect to see proposals that attempt to balance global and local information to increase their performance in downstream tasks.\par

Text can also be used in \acrshort*{avcl} to provide auxiliary information to learn correlations between audio and video. They can primarily be used to introduce semantic meaning to the learning process. Nevertheless, this implies that instead of relying on audio-visual pairs, we must consider information shared between audio and text and between video and text. Consequently, the learning objectives used by these approaches are more complex due to the extra hyperparameters and loss components \cite{ye2022hitea,huang2022clover}. Moreover, having extra components in the objective function makes the training process more prone to instability.\par

\section{Conclusion}
\label{section:conclusion}

This survey paper is a resource for researchers, practitioners, and industry professionals who wish to stay updated on the latest developments in \acrshort*{avcl}. It compares and categorizes different methodologies to identify trends, commonalities, and gaps in the field. It provides a comprehensive understanding of existing approaches, facilitates collaboration between researchers, and guides future research efforts. As the field evolves rapidly, we hope it serves as a vital compass, helping navigate the vast landscape of \acrshort*{avcl} and paving the way for breakthroughs and discoveries.\par

More specifically, this survey paper aims to compare, categorize, and identify shared properties between different methodologies used in AV learning over all its application use cases. We determined that \acrshort*{soa} methodologies that target different applications share properties and elaborated this document to group and summarize them. As reflected in Table \ref{tab:methods-final}, recent research has intensively demonstrated that i) self-supervised learning methods can achieve more discriminative features as they allow leveraging large volumes of (unlabeled) data, and ii) attention-based methods are designed to improve the synchronization between audio and visual sequences.\par


This paper focuses on different methodologies for representing knowledge in the audio-visual domain. Our review lies within the field of Multiple Knowledge Representation \cite{yang2021multiple}. Following their taxonomy, the revised methods lie within the deep knowledge representation, which focuses on encoding raw signals into abstract representations. The outputs from these methods are not explainable. Representations extracted using \acrshort*{soa} models cannot present information in understandable terms to humans. Therefore, we firmly believe this issue could be relieved by introducing human-understandable mechanisms that reflect comprehensible knowledge and guide the learning process (i.e., similar to structured knowledge used in self-supervised learning settings). In due course, we expect to see more complex proposals on selecting and defining structured knowledge as input to deep models to fuel representation learning.\par

\section{Research Challenges and Directions}
\label{section:challenges}

Multimodal correlation learning aims to capture inter-modality information that complements each other between different modalities. However, learning an ideal unified representation of paired audio-visual sequences that can extract semantic inter-modality information and preserve intra-modality structures is challenging. Considering local or global sequence embedding methods with \acrshort*{dl} is promising to improve the capability of correlation learning between different modalities \cite{chen2023valor,denize2022similarity}. Additionally, we believe that self-supervised methodologies for selecting pseudo-labels (and their corresponding proxy tasks) can be extended to simultaneously consider multiple groups of labels. For instance, information from the timeline (temporal cohesion) and labels for each video in the same loss function.\par

Recent research progress has shown that self-supervised learning methods can achieve more discriminative features when trained on sufficient data, which has encouraged to pretrain deep models, e.g. Flamingo \cite{alayrac2022flamingo}, on vast amounts of data and then fine-tune them on much smaller datasets to realise specific tasks. Therefore, the investigation of standard settings for multimodal audio-visual pretrained models is indispensable for the research community, which will be very helpful for obtaining more knowledge about the pretrained mechanism. Regarding the benchmark settings with pretraining proposals, we believe that the \acrshort*{avcl} field needs standard protocols (i.e., datasets and tasks) to better compare them.\par

Interpretability and reliability are essential properties of audio-visual correlation learning models. However, many of these models need help explaining their predictions and demonstrating their reliability. Therefore, it is crucial to develop interpretable multimodal correlation learning methods. These improved methods would better extract and understand the characteristics and correlations of audio and visual sequence data. They would also help explain when a model functions correctly, why it might fail, and ultimately increase the reliability of multimodal intelligence in various applications.\par

In an attempt to tackle the previous issue (e.g., interpretability of latent spaces), text can be a practical choice for providing comprehensible descriptions. Previous proposals use \acrshort*{avcl} models to condition text generation, allowing the alignment of audio-visual representations with the semantic meaning provided through text. However, incorporating \acrshort*{llm}s into \acrshort*{avcl} amplifies its computational demands, which are already heightened by the mini-batch requirements of self-supervised learning frameworks. Consequently, we must thoughtfully evaluate their integration into newer proposals. Consequently, graphs have been introduced to capture detailed semantics and decrease the computations performed by language models \cite{liu2022cross,chen2020fine,saddik2023graph}. They represent the text input space by exploiting lexical relationships between words. These proposals utilised semantic graphs of text captions from each video/image to align visual and text embeddings, allowing cross-modal retrieval. Thus, we expect to observe in the future proposals that use graph neural networks as a way to represent text and alleviate the amount of connections typically used between tokens on transformer-based encoders. Nevertheless, the semantic graph's link structure could be used to identify positive and negative elements for self-supervised learning frameworks, simplifying the selection of negative samples.\par

\begin{acks}
    Luís Vilaça, was partially supported by Fundação para a Ciência e a Tecnologia (FCT) under Grant No.: 2022.11905.BD. Yi Yu was partially supported by \grantsponsor{}{JSPS Scientific Research (C)}{} under Grant No.:\grantnum{19K11987}{19K11987}. This work was partially supported by \grantsponsor{}{European Union's Horizon Europe Research and Innovation Program}{}, within project Telecommunications and Computer Vision Convergence Tools for Research Infrastructures (CONVERGE) under Grant \grantnum{101094831}{101094831}.
    The authors also thank the anonymous referees for their valuable comments and helpful suggestions.
\end{acks}

\bibliographystyle{ACM-Reference-Format}
\bibliography{sample-base.bib}







    \clearpage
    \appendix


    
\begin{landscape}
    \section{AudioSet benchmark results (Audio Classification)}
    \label{appendix:audioset-proposals}
    \begin{table}[H]
        \caption{AudioSet benchmark results (Audio Classification). For the acronyms, U denotes Unsupervised, SS denotes Self-Supervised, S denotes Supervised, LMS denotes Log-Mel Spectrogram, V denotes Video (Visual data), T denotes Text, and AW denotes Audio Waveform.}
        \label{tab:audiosetTest}
        \resizebox{0.85\linewidth}{!}{
            \begin{tabular}{@{}lccccccccccccccl@{}}
                \toprule
                Method                                                                                                    & Learning & Train Inputs & Kernel-AW & Stride-AW & FFT Win. & FFT Hop & Mel-B. & Mel W. & MFCCs & $F_s$    & Emb. Size     & Val Inputs & mAP   & Reference                   \\ \midrule
                Triplet                                                                                                   & U        & LMS          & -         & -         & 25 ms    & 10 ms   & 64     & 0.96 s & -     & 16 kHz   & 128           & LMS        & 0.244 & \cite{triplet2020}          \\
                L3-Net                                                                                                    & SS       & LMS + V      & -         & -         & 10 ms    & 5 ms    & 257    & 1 s    & -     & 48 kHz   & 512           & LMS        & 0.249 & \cite{l3net2017}            \\
                CPC                                                                                                       & U        & AW           & 10        & 5         & -        & -       & -      & -      & -     & 16 kHz   & 512           & AW         & 0.277 & \cite{cpc2020}              \\
                C3                                                                                                        & U        & LMS + V      & -         & -         & 25 ms    & 10 ms   & 64     & 0.96 s & -     & 1 Hz     & 128           & LMS        & 0.285 & \cite{c32020}               \\
                MMV                                                                                                       & U        & LMS + V + T  & -         & -         & -        & -       & 80     & -      & -     & 10 Hz    & 2048          & LMS        & 0.309 & \cite{mm-versatile2020}     \\
                A. Contrastive L.                                                                                         & SS       & LMS          & 10        & 5         & 20 ms    & 10 ms   & 80     & -      & 13    & -        & 512           & LMS        & 0.329 & \cite{cl-audio2021}         \\
                A. Contrastive L.                                                                                         & SS       & AW           & 10        & 5         & 20 ms    & 10 ms   & 80     & -      & 13    & -        & 512           & AW         & 0.336 & \cite{cl-audio2021}         \\
                A. Contrastive L.                                                                                         & SS       & LMS + AW     & 10        & 5         & 20 ms    & 10 ms   & 80     & -      & 13    & -        & 512           & AW         & 0.355 & \cite{cl-audio2021}         \\
                Res1dNet31                                                                                                & S        & AW           & 10        & 5         & 32 ms    & 10 ms   & 64     & 10 s   & -     & 32 kHz   & -             & AW         & 0.365 & \cite{panns2020}            \\
                A. Contrastive L.                                                                                         & SS       & LMS + AW     & 10        & 5         & 20 ms    & 10 ms   & 80     & -      & 13    & -        & 512           & LMS        & 0.368 & \cite{cl-audio2021}         \\
                A. Contrastive L.                                                                                         & SS       & LMS + AW     & 10        & 5         & 20 ms    & 10 ms   & 80     & -      & 13    & -        & 512           & LMS + AW   & 0.376 & \cite{cl-audio2021}         \\
                A-V Contrastive L.                                                                                        & SS       & LMS + AW + V & 10        & 5         & 20/25 ms & 10 ms   & 80/64  & -      & -     & 16 kHz   & 1280/512      & LMS        & 0.377 & \cite{avcl2021}             \\
                MobileNetV1                                                                                               & S        & LMS          & -         & -         & 32 ms    & 10 ms   & 64     & 10 s   & -     & 32 kHz   & 1024          & LMS        & 0.389 & \cite{panns2020}            \\
                Wavegram-CNN                                                                                              & S        & AW           & 10        & 5         & 32 ms    & 10 ms   & 64     & 10 s   & -     & 32 kHz   & 2048          & AW         & 0.389 & \cite{panns2020}            \\
                A-V Contrastive L.                                                                                        & SS       & LMS + AW + V & 10        & 5         & 20/25 ms & 10 ms   & 80/64  & -      & -     & 16 kHz   & 1280/512      & AW         & 0.397 & \cite{avcl2021}             \\
                Multimodal Bottleneck Transf.                                                                             & S        & LMS + V      & -         & -         & 25 ms    & 10 ms   & 128    & 10 s   & -     & 16 kHz   & 768 (/ patch) & LMS + V    & 0.415 & \cite{mbt2021}              \\
                CNN14                                                                                                     & S        & LMS          & -         & -         & 32 ms    & 10 ms   & 64     & 10 s   & -     & 32 kHz   & 2048          & LMS        & 0.431 & \cite{panns2020}            \\
                Perceiver                                                                                                 & S        & LMS          & -         & -         & 20 ms    & 10 ms   & 128    & -      & -     & 48 kHz   & -             & LMS + V    & 0.432 & \cite{perceiver2021}        \\
                Resnet38                                                                                                  & S        & LMS          & -         & -         & 32 ms    & 10 ms   & 64     & 10 s   & -     & 32 kHz   & 2048          & LMS        & 0.434 & \cite{panns2020}            \\
                Perceiver                                                                                                 & S        & AW           & -         & -         & -        & -       & -      & -      & -     & 48 kHz   & -             & AW + V     & 0.435 & \cite{perceiver2021}        \\
                Wavegram-Logmel-CNN                                                                                       & S        & LMS + AW     & 10        & 5         & 32 ms    & 10 ms   & 64     & 10 s   & -     & 32 kHz   & 2048          & LMS + AW   & 0.439 & \cite{panns2020}            \\
                ERANNs-2-5 \footnote{Efficient Residual Audio Neural Networks (Wide ResNet w/ Time Frequency Blocks)}     & S        & LMS          & -         & -         & 31 ms    & 8 ms    & 128    & 10 s   & -     & 44.1 kHz & 640           & LMS        & 0.446 & \cite{eranns2021}           \\
                ERANNs-1-6                                                                                                & S        & LMS          & -         & -         & 31 ms    & 8 ms    & 128    & 10 s   & -     & 44.1 kHz & 768           & LMS        & 0.450 & \cite{eranns2021}           \\
                UAVM                                                                                                      & S        & LMS          & -         & -         & 25 ms    & 10 ms   & 128    & -      & -     & 16 kHz   & 1024          & LMS        & 0.456 & \cite{glass2022uavm}        \\
                Audio Spec. Transf.                                                                                       & S        & LMS          & -         & -         & 25 ms    & 10 ms   & 128    & 10 s   & -     & 16 kHz   & 768 (/ patch) & LMS        & 0.459 & \cite{ast2021}              \\
                CAV-MAE                                                                                                   & SS       & LMS          & -         & -         & 25 ms    & 10 ms   & 128    & -      & -     & 16 kHz   & -             & LMS        & 0.466 & \cite{glass2022contrastive} \\
                Audio-Visual Masked Encoders                                                                              & SS       & LMS          & -         & -         & 25 ms    & 10 ms   & 128    & -      & -     & 16 kHz   & -             & LMS        & 0.466 & \cite{arnab2022audiovisual} \\
                Transformer-to-CNN                                                                                        & S        & LMS          & -         & -         & 25 ms    & 10 ms   & 128    & -      & -     & 32 kHz   & -             & LMS        & 0.470 & \cite{schmid2022efficient}  \\
                HTS-AT                                                                                                    & SS       & LMS          & -         & -         & 25 ms    & 10 ms   & 128    & -      & -     & 16 kHz   & -             & LMS        & 0.471 & \cite{hts-at2021}           \\
                PaSST                                                                                                     & SS       & LMS          & -         & -         & 25 ms    & 10 ms   & 128    & -      & -     & 32 kHz   & -             & LMS        & 0.471 & \cite{passt2021}            \\
                BEATs                                                                                                     & SS       & LMS          & -         & -         & 25 ms    & 10 ms   & 128    & -      & -     & 16 kHz   & -             & LMS        & 0.486 & \cite{chen2022beats}        \\
                MAViL                                                                                                     & SS       & LMS          & -         & -         & 25 ms    & 10 ms   & 128    & -      & -     & 16 kHz   & 512           & LMS        & 0.487 & \cite{huang2022mavil}       \\
                HTS-AT (emsemble)                                                                                         & SS       & LMS          & -         & -         & 25 ms    & 10 ms   & 128    & -      & -     & 16 kHz   & -             & LMS        & 0.487 & \cite{hts-at2021}           \\
                Multimodal Bottleneck Transf. (emsemble)                                                                  & S        & LMS + V      & -         & -         & 25 ms    & 10 ms   & 128    & 10 s   & -     & 16 kHz   & 768 (/ patch) & LMS + V    & 0.496 & \cite{mbt2021}              \\
                PaSST (emsemble)                                                                                          & SS       & LMS          & -         & -         & 25 ms    & 10 ms   & 128    & -      & -     & 32 kHz   & -             & LMS        & 0.496 & \cite{passt2021}            \\
                UAVM                                                                                                      & S        & LMS + V      & -         & -         & 25 ms    & 10 ms   & 128    & -      & -     & 16 kHz   & 1024          & LMS + V    & 0.504 & \cite{glass2022uavm}        \\
                BEATs (emsemble)                                                                                          & SS       & LMS          & -         & -         & 25 ms    & 10 ms   & 128    & -      & -     & 16 kHz   & -             & LMS        & 0.506 & \cite{chen2022beats}        \\
                Audio-Visual Masked Encoders                                                                              & SS       & LMS + V      & -         & -         & 25 ms    & 10 ms   & 128    & -      & -     & 16 kHz   & -             & LMS + V    & 0.518 & \cite{arnab2022audiovisual} \\
                MAViL                                                                                                     & SS       & LMS + V      & -         & -         & 25 ms    & 10 ms   & 128    & -      & -     & 16 kHz   & 512           & LMS + V    & 0.533 & \cite{huang2022mavil}       \\
                AudioFormer                                                                                               & SS       & AW           & -         & -         & -        & -       & -      & -      & -     & 24 kHz   & 750           & AW         & 0.539 & \cite{li2023audioformer}    \\
                COLA\footnote{Google's research: Contrastive learning with AW; Results were surpassed by \cite{avcl2021}} & SS       & LMS          & -         & -         & 25 ms    & 10 ms   & 64     & 0.96 s & -     & -        & 512           & AW         & -     & \cite{cola2021}             \\
                \bottomrule
            \end{tabular}
        }
    \end{table}
\end{landscape}

    \clearpage

    \section{Proposals for Action Recognition (``Video Frame Sequence Analysis'')}
\label{appendix:actionrecognition-proposals}
{
    \footnotesize
    \begin{longtable}[!htc]{llcccc}
        \caption{Proposals for Action Recognition (``Video Frame Sequence Analysis''). $\dagger$ indicates performance with fine-tunning. $\ddagger$ indicates performance on linear evaluation.}
        \label{tab:actionRecMethods}                                                                                                                                                            \\
        \hline
        Methods                                                    & HMDB-51 & UCF-101 & Modalities                 & Input Size (frames) & Paper                                               \\ \hline
        \endhead
        \hline
        \endfoot
        \endlastfoot
        MoCo $\ddagger$                                            & -       & 94.4    & RGB                        & 8                   & \cite{feichtenhofer2021large,girshick2020momentum}  \\
        BYOL $\ddagger$                                            & -       & 93.4    & RGB                        & 8                   & \cite{feichtenhofer2021large,grill2020bootstrap}    \\
        SwAV $\ddagger$                                            & -       & 89.4    & RGB                        & 8                   & \cite{feichtenhofer2021large,caron2020unsupervised} \\
        SimCLR $\ddagger$                                          & -       & 87.9    & RGB                        & 8                   & \cite{feichtenhofer2021large,chen2020simple}        \\
        CLIP (ViT-L/14) $\ddagger$                                 & -       & 80.3    & RGB                        & 1                   & \cite{radford2021learning}                          \\
        CVRL (R3D-50)$\ddagger$                                    & 59.7    & 90.8    & RGB                        & 32                  & \cite{qian2021spatiotemporal}                       \\
        XKD$\ddagger$                                              & 57.4    & 83.9    & RGB                        & 32                  & \cite{sarkar2022xkd}                                \\
        MOV$\ddagger$                                              & 57.8    & 80.9    & RGB+Flow                   & 16                  & \cite{qian2022multimodal}                           \\
        MAXI$\ddagger$                                             & 52.3    & 78.2    & RGB                        & 16                  & \cite{lin2023match}                                 \\
        ViT-L/14$\ddagger$ (CLIP)                                  & 49.8    & 79.6    & RGB                        & 8                   & \cite{wu2022revisiting}                             \\
        ViCC$\ddagger$ (S3D)                                       & 47.9    & 78      & RGB+Flow                   & 32                  & \cite{hu2022self}                                   \\
        VicTR$\ddagger$                                            & 72.4    & 51      & RGB                        & 32                  & \cite{kahatapitiya2023victr}                        \\
        ViCC$\ddagger$ (S3D)                                       & 38.5    & 72.2    & RGB                        & 16                  & \cite{hu2022self}                                   \\ \midrule
        VideoMAE v2$\dagger$                                       & 88.1    & 99.6    & RGB                        & 16                  & \cite{wang2023videomaev2}                           \\
        R(2+1)D BERT $\dagger$                                     & 85.1    & 98.69   & RGB                        & 64                  & \cite{kalfaoglu2020late}                            \\
        SMART (TSN) $\dagger$                                      & 84.36   & 98.64   & Video                      & 10                  & \cite{gowda2021smart}                               \\
        R(2+1)D BERT $\dagger$                                     & 83.9    & 98.65   & RGB                        & 32                  & \cite{kalfaoglu2020late}                            \\
        OmniSource (SlowOnly-8x8-R101 + Kinetics) $\dagger$        & 83.8    & 98.6    & RGB+RGB                    & 64                  & \cite{duan2020omni}                                 \\
        PERF-Net $\dagger$                                         & 83.2    & 98.6    & Flow+Pose+RGB              & 64                  & \cite{li2022perf}                                   \\
        ResNeXt BERT $\dagger$                                     & 83.55   & 97.87   & RGB+Flow                   & 64                  & \cite{kalfaoglu2020late}                            \\
        BubbleNET $\dagger$                                        & 82.6    & 97.2    & RGB+Flow                   & 16                  & \cite{o2020bubblenet}                               \\
        CLIP (ViT-L/14) $\dagger$                                  & -       & 92      & RGB                        & 1                   & \cite{radford2021learning}                          \\
        Two-Stream I3D (RGB + Flow + Kinetics) $\dagger$           & 81.9    & 97.8    & RGB+Flow                   & 64                  & \cite{carreira2017quo}                              \\
        CCS + TSN (ImageNet + Kinetics) $\dagger$                  & 81.9    & 97.4    & RGB+Flow                   & 25                  & \cite{zhang2019cooperative}                         \\
        Image-Based $\dagger$                                      & -       & 89.6    & RGB                        & - (Avg. Fusion)     & \cite{zha2015exploring}                             \\
        MARS+RGB+Flow (Kinetics) $\dagger$                         & 80.9    & 98.1    & RGB+Flow                   & 64                  & \cite{crasto2019mars}                               \\
        I3D-(RGB and Flow and Pairwise and Pose) $\dagger$         & 80.92   & 98.02   & Pose+Flow+RGB              & 64                  & \cite{hong2019contextual}                           \\
        LGD-3D Two-stream (ImageNet+Kinetics) $\dagger$            & 80.5    & 98.2    & RGB                        & 16                  & \cite{qiu2019learning}                              \\
        Two-Stream I3D (Imagenet + Kinetics) $\dagger$             & 80.7    & 98      & RGB+Flow                   & 64                  & \cite{carreira2017quo}                              \\
        D3D + D3D $\dagger$                                        & 80.5    & 97.6    & RGB+Flow                   & 64                  & \cite{stroud2020d3d}                                \\
        MVD-B $\dagger$                                            & 79.7    & 97.5    & RGB                        & 16                  & \cite{wang2023masked}                               \\
        LSTM $\dagger$                                             & -       & 88.6    & RGB+Flow                   & 30                  & \cite{ng2015beyond}                                 \\
        OmniSource (SlowOnly-8x8-R101 + Flow + Kinetics) $\dagger$ & 79      & 97.3    & RGB+Flow                   & 64                  & \cite{duan2020omni}                                 \\
        R(2+1)D - TwoStream (Kinetics) $\dagger$                   & 78.7    & 97.3    & RGB                        & 16                  & \cite{tran2018closer}                               \\
        LGD-3D Flow (ImageNet+Kinetics) $\dagger$                  & 78.9    & 96.8    & Flow                       & 16                  & \cite{qiu2019learning}                              \\
        Deep-HAL+W+G+ODF+SDF $\dagger$                             & 87.56   & -       & RGB+Flow+Mult. Predictions & -                   & \cite{shen2021self}                                 \\
        HAF+BoW/FV halluc $\dagger$                                & 82.48   & -       & RGB+Flow                   & 10                  & \cite{wang2019hallucinating}                        \\
        RepFlow-50 $\dagger$                                       & 81.1    & -       & RGB                        & 16                  & \cite{piergiovanni2019representation}               \\
        MME $\dagger$                                              & 78      & 96.5    & RGB+Flow                   & 16                  & \cite{sun2023masked}                                \\
        HATNet (Multistream 2D + 3D)                               & 76.5    & 97.8    & RGB                        & 32                  & \cite{diba2020large}                                \\
        LGD-3D RGB (ImageNet+Kinetics) $\dagger$                   & 75.7    & 97      & RGB                        & 16                  & \cite{qiu2019learning}                              \\
        CCS + TSN (ImageNet) $\dagger$                             & 77.2    & 95.3    & RGB+Flow                   & 25                  & \cite{zhang2019cooperative}                         \\
        R(2+1)D-Flow (Kinetics) $\dagger$                          & 76.4    & 95.5    & Flow                       & 16                  & \cite{tran2018closer}                               \\
        R(2+1)D-RGB (Kinetics) $\dagger$                           & 74.5    & 96.8    & RGB                        & 16                  & \cite{tran2018closer}                               \\
        VidTr-L $\dagger$                                          & 74.4    & 96.7    & RGB                        & 32                  & \cite{tighe2021vidtr}                               \\
        SMART (TSN + Kinetics) $\dagger$                           & 74.6    & 95.8    & RGB+Flow                   & 10                  & \cite{gowda2021smart}                               \\
        SCE (R3D-50)$\dagger$                                      & 74.7    & 95.3    & RGB                        & 16                  & \cite{denize2022similarity}                         \\
        VideoMAE$\dagger$                                          & 73.3    & 96.1    & RGB                        & 16                  & \cite{tong2022videomae}                             \\
        VicTR$\dagger$                                             & 70.7    & 95.8    & RGB                        & 32                  & \cite{kahatapitiya2023victr}                        \\
        CVRL (R3D-152) $\dagger$                                   & 70.6    & 94.4    & RGB                        & 32                  & \cite{qian2021spatiotemporal}                       \\
        XDC (R(2+1)D-18 \& IG-Kinetics) $\dagger$                  & 68.9    & 95.5    & RGB                        & 8                   & \cite{alwassel2019self}                             \\
        Temporal Segment Networks (TSN) $\dagger$                  & 69.4    & 94.2    & RGB+Flow                   & 25                  & \cite{gool2016temporal}                             \\
        XKD$\dagger$                                               & 69      & 94.1    & RGB                        & 32                  & \cite{sarkar2022xkd}                                \\
        CVRL (R3D-50) $\dagger$                                    & 69.4    & 93.6    & RGB                        & 32                  & \cite{qian2021spatiotemporal}                       \\
        Convolutional Two-Stream $\dagger$                         & 69.2    & 93.5    & RGB+Flow                   & 5                   & \cite{feichtenhofer2016convolutional}               \\
        CCS + I3D (ImageNet) $\dagger$                             & 68.2    & 93.8    & RGB+Flow                   & 25                  & \cite{zhang2019cooperative}                         \\
        Key-Volume Mining $\dagger$                                & 67.2    & 92.7    & RGB+Flow                   & 10                  & \cite{zhu2016key}                                   \\
        Two-Stream I3D (RGB + Flow) $\dagger$                      & 66.4    & 93.4    & RGB+Flow                   & 64                  & \cite{carreira2017quo}                              \\
        CrissCross$\dagger$                                        & 67.4    & 92.4    & RGB                        & 32                  & \cite{sarkar2021self}                               \\
        Transformation CNN $\dagger$                               & 63.4    & 92.4    & RGB+Flow                   & 20                  & \cite{wang2016actions}                              \\
        ViCC$\dagger$ (S3D)                                        & 62.2    & 90.5    & RGB+Flow                   & 32                  & \cite{hu2022self}                                   \\
        Two-Stream I3D (Flow) $\dagger$                            & 61.9    & 90.6    & Flow                       & 64                  & \cite{carreira2017quo}                              \\
        MOV$\dagger$                                               & 64.7    & 87.1    & RGB+Flow                   & 16                  & \cite{qian2022multimodal}                           \\
        TCLR$\dagger$ (R(2+1)D-18)                                 & 60      & 88.2    & RGB+Flow                   & 16                  & \cite{dave2022tclr}                                 \\
        Two-Stream $\dagger$                                       & 59.4    & 88      & RGB+Flow                   & 25                  & \cite{simonyan2014two}                              \\
        Two-Stream I3D (RGB) $\dagger$                             & 49.8    & 84.5    & RGB                        & 64                  & \cite{carreira2017quo}                              \\
        C3D $\dagger$                                              & 51.6    & 82.3    & RGB                        & 16                  & \cite{tran2015learning}                             \\
        ViCC $\dagger$ (S3D)                                       & 47.9    & 84.3    & RGB                        & 16                  & \cite{hu2022self}                                   \\
        IIC $\dagger$                                              & 38.3    & 74.4    & RGB+Flow                   & 16                  & \cite{tao2020self}                                  \\
        IIC (shuffle + res) $\dagger$                              & 74.4    & 38.3    & RGB                        & 16                  & \cite{tao2020self}                                  \\
        C3D VCP $\dagger$                                          & 32.5    & 68.5    & RGB                        & 16                  & \cite{luo2020video}                                 \\
        R(2+1)D VCP $\dagger$                                      & 32.2    & 66.3    & RGB                        & 16                  & \cite{luo2020video}                                 \\
        R3D VCP $\dagger$                                          & 31.5    & 66      & RGB                        & 16                  & \cite{luo2020video}                                 \\ \hline
    \end{longtable}
}

    \section{Datasets for AVCL}
\label{appendix:datasets}

From the taxonomy of applications provided by \cite{DAVLearningSurvey}, we observed that most representation learning methods (as expected) pre-train their models on large databases and fine-tune/benchmark using specific datasets for the various downstream tasks. The datasets are specific to the task at hand, and, generally speaking, their main limitations/issues are related to the data gathering and annotation process. The most important limitations are: lack synchronization between audio and visual data, lack of ``natural'' semantic relations between audio and visual information\footnote{Audio channel might not be related with the video channel.}, overlapping sound sources, and background noise (audio channel). On another note, the type of annotations and the depth of description currently available depends on the type of task. For instance, some tasks require fine-grained temporal annotations, while others only require semantic labels. This semantic description is available in different formats. Some efforts have been made to improve the description level by creating ontologies for representing the hierarchy of high-level semantic information (e.g., Audio Set). In appendix \ref{appendix:datasets}, we present Table \ref{tab:datasets} with the most used datasets we found for \acrshort*{avcl}. Each dataset can be used for different tasks. Therefore, in addition, we provide a summary of the most used evaluation metrics for each task in appendix \ref{appendix:metrics}.\par

\paragraph{\textbf{Discussion:}} We identified that there exists a lack of consensus in \acrshort*{soa} methods regarding the datasets used for benchmarking; thus, for the majority of them, it is not possible to make a direct quantitative comparison between them for a wide range of tasks. In fact, from Table \ref{tab:methods-final}, we can observe that most representation learning methods only provide benchmarks for classification, and only 1 or 2 results allow comparing different methods without running any code. Therefore, the research community should define a set of data to create a baseline comparison. Moreover, it should be noted that representation learning proposals should be benchmarked across a wide range of \acrshort*{av} applications, which is currently not done. All in all, the analyzed datasets (Table \ref{tab:datasets}) provide \acrshort*{av} pairs and can be used/adapted for a wide range of tasks. As such, given an application use-case, the datasets must be chosen by taking into account the level (depth) of description (labels) required for it (e.g., temporal segmentation annotations).\par

{
\scriptsize
\begin{longtable}{@{}p{1.5cm}p{1.25cm}p{1.25cm}p{0.75cm}p{0.75cm}p{0.75cm}p{6.5cm}@{}}
    \caption{Datasets for \acrlong*{avcl}}
    \label{tab:datasets}                                                                                                                                                                                                                                                                                                                                                                                                                                                                                                                                        \\
    \hline
    Name              & Category       & Target Task                & Env. & Classes & \# clips         & Description                                                                                                                                                                                                                                                                                                                                                                                                                                           \\
    \hline
    %
    \endhead
    %
    \endfoot
    \endlastfoot
    ENST-Drums        & Music          & Event Rec.                 & Lab  & 10      & 456              & Comprises 8 audio channels and 2 video channels with a resolution of 720x576. It includes a variety of music genres and features temporal annotations for each drum event, categorized into 10 distinct classes.                                                                                                                                                                                                                                                                                                                  \\
    ACAV100M          & Real Event     & Repres. Lear.              & Wild & -       & 100M             & Consists of 10-second clips at 360p resolution, featuring multiple sound sources that may appear simultaneously. It is automatically curated and designed to ensure a high level of audio-visual correspondence.                                                                                                                                                                                                                                                                                                                                 \\
    VGG-SS            & Real Event     & AV Localization            & Wild & 220     & 5158             & Based on VGG-Sounds and includes explicit bounding box annotations that mark the visible sound sources in each video clip.                                                                                                                                                                                                                                                                                                                                    \\
    ACIVW             & Real Event     & Repres. Lear.              & Wild & 10      & 21000            & Offers synchronized data across three modalities: acoustic images, audio, and RGB frames. The audio and video are aligned both temporally and spatially, allowing for the extraction of information into monaural audio and audio-visual representations.                                                                                                                                                                              \\
    Countix           & Real Event     & Video Repetition Counting  & Wild & 100     & 8757             & Includes timecode annotations that mark repetitions and specify the kinetics action class for each segment.                                                                                                                                                                                                                                                                                                                                                                                \\
    Kinetics 700-2020 & Real Event     & Video Action Rec.          & Wild & 700     & 647907           & Consists of approximately 10-second clips where multiple sound sources may appear, and the audio may not always be related to the actions occurring in the video.                                                                                                                                                                                                                                                                                                                              \\
    LLP               & Real Event     & AV Localization            & Wild & 25      & 11849            & Features 10-second clips where each sound source and event in the video is annotated, with a minimum duration of 1 second per clip. The test set focuses on the temporal localization of sounds, and overlapping events are included.                                                                                                                                                                                                                                                            \\
    VGG Sound         & Real Event     & Audio-Video Matching       & Wild & 309     & 199196           & Consists of 10-second clips where audio and video are always semantically correlated, verified manually. It features audio-visual correspondences, with sounds extracted from short clips of YouTube videos.                                                                                                                                                                                                                                          \\
    Youtube-ASME-300K & Real Event     & Spatial Correspondence     & Wild & -       & 33725            & Includes 10-second 360º video clips with ambisonic audio, featuring multiple moving sound sources. It comprises ASMR videos collected from YouTube, which contain stereo audio.                                                                                                                                                                                                                                                                                                             \\
    MEAD              & Speech         & Emotion Rec.               & Lab  & 8+7     & 281400           & Features 1080p clips at 30 fps with multi-view data, including annotations for emotion and intensity in each speech segment. Audio and video are paired within a 1-second temporal window.                                                                                                                                                                                                                                                                                                     \\
    FAIR-Play         & Music          & AV Source Separation       & Lab  & -       & 1871             & Consists of 10-second clips recorded in various spatial contexts with binaural audio. The video and corresponding binaural audio clips are roughly aligned by index. The objective is to enhance single-channel soundtracks by incorporating visual information to create dual-channel audio.                                                                                                                                                                                                                \\
    AVIA              & Real Event     & AV Localization            & Lab  & 14      & 378              & Contains sequences ranging from 30 to 60 seconds, featuring varying noise conditions for each class. Both the visual and acoustic images are aligned spatially and synchronized temporally.                                                                                                                                                                                                                                                                                    \\
    CrossTask         & Real Event     & Video Event Rec.           & Wild & 83      & 4700             & The dataset is split into two sections: 18 primary tasks and 65 related tasks. Videos for primary tasks are manually collected and annotated with temporal step boundaries, while videos for related tasks are automatically gathered and lack annotations. The videos are generally long, with an average length of 4 minutes and 57 seconds.                                                                                                                                 \\
    HowTo100M         & Real Event     & Video-Language RL          & Wild & -       & 136M             & Includes narrated audio in full clips of 6.5 minutes each. After parsing, the data is segmented into 4-second clips and categorized into 4-word segments.                                                                                                                                                                                                                                                                                                                                                                               \\
    AVA-ActiveSpeaker & Speech         & Speaker Detection          & Wild & 3       & 38500            & Provides frame-level annotations where the video track includes the location of each face and active speaker annotations, while the audio track offers temporal and categorical annotations for active speaker detection. It is noted that missed face tracks and overlapping active speakers may occur.                                                                                                                                                                                  \\
    SEWA              & Speech         & Human-Activity Underst.    & Lab  & -       & 1990             & Consists of 10-30 second clips capturing the spontaneous behavior of volunteers in real-world environments using standard webcams and microphones. The SEWA DB includes richly annotated audio-visual recordings with detailed information on facial action units (FAUs), facial landmarks, vocal and verbal cues, and continuous emotion dimensions such as valence, arousal, and liking, as well as social signals like agreement and mimicry. \\
    MUSIC             & Music          & AV Source Separation       & Lab  & 11      & 685              & Consists of clean, synchronized audio-visual recordings where each audio sample is 6 seconds long.                                                                                                                                                                                                                                                                                                                                                                              \\
    MV-10K            & Music          & AV Retrieval               & Wild & -       & 10000            & Contains clips ranging from 213 to 219 seconds in length, divided into segments or "chunks" of 3 seconds each.                                                                                                                                                                                                                                                                                                                                                                                                                       \\
    AVA Actions       & Real Event     & Human-Activity Underst.    & Wild & 80      & 430              & Features videos from movies with multiple labels assigned to individuals, occurring frequently throughout. It includes spatio-temporal annotations and consists of 15-minute clips, each partitioned into 897 overlapping 3-second segments with a 1-second stride.                                                                                                                                                                                                                                              \\
    AVE               & Real Event     & AV Localization            & Wild & 28      & 4143             & Consists of 10-second clips where multiple sound sources may be present. The test set focuses solely on the temporal localization of these sounds and includes annotations for temporal boundaries.                                                                                                                                                                                                                                                                                                                \\
    EPIC-Kitchens     & Real Event     & Event Rec.                 & Wild & 149     & 39594            & Features first-person point-of-view actions in kitchen environments, with temporal annotations for each action. It includes overlapping actions and provides object bounding boxes for detailed analysis.                                                                                                                                                                                                                                                                                      \\
    Moments in Time   & Real Event     & Event Underst.             & Wild & 339     & 791297           & Consists of 3-second clips that may feature multiple sound sources. These clips are sourced from various video hosting sites, exhibiting strong inter- and intra-variations in the number of events depicted in each video.                                                                                                                                                                                                                                                  \\
    VEGAS             & Real Event     & Video-Audio Gen.           & Wild & 10      & 28109            & Features approximately 7-second clips that are unbalanced and include 10 categories manually curated from AudioSet.                                                                                                                                                                                                                                                                                                                                                                             \\
    AVA-Speech        & Speech         & Speaker Detection          & Wild & 4       & 40000            & Includes temporal and categorical annotations for speech detection, with the possibility of overlapping active speakers. Categorical labels also cover background noise, and only the audio track is annotated.                                                                                                                                                                                                                                                \\
    AVSpeech          & Speech         & Speaker Detection          & Wild & -       & 290000           & Consists of segments ranging from 3 to 10 seconds, featuring only one speaker in each segment. Collected automatically from YouTube videos, it includes clean and non-interfering speech. The visual inputs are limited to facial crops, while the audio signal provides the accompanying sound.                                                                                                                                                                                                                 \\
    RAVDESS           & Speech         & AV Vocal Expressions Rec.  & Lab  & 8+6     & 7356             & Includes 4,320 speeches and 3,036 songs, created with induced emotional expressions. It provides data in three modalities: Audio-only (16-bit, 48kHz .wav), Audio-Video (720p H.264, AAC 48kHz .mp4), and Video-only (no sound).                                                                                                                                                                                                               \\
    VoxCeleb2         & Speech         & AV Speaker Rec.            & Wild & 6112    & 1128246          & Comprises approximately 7.8-second clips for each phrase or "utterance." The audio segments include various degradations such as background chatter, laughter, overlapping speech, and varying room acoustics.                                                                                                                                                                                                                                                                  \\
    C4S               & Music          & AV Localization            & Lab  & -       & 39000            & Consists of 54 videos, each with labels for onset, salient points, regions of interest (ROIs), and the corresponding sound wave.                                                                                                                                                                                                                                                                                                                                                \\
    HIMV-200K         & Music          & Repres. Lear.              & Wild & -       & 200500           & Consists of video-music pairs, with labels extracted from YouTube-8M, specifically from videos classified as "music video."                                                                                                                                                                                                                                                                                                                                                           \\
    URMP              & Music          & Cross-modal AV Gen.        & Lab  & 44      & 17555            & Features high-quality, synchronized 0.5-second clips of multi-instrument recordings, captured in a studio setting. It provides ground truth labels for each sound source, along with the musical score in MIDI format, individual instrument audio recordings, and videos of the assembled pieces.                                                                                                                           \\
    AudioSet          & Real Event     & Audio Event Rec.           & Wild & 485     & 1789621          & Consists of 10-second clips sampled from 2 million YouTube videos. It includes audio events with their corresponding categories and the associated videos.                                                                                                                                                                                                                                                                                                                                               \\
    Kinetics-Sounds   & Real Event     & Video Action Rec.          & Wild & 34      & 19000            & Consists of 10-second clips where audio and video are not always semantically correlated, and background noise may be present in the recordings.                                                                                                                                                                                                                                                                                                                                                        \\
    YouCook2          & Real Event     & Procedure Segmentation     & Wild & 89      & 2000             & Features videos annotated with temporal boundaries and described by imperative English sentences detailing the procedure steps. Sourced from YouTube, these videos are all in a third-person viewpoint and are recorded unconstrained, with individuals performing tasks in their homes using fixed cameras. On average, each video contains 7.7 segments.                                                              \\
    LRS               & Speech         & Lip Reading                & Wild & -       & 11816            & The dataset's metadata includes bounding boxes for each face and annotations for spoken sentences, with each sentence up to 100 characters or 10 seconds long. Each word is also associated with a specific temporal annotation.                                                                                                                                                                                                                                                             \\
    Flickr-SoundNet   & Ambient Sounds & Repres. Lear.              & Wild & -       & \textgreater{}2M & Consists of unlabeled audio-video pairs, thus providing paired audio and visual content without specific annotations.                                                                                                                                                                                                                                                                                                                                                                                                                 \\
    MVD               & Music          & AV Repres. Lear.           & Wild & 8       & 2212             & This manually curated dataset includes video clips with resolutions ranging from 320x240 to 640x480, each lasting 4 minutes. It excludes live performances, abstract content, and animated videos.                                                                                                                                                                                                                                                                                                                           \\
    Charades          & Real Event     & Audio Event Rec.           & Wild & 157     & 9848             & Contains video clips of approximately 30 seconds each, labeled with multiple activities. Users were given sentences featuring objects and actions from a fixed vocabulary and recorded themselves acting out these sentences using a charades-like method. The dataset includes categorical temporal annotations for the recorded activities.                                                                                                                                                                                     \\
    MSR-VTT           & Real Event     & Event Underst.             & Wild & 20      & 7180             & Consists of video and sentence pairs, with videos categorized into 20 general categories. Each video is accompanied by textual descriptions in the metadata, and the duration of each clip ranges from 10 to 30 seconds.                                                                                                                                                                                                                                     \\
    Youtube-8M        & Real Event     & Video Underst.             & Wild & 4800    & 8264650          & Features clips with an average duration of 229.6 seconds, each containing both audio and visual modalities. Videos are categorized into 24 topics based on visual information, including categories such as sports, games, and arts \& entertainment.                                                                                                                                                                                                                                                         \\
    LRW               & Speech         & Lip Reading                & Wild & 500     & 1000             & Includes bounding boxes for each face and annotations for spoken sentences in videos that are 29 frames long (approximately 1.16 seconds). Each word in the sentences has a corresponding temporal annotation.                                                                                                                                                                                                                                                                                                 \\
    ActivityNet       & Real Event     & Human-Activity Underst.    & Wild & 200     & 14950            & Features both untrimmed and trimmed activity recognition scenarios. In the untrimmed scenario, the goal is to predict activities throughout a full video, which may include multiple activities and is annotated with timecodes. In the trimmed scenario, the task is to predict the label of a video. The dataset offers a diverse and rich taxonomy of activities, with audio and video not always being semantically correlated.                                                                                                        \\
    ESC-50            & Real Event     & Audio Event Classification & Wild & 50      & 2000             & Consists of 5-second recordings organized into 50 semantic classes, with approximately 40 examples per class. These classes are loosely grouped into 5 major categories.                                                                                                                                                                                                                                                                                                                  \\
    Vis. Ind. Sounds  & Real Event     & Video-Audio Gen.           & Wild & 17      & 977              & Features approximately 35-second clips designed to study physical interactions within visual scenes by predicting sounds from videos. It includes videos of interactions with objects, along with the corresponding matching sounds                                                                                                                                                                                                                                                         \\
    OULUVS2           & Speech         & Mouth Motion Analysis      & Lab  & 52      & 3640             & Includes utterances captured from different points of view, encompassing continuous digits, phrases, and sentences.                                                                                                                                                                                                                                                                                                                                                                         \\
    TCD-TIMIT         & Speech         & AV Speech Rec.             & Lab  & 6913    & 13826            & Features high-quality audio and video of 62 speakers reading 6,913 phonetically rich sentences. The video footage is recorded from two angles: directly from the front and at a 30-degree angle.                                                                                                                                                                                                                                             \\
    UCF-101           & Real Event     & Video Action Rec.          & Wild & 101     & 13320            & Consists of approximately 7.2-second clips with a resolution of 320x240 pixels. It features dynamic backgrounds and camera motion, with audio and video not always being semantically correlated.                                                                                                                                                                                                                                                                                                                             \\
    HMDB-51           & Real Event     & Video Action Rec.          & Wild & 51      & 6849             & Includes video and audio pairs with associated action category labels. Metadata provides additional details such as visible body parts or occlusions, camera motion, camera viewpoint, and the number of people involved in each action.                                                                                                                                                                                                                                              \\* \bottomrule
\end{longtable}
}

    \clearpage

    \section{Evaluation Metrics for AVCL}
\label{appendix:metrics}

Table \ref{tab:methods-final} illustrate the methodologies covered in this survey categorized by their main task. However, most of them provide benchmarks with different evaluation metrics, increasing the difficulty of making appropriate comparisons between methods. This section gives an overview of the evaluation metrics/procedures and discusses their usage.\par

\subsection{Generation}
\label{subsection:generation-metrics}

In \acrshort*{avcl}, generation consists of translating/generating synthetic samples of one modality given another, e.g., generating audio from visual data. Therefore, the output can be evaluated considering the information it contains (i.e., diversity and consistency) and the similarity between generated samples of the same class. Our research analyzed papers covering generation with different modalities and found little consensus. In fact, all the generation methodologies studied in Table \ref{tab:methods-final} provide results with different metrics.\par

\acrfull*{i} \cite{fanzeres2021sound} provides a measure of information for translations, and it is used to infer if the results are semantically coherent. It consists of training classifiers with synthetic data for predicting the outputs as informative or uninformative. This is a relevant attempt to infer the quality of translations because the outputs often do not correspond with real data and are not interpretable. However, this evaluation protocol consists of gathering training data with other translation methods and training different classifiers, which is time-consuming. Moreover, the risk of obtaining biased results is high.\par

\acrfull*{mrr} \cite{zhu2021learning} is an evaluation metric used for retrieval. It consists of the average value of the rank of the first correct answer in a query. In generation, the authors used it to evaluate the degree of similarity between generated samples of the same class.\par

\acrfull*{wer} \cite{mira2021end} is used to measure the accuracy of speech recognition methodologies. It consists of a numerical calculation that considers the number of substitutions, deletions, insertions, and the total number of words in an utterance for obtaining the overall word-level accuracy of reconstructed speech. In \cite{mira2021end}, it was mainly used to evaluate sound-to-text translations (i.e., speech recognition). This is a standard metric for text-level evaluation.\par

\acrfull*{ndb} \cite{gan2020foley} is a measure used to evaluate the diversity of generated sounds. It consists of clustering each cell in the log-mel spectrograms of all elements in the training set into different Voronoi cells (i.e., partition of a plane into polygons). The final score for each generated sample is calculated by counting the number of cells in which the training samples differ from the testing samples.\par

\acrfull*{fid} \cite{athanasiadis2020audio} is an evaluation metric used to compare the statistics of the generated samples with the real ones. It uses the embeddings from the last layer before the classification of deep models and returns a score that reflects the quality and diversity of generated data.\par

\subsection{Recognition}
\label{subsection:recognition-metrics}

The majority of methods for recognition tasks are based on, but not restricted to, self-supervised learning because it allows to indirectly approximate the separability of classes \cite{Reed2022WACV}. The focus of these works is the development of representation learning methods for feature extraction. Afterward, benchmarks are reported in linear and fine-tuned evaluation protocols. In the first setting, a linear classifier is trained on top of the frozen features, which is used to evaluate their separability. In the fine-tuned setting, the classifier is usually part of the feature extraction model, and the architecture is fine-tuned end-to-end. In both cases, the standard evaluation metric is the Accuracy (ACC) \cite{recasens2021broaden,ma2020active,akbari2021vatt,patrick2020multi,mm-versatile2020}, which measures the classification performance of the model.\par

\subsection{Retrieval}
\label{subsection:retrieval-metrics}

Retrieval methods report their results with standard information retrieval metrics. After submitting a query (e.g., using visual data from a given class), the goal is to evaluate the quality of the ranked list returned by the algorithms (e.g., a list of sound clips related to a visual query). The most common used metrics are \acrfull*{map} \cite{afouras2021self,cao2016correlation,zhang2021variational,zhen2019deep} and Recall@10 \cite{rouditchenko2020avlnet,wu2019unified} (recall in the top-10 retrieved items from the ranked list). Respectively, they provide standardized calculations to understand how much relevant items are concentrated in the top-ranked predictions across all queries and the proportion of relevant elements in the top K predictions of the ranked list. From Table \ref{tab:methods-final}, we observed that these methodologies provide benchmarks in supervised datasets where the goal is to either obtain relevant items from one modality in the other or to perform retrieval only using the class labels (i.e., without considering different modalities).\par

\subsection{Sound localization}
\label{subsection:slocalization-metrics}

Sound localization is the task of obtaining the location of a sound source in video frame sequences/images, where the output is usually a binary mask (i.e., activation map) indicating the location. To evaluate the performance of these methods, this location is matched with the ground truth using \acrfull*{iou}, and positive matches are usually considered to be above 50\%. After obtaining these matches, the methodologies analyzed in Table \ref{tab:methods-final} report their results using the \acrfull*{auc} and \acrfull*{ciou} \cite{chen2021localizing,hu2020curriculum}, and frame-wise accuracy \cite{lin2020audiovisual,morgado2020learning,ramaswamy2020see}.\par

\acrshort*{auc} and \acrshort*{ciou} are usually used together in a single validation procedure \cite{chen2021localizing,hu2020curriculum}. AUC is used to evaluate the performance of a classifier between classes and cIoU as a measure of the dispersion of activations in the predicted activation map that matches the ground truth (i.e., quality of the map). By assuming two classes (i.e., positive and negative) in \acrshort*{auc} and using \acrshort*{ciou} as a confidence/prediction score of the activation map, it is possible to benchmark the performance of sound localization methods. However, computing these metrics in scenarios that aim to provide performance in video frame sequences can be slow and expensive. In this case, the most commonly used metric is frame-wise ACC, obtained by computing the correct matching, using \acrshort*{iou}, over all input frames.\par

\subsection{Sound separation}
\label{subsection:sseparation-metrics}

Sound separation is the task of creating different sound representations for each sound source, thus separating one audio representation into multiple ones. In other words, starting from visual (i.e., video) or audio feeds, the goal is to obtain separate audio intermediate representations (e.g., spectrograms) for each sound source \cite{gao2019co,zhao2019sound,lin2021exploiting}. As such, the typically used benchmark metrics in this scenario calculate the distance or distortion between the ground truth representations and the predicted ones. More concretely, the most commonly used metrics are the \acrshort*{sdr} \cite{gao2019co,zhao2019sound} and the \acrfull*{stft} distance \cite{lin2021exploiting}. Respectively, they measure the Euclidean distance between representations and the amount of background noise contained in the prediction concerning the ground truth signal.\par
    \clearpage


\end{document}